\documentclass[12pt,document,nofootinbib,superscriptaddress, onecolumn,preprintnumbers,balancelastpage]{revtex4}
\pdfoutput=1
\usepackage{graphicx}% Include figure files
\usepackage{dcolumn}% Align table columns on decimal point
\usepackage{bm}% bold math
\usepackage{verbatim}
\usepackage{tabularx}
\usepackage{hyperref}
\usepackage{comment,amssymb,amsmath,latexsym} 
\usepackage{color}
\usepackage{slashed}

\def\be{\begin{equation}}
\def\ee{\end{equation}}
\newcommand{\beq}{\begin{equation}}
\newcommand{\eeq}{\end{equation}}
\def\bea{\begin{eqnarray}}
\def\eea{\end{eqnarray}}

\newcommand{\Eref}[1]{Eq.~(\ref{#1})}

\newcommand{\lsim}{ \mathop{}_{\textstyle \sim}^{\textstyle <} }
\newcommand{\vev}[1]{ \left\langle {#1} \right\rangle }

\newcommand{\GeV}{{\text{ GeV}}}
\newcommand{\TeV}{{\text{ TeV}}}
\newcommand{\half}{{\frac{1}{2}}}

\newcommand{\ifb}{{{\text{ fb}}^{-1}}}

\newcommand{\refsec}[1]{Sec.~\ref{Sec: #1}}

\newcommand{\Cmodel}[1]{{\bf {\color{#1}Benchmark ~\theregion.\themodel~}}}

\newcommand{\cc}{\text{c.c. }}
\newcommand{\fb}{\text{ fb}}
\newcommand{\pb}{\text{ pb}}
\newcommand{\cmcs}{\text{ cm}^3/\text{s}}
\newcommand{\mr}[1]{\text{#1}}
\newcommand{\st}{$^\text{st}\,$}
\newcommand{\nd}{$^\text{nd}\,$}
\newcommand{\rd}{$^\text{rd}\,$}
\newcommand{\nth}{$^\text{th}\,$}
\newcommand{\MET}{\slashed{E}_T\,\,}
\newcommand{\eg}{{\it e.g. }}
\newcommand{\OO}{\mathcal{O}}
\newcommand{\sign}{\text{ sign}}

\newcommand{\nocontentsline}[3]{}
\newcommand{\tocless}[2]{\bgroup\let\addcontentsline=\nocontentsline#1{#2}\egroup}

\definecolor{LightChi}{rgb}{0.5,0.5,0.5}
\definecolor{WT}{rgb}{0.0,0.8,0.0}
\definecolor{A}{rgb}{1.0,0.0,0.0}
\definecolor{STAU}{rgb}{0.0,0.0,1.0}
\definecolor{STOP}{rgb}{0.7,0.7,0.0}

\begin{document}
\begin{flushright}
\text{\normalsize SLAC-PUB-15468}
\end{flushright}
\vskip 65 pt

\title{Here be Dragons: The Unexplored Continents of the CMSSM}
 
\author{Timothy Cohen}
\author{Jay G. Wacker}
\affiliation{
Theory Group, SLAC National Accelerator Laboratory\\
\vskip -4 pt
Menlo Park, CA, 94025}


\begin{abstract}
\vskip 15 pt
\begin{center}
{\bf Abstract}
\end{center}
\vskip -8 pt
$\quad$
The Higgs boson mass and the abundance of dark matter constrain the CMSSM/mSUGRA supersymmetry breaking inputs.  A complete map of the CMSSM that is consistent with these two measured quantities is provided.  Various ``continents," consisting of non-excluded models, can be organized by their dark matter dynamics.  The following mechanisms manifest: well-tempering, resonant pseudo-scalar Higgs annihilation, neutralino/stau coannihilations and neutralino/stop coannihilations.  Benchmark models are chosen in order to characterize the viable regions.  The expected visible signals of each are described, demonstrating a wide range of predictions for the 13 TeV LHC and a high degree of complementarity between dark matter and collider experiments.  The parameter space spans a finite volume, which can be probed in its entirety with experiments currently under consideration.
\end{abstract}

%\pacs{XX}% PACS, the Physics and Astronomy
\maketitle

\newpage

\setcounter{tocdepth}{0}

\tableofcontents

\renewcommand{\thesection}{\arabic{section}}
\renewcommand{\thesubsection}{\arabic{section}.\arabic{subsection}}
\renewcommand{\thesubsubsection}{\arabic{section}.\arabic{subsection}.\arabic{subsubsection}}
\renewcommand{\thefigure}{\thesection.\arabic{figure}}
\renewcommand{\thetable}{\thesubsection.\arabic{table}}
\newcounter{region}
\setcounter{region}{0}
\newcounter{model}

\section{Introduction}
\label{Sec: Intro} 
The discovery of the Higgs boson \cite{Aad:2012tfa, Chatrchyan:2012ufa} confirmed the Standard Model.  But open questions remain: Why is the $W$-boson mass so far below the Planck scale? What non-baryonic substance makes up  roughly $80\%$ of the matter content for our Universe \cite{Ade:2013lta}?  Is the Standard Model prediction that the gauge couplings nearly unify at high scales a hint of new physics?

The leading framework that can address all of these outstanding issues with the Standard Model is supersymmetry (SUSY).  The Minimal Supersymmetric Standard Model (MSSM)~\cite{Dimopoulos:1981zb} predicts precision gauge coupling unification~\cite{Dimopoulos:1981yj}, has a stable particle which can freeze-out to the observed abundance of dark matter (under the assumption of a simple (thermal) cosmological history)~\cite{Drees:1992am, Barger:1997kb}, and provides a TeV scale cutoff for quadratic divergences in the Higgs mass.  

Despite all of its theoretical successes, the observed value of the Higgs mass is a challenge to accommodate inside the MSSM.  At the same time the non-observation of superpartners at the LHC calls into question the existence of low-scale supersymmetry.  Arguably, these two lessons from the LHC may be related to each other.  The tree level prediction for the Higgs boson mass $m_h$ in the MSSM is $m_{h} \le m_{Z}$, where $m_Z$ is the $Z^0$-boson mass.  However, as the superpartners become heavy there are sizable one-loop radiative corrections \cite{Okada:1990vk,Okada:1990gg,Haber:1993an,Haber:1996fp}:
\begin{eqnarray}\label{Eq: 1 loop Higgs mass}
m_{h}^2 \simeq m_{Z}^2 \cos^22\,\beta + \frac{3\,g^2\, m_t^4}{8\,\pi^2 \,m_{W}^2}\left[ \log 
\left(\frac{ 
m_{\tilde{t}_1}\,
m_{\tilde{t}_2}
}
{m_t^2}\right) + 
\frac{A_t^2}{m_{\tilde{t}_1}\,m_{\tilde{t}_2}}
\left( 1 - \frac{A_t^2}{12\,m_{\tilde{t}_1} \,m_{\tilde{t}_2}}\right)\right]
\end{eqnarray}
where $m_W$ is the $W^\pm$-boson mass, $g$ is the $SU(2)$ standard model gauge coupling, $\tan\beta$ is the ratio of the Higgs vevs, $m_t$ is the top quark mass, $m_{\tilde{t}_{i}}$ are the physical stop masses, and $A_t$ is the stop-Higgs soft SUSY breaking trilinear.

From Eq.~(\ref{Eq: 1 loop Higgs mass}), the Higgs boson mass depends on the logarithm of the stop masses; for fixed $A_t$, increasing $m_h$ requires an exponential increase in the top squark masses.  If the Higgs mass is raised from $m_{h}$ to $m_{h'}$ while keeping $A_t$ fixed,
\begin{eqnarray}
 m_{h'} - m_{h}\simeq \frac{ 3\,g^2\, m_t^4}{16\,\pi^2\, m_{h}\,m_{W}^2  } \log \frac{m_{\tilde{t}'_1} \,m_{\tilde{t}'_2}}{m_{\tilde{t}_1}\,m_{\tilde{t}_2}}
\quad \Longrightarrow  \quad 
m_{\tilde{t}'_1}\,m_{\tilde{t}'_2} \simeq m_{\tilde{t}_1}\, m_{\tilde{t}_2} \,  2^{ \frac{ \Delta m_{h}}{5.6 \GeV}}
\end{eqnarray}
This demonstrates that going from the LEP2 limit on the Higgs mass of $m_{h} \ge 114.4\GeV$ \cite{Barate:2003sz} to the observed value of $m_{h} \simeq 125\GeV$ \cite{Aad:2012tfa, Chatrchyan:2012ufa} requires quadrupling the top squark masses.  Taken at face value, the Higgs discovery has profound implications on the expectation of the mass scale for the supersymmetric particles.\footnote{An arbitrarily light $\widetilde{t}_1$ is possible with a 125~GeV Higgs with careful choices of $A_t$, $m_{q_3}^2$ and $m_{u_3^c}^2$. \cite{Hall:2011aa, Carena:2011aa}.}

One goal of the SUSY phenomenology community is to understand the consequences of $m_h \simeq 125 \GeV$ on the $\sim 120$ dimensional MSSM parameter space.  In practice this is intractable due to the immense size of the MSSM, not to mention all the possible extensions.   The resulting space of phenomenological signatures is enormous.  It is an unrealistic task to ``exclude the MSSM."   

The desire to chart all possible experimental implications of the MSSM has motivated many different approaches.  Evoking a top-down perspective, many models of the SUSY breaking parameters have been constructed.  These proposals derive the low energy parameters from far fewer inputs.  Specific frameworks tend to be highly predictive; large classes of SUSY signatures can be forbidden.  In some cases, it is conceivable to test the full parameter space.  In contrast, a bottom-up motivated reduction of the full MSSM parameters to a set of 19 phenomenologically motivated inputs was proposed \cite{Berger:2008cq, CahillRowley:2012rv}.  Even with this dramatic decrease in complexity, it is not possible to map all possible signals.  

There is a more restrictive choice which is often made when attempting to understand ``common" SUSY signatures.   This 4 dimensional slice of parameter space is known as the Constrained MSSM (CMSSM or mSUGRA) \cite{Chamseddine:1982jx, Barbieri:1982eh, Hall:1983iz}.  For some studies of the CMSSM in light of the Higgs discovery, see \cite{Baer:2011ab, Baer:2012mv, Baer:2012uya, Arbey:2011ab, Kadastik:2011aa, Buchmueller:2011ab, Buchmueller:2012hv,  Ellis:2012aa, Cao:2011sn, Akula:2012kk, Fowlie:2012im, Kowalska:2013hha, Bertone:2011nj, Strege:2011pk, Strege:2012bt, Ghosh:2012dh, Dighe:2013wfa}.  This {\em ansatz} is defined by four parameters and a sign which are delineated at the scale $M_\mr{GUT}$ where the gauge couplings unify:  a universal scalar mass $M_0$, a universal gaugino mass $M_\half$, a universal scalar-trilinear coupling $A_0$, and the $B_\mu$-term (usually set by choosing $\tan \beta$) along with the sign of $\mu$.   These high scale inputs are evolved to the weak scale using the renormalization group and the $\mu$-term is chosen to reproduce the measured value of the $Z^0$-boson mass.

The Higgs boson mass $m_h$ and the dark matter thermal relic density $\Omega\,h^2$ can be calculated in this framework.  Matching the predictions for $m_h$ and $\Omega\,h^2$ to the measured values constrains the four-dimensional parameter space.  The goal of this article is to provide a map of the CMSSM regions\footnote{For a previous attempt to map the full CMSSM, see \cite{Baltz:2004aw}.} which are consistent with these two requirements.  Additional constraints such as evading LHC searches and limits from direct and indirect dark matter detection experiments will also be discussed.  

The CMSSM parameter space is compact.  Stated precisely, every direction in parameter space is bounded using four minimal assumptions \cite{Kane:1993td}: 
\begin{itemize}
\item The Higgs mass is less than $128\GeV$.
\item The lifetime of the Universe is longer than the observed value.
\item The bottom quark Yukawa coupling remains perturbative to the unification scale.
\item The LSP is a thermal relic that does not overclose the Universe.
\end{itemize}
One purpose of this article is to quantify the extent of the CMSSM.

With the exception of the last, these are unarguable constraints on the CMSSM.  There are several possibilities beyond a neutralino WIMP $\chi$ --- $R$-parity violation would cause the Lightest SuperPartner (LSP) to decay \cite{Barbier:2004ez}; a non-trivial cosmological history such as a late-stage entropy production from moduli decay or a low reheat temperature could alter the freeze-out prediction \cite{Moroi:1999zb, Gelmini:2006pw, Acharya:2009zt}.  %Nevertheless, the last constraint is a plausible possibility that does not require altering Big Bang cosmology after the inflationary reheating of the Universe.
Nevertheless, requiring a thermal history and stable $\chi$ require no additional assumptions and together stand as one of the main motivators for the MSSM in the first place.  This scenario is incorporated in all that follows. 

The full extent of the CMSSM will be demonstrated.  It will be argued that it is in principle possible to experimentally access all of it.  However, the range of allowed masses for the supersymmetric particles extends significantly further than is usually discussed.  While tremendous progress toward excluding this slice of the MSSM (or discovering something like it) will be made, there exist regions which will remain beyond the combined reach of the 13~TeV LHC, 1 ton scale direct detection experiments, and telescopes which target gamma rays from dark matter annihilations.  One goal of this work will be to enumerate what will remain of the CMSSM once near-term experiments have competed their searches.  The entire CMSSM parameter space can be probed using experiments currently under consideration.  Most of the parameter space can be reach with  the 33~TeV HE-LHC and the remaining regions can be completely covered with the 100~TeV VHE-LHC.  

As is often done, the different islands of the CMSSM will be classified by the (dominant) process which sets the relic density.  Several mechanisms manifest:
\begin{itemize}
\item Light $\chi$ --- $Z^0$ and $h$ pole annihilation determines the relic density.  This channel is active for dark matter masses $m_\chi \lsim 70 \GeV$.  Since the LSP is dominantly bino and gaugino masses unify, there is a corresponding bound on the gluino mass of $ m_{\tilde{g}}\simeq 450 \GeV$.  This region has been excluded by LHC7 searches for gluinos.
\item Well-tempered --- the dark matter has a non-trivial Higgsino component \cite{ArkaniHamed:2006mb}.  This is the home of the ``focus point" supersymmetry region \cite{Feng:1999mn,Feng:2000gh,Feng:2011aa,Feng:2012jfa}.  Most of this region can be probed using 1 ton scale direct detection \cite{Cohen:2010gj, Cheung:2012qy, Draper:2013cka}.  Once the lightest neutralino masses reach $\OO(1\TeV)$, the relic density requirement forces a dominantly Higgsino ad-mixture; in this limit, the tree-level direct detection cross section becomes suppressed, although it remains within reach of multi-ton scale experiments.
\item $A^0$-pole annihilation\footnote{Throughout the universal $A$-term CMSSM input parameter will be referred to as $A_0$ and the pseudo-scalar Higgs as $A^0$.} --- the strongest dark matter annihilation channel is through an $s$-channel pseudo-scalar Higgs resonance.  This region tends to have heavy colored superpartners which limits the ability of the 13~TeV LHC to explore this entire region.  Direct detection can also be highly suppressed.  There is some hope for indirect detection since the annihilation cross section today is also dominated by $s$-channel $A^0$ exchange yielding a $b\,\overline{b}$ final state.
\item Stau coannihilation --- this is being tested by a combination of searches for colored particles and direct detection.  In some of this parameter space the staus decay length becomes macroscopic; it can be useful to search for charged tracks or displaced vertices to test these models~\cite{Citron:2012fg}.
\item Stop coannihilation \cite{Ellis:2001nx} --- this region is largely untested.  Furthermore, much of this parameter space will remain even after the full run of the 13~TeV LHC.
\end{itemize}

All of these annihilation mechanisms have been previously discovered within the CMSSM parameter space;  however, the number of disconnected regions existing in the post-Higgs discovery era has not been discussed.  Furthermore, many studies are based on specific slices (one common strategy is to fix $A_0$ and $\tan \beta$ and explore the $M_\half - M_0$ plane).  While this can be an instructive exercise, it can lead to incorrect inferences about the general predictions of the CMSSM {\em ansatz}.  This work will serve to clarify many of these issues.

There exists a large literature on mapping the CMSSM where a likelihood value is applied to each point in parameter space, see for example \cite{Baltz:2004aw, Allanach:2005kz, deAustri:2006pe, Akrami:2009hp, Trotta:2008bp, Bridges:2010de, Feroz:2011bj, Buchmueller:2009fn}.  These approaches are extremely powerful and can allow for exhaustive studies of high dimensional parameter spaces.  In order to compute the likelihood,  a variety constraints with associated error bars must be compiled.  This can lead to conclusions about the ``allowed" parameter space which are driven by measurements on quantities such as $(g-2)$ of the muon, $b\rightarrow s\,\gamma$, and so on.  For example, stop coannihilation was shown to be strongly disfavored using these methods \cite{Allanach:2005kz}.  The approach taken here does not attempt to assign any statistical significance to any point in parameter space.  This draws a distinction between the results presented here and these other studies, \emph{e.g.} we determine that large regions of stop coannihilation are allowed.  Therefore, this work provides a compliment to the existing literature.

The organization of the paper is as follows. Sec.~\ref{Sec: Cartography}  provides our map of the viable CMSSM parameter space.  Sec.~\ref{Sec: Method} discusses the specifics of how the spectra and processes are calculated from the CMSSM inputs.  Sec.~\ref{Sec: Regions} provides an in depth look at the separate regions and discusses their properties including detailed descriptions of of the expected first signals. Sec.~\ref{Sec: Discussion} summarizes our findings and gives a rough idea of what regions will remain unexplored after the full run of the 13~TeV LHC and ton scale direct detection experiments.  An appendix is given which provides the details of how to reproduce our maps of the CMSSM using files which are available on the arXiv.  Also available on the arXiv are the relevant cross sections and decay tables for all of the presented benchmark models.

\section{CMSSM Cartography}
\label{Sec: Cartography}

This section  presents four two-dimensional slices of the CMSSM parameter space which allow one to infer the location of each continent in the coordinates $M_0$, $M_\half$, $A_0$, $\tan\beta$, and the sign of $\mu$.  

These figures impose the constraints $122\GeV < m_h < 128 \GeV$ and $0.08  < \Omega\,h^2 < 0.14$.  A 3 GeV error bar for the Higgs mass is used; this is an estimate of the uncertainty in the theoretical calculation \cite{Allanach:2004rh}.  The spread in allowed dark matter relic density is taken to account for the $\OO(10\%)$ uncertainty in the calculation of $\Omega \,h^2$,  \eg from the fact that only two-to-two tree processes are included when computing this quantity.  
A naive bound on charginos, $m_{\chi^\pm} > 100 \GeV$ \cite{LEPSUSYWG/02-05.1} is imposed to avoid issues when there are multiple  $\mu$ solutions to the $m_Z(\mu) = m_{Z_\text{SM}}$  \cite{Allanach:2013cda}  (see Sec.~\ref{Sec: Scan Strategy}).   Currently \texttt{SoftSUSY v3.3.7} typically picks the larger values, but  extremely small values of $\mu \simeq 3\GeV$ are sporadically found resulting from these multiple solutions to the $\mu$ term.   These disconnected solutions are safely excluded by the LEP2 constraint and are uninteresting from a phenomenological point of view.  

\begin{figure}[h!]
\centering
\includegraphics[width=0.6\textwidth]{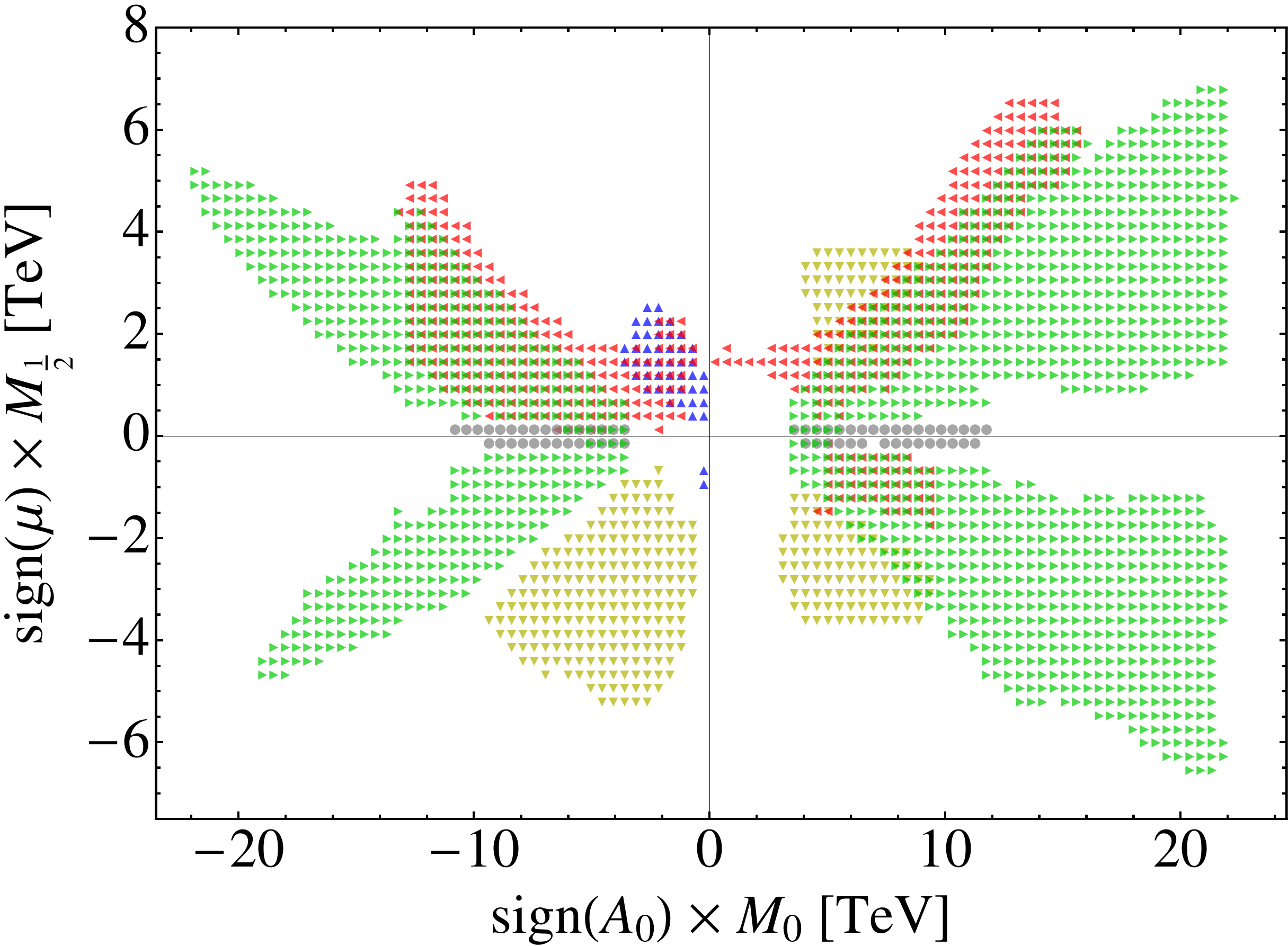}
\caption{
A map of the full CMSSM projected into the $\text{sign}(\mu) \times M_\half$ versus $\text{sign}(A_0)\times M_0$  plane.  The SM-like Higgs boson mass and dark matter relic density are constrained to their measured values.  \emph{No LHC or direct detection bounds have been applied.}  The regions are demarcated by their dominant dark matter annihilation channel: light $\chi$ [grey; circles], well-tempered [green; right pointing triangles], stau-coannihilation [blue; upward pointing triangles], $A^0$-pole annihilation [red; left pointing triangles], stop-coannihilation [yellow; downward pointing triangles].
}
\label{Fig: FullSpaceA}
\vspace{22pt}
\centering
\includegraphics[width=0.6\textwidth]{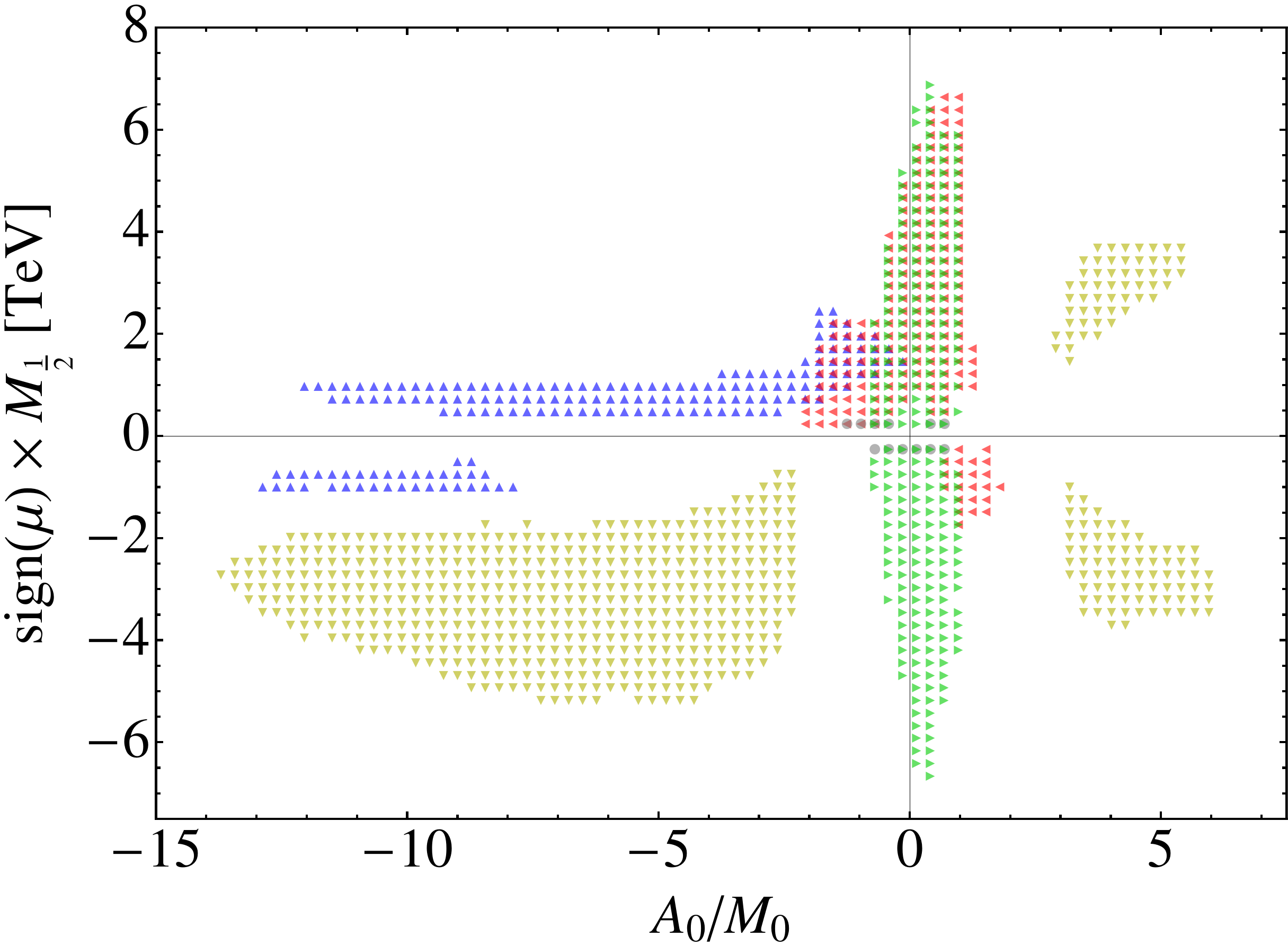}
\caption{A map of the full CMSSM projected into the $\text{sign}(\mu) \times M_\half$ versus $A_0/M_0$ plane.  See the caption of Fig.~\ref{Fig: FullSpaceA} for details.}
\label{Fig: FullSpaceB}
\end{figure}

\begin{figure}[h!]
\centering
\includegraphics[width=0.64\textwidth]{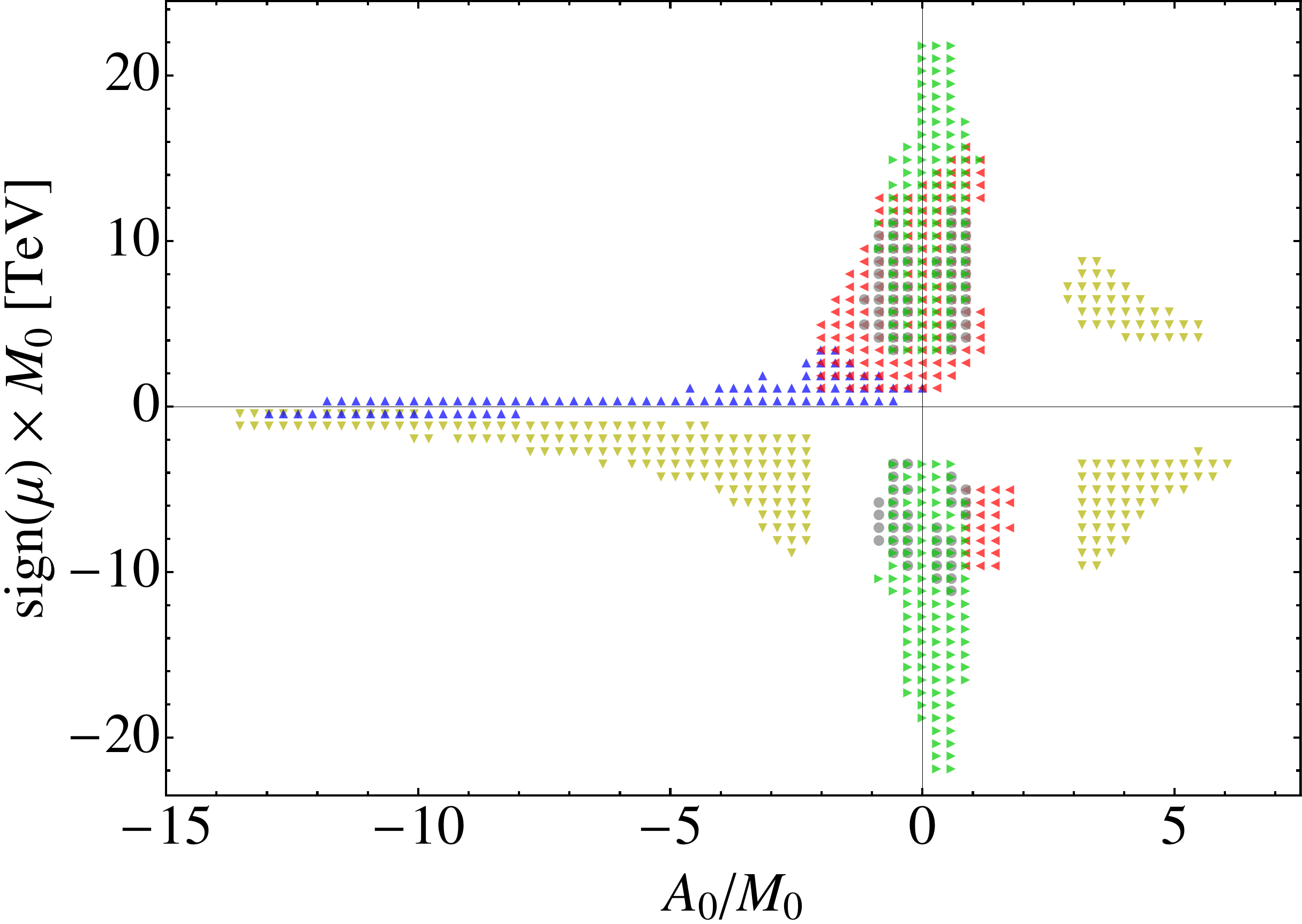}
\caption{
A map of the full CMSSM projected into the $\text{sign}(\mu) \times M_0$ versus $\text{sign}(A_0)\times M_0$ plane. The SM-like Higgs boson mass and dark matter relic density are constrained to their measured values.  \emph{No LHC or direct detection bounds have been applied.}  The regions are demarcated by their dominant dark matter annihilation channel: light $\chi$ [grey; circles], well-tempered [green; right pointing triangles], stau-coannihilation [blue; upward pointing triangles], $A^0$-pole annihilation [red; left pointing triangles], stop-coannihilation [yellow; downward pointing triangles].
}
\label{Fig: FullSpaceC}
\end{figure}

\begin{figure}[h!]
\centering
\includegraphics[width=0.64\textwidth]{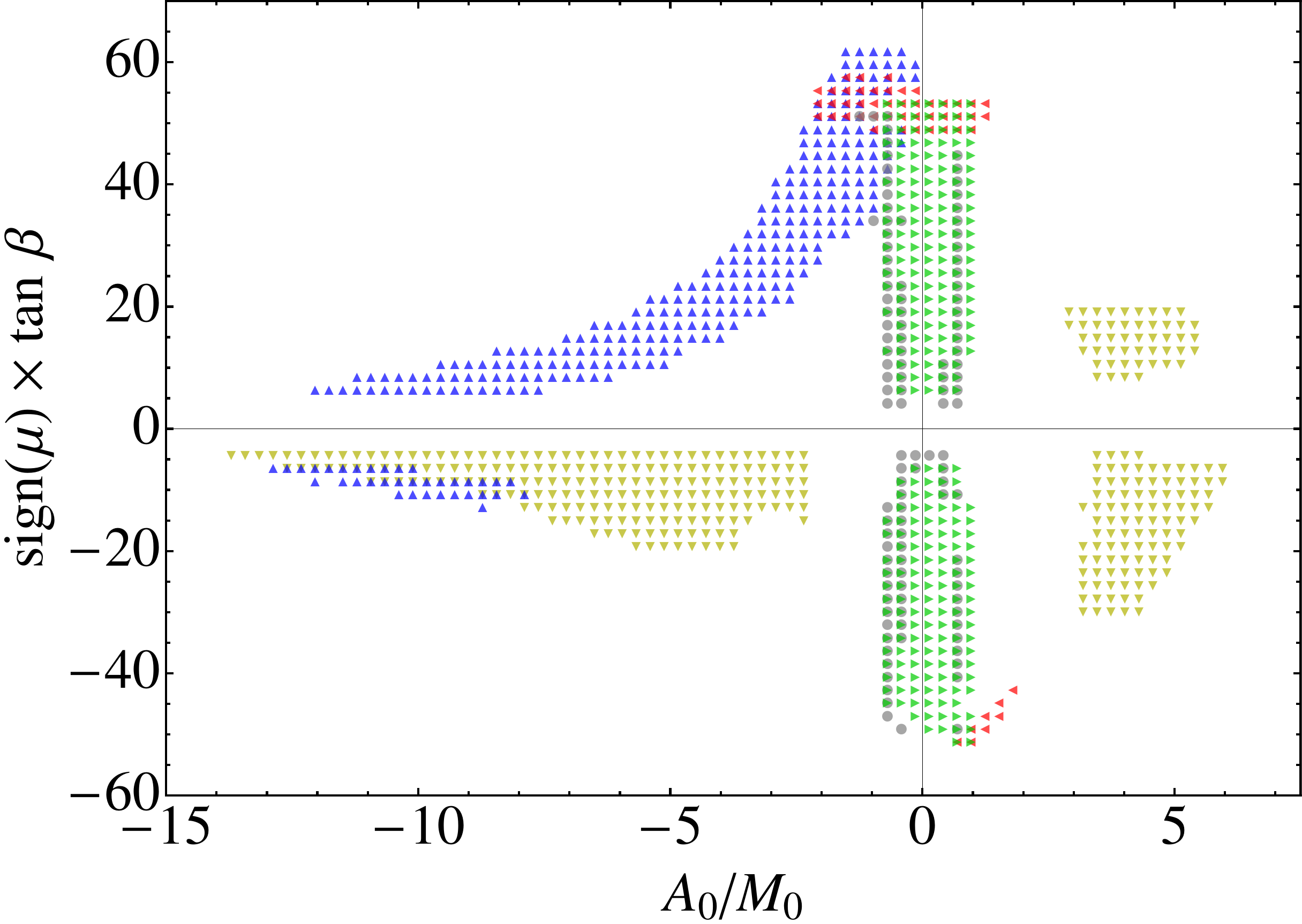}
\caption{A map of the full CMSSM projected into the $\sign(\mu) \times \tan \beta$ versus $A_0/M_0$ plane.  See the caption of Fig.~\ref{Fig: FullSpaceC} for details.}
\label{Fig: FullSpaceD}
\end{figure}

 Fig.~\ref{Fig: FullSpaceA}  presents the canonical $M_\half$ versus $M_0$ plane unfolded into four quadrants:
\begin{eqnarray}
\renewcommand{\arraystretch}{1.1}
\begin{array}{c|c}
\quad\quad  \text{Quadrant 2:  } A_0 <0 \text{ and } \mu >0 \quad\quad &\quad \text{Quadrant 1: } A_0 >0 \text{ and }  \mu >0 \quad\quad \\
\hline
\quad\quad \text{Quadrant 3: } A_0 <0 \text{ and } \mu <0 \quad\quad & \quad \text{Quadrant 4: } A_0 >0  \text{ and } \mu <0 \quad\quad \nonumber
\end{array}
\end{eqnarray}
This classification will be utilized for the rest of this paper and will be motivated physically in \refsec{Method}.  Fig.~\ref{Fig: FullSpaceB}  shows the full parameter space in the $\text{sign}(\mu) \times M_\half$ versus $A_0/M_0$ plane.  This plot illustrates that a wide range of viable parameter space exists for large values of $A_0$ with respect to $M_0$.  Figs.~\ref{Fig: FullSpaceC} and \ref{Fig: FullSpaceD} plot the complimentary planes involving $M_{0}$ versus $A_0/M_0$ and $\tan \beta$ versus $A_0/M_0$ respectively.

As described in \refsec{Intro}, the regions  are separated by their dominant dark matter annihilation channel.  Throughout this article, these continents are denoted in all figures in the paper using the following scheme:  light $\chi$ [grey; circles], well-tempered [green; right pointing triangles], stau coannihilation [blue; upward pointing triangles], $A^0$-pole annihilation [red; left pointing triangles], stop coannihilation [yellow; downward pointing triangles].  

In practice, to determine if a point should be classified into one of these categories, the following scheme is employed:
\begin{enumerate}
\item if $m_\chi < 70 \text{ GeV} \quad \Longrightarrow\quad$ light $\chi$
\item else if $\text{Diff}(m_\chi,\,m_{\tilde{t}_1}) < 0.2 \quad \Longrightarrow\quad$ stop coannihilation
\item else if $\text{Diff}(2\times m_\chi,\,m_{A^0}) < 0.4 \text{ and } \sigma_\text{ann} v> 2 \times 10^{-27}\cmcs \;\;  \Longrightarrow\;\;$ $A^0$-pole annihilation
\item else if $\text{Diff}(m_\chi,\,m_{\tilde{\tau}_1}) < 0.2 \quad \Longrightarrow\quad$ stau coannihilation
\item else if $|Z_B|^2 < 0.9 \quad \Longrightarrow\quad$ well-tempered
\end{enumerate}
where $Z_B$ is the bino-LSP mixing angle and
\be
\text{Diff}(m_a,\,m_b) \equiv \frac{| m_a - m_b |}{\text{min}(m_a,\,m_b)}.
\ee
All CMSSM points which were generated and satisfied the Higgs mass and relic density constraints fall into one of these categories.  While we have tested that this scheme matches closely with the actual processes that contribute to the neutralino annihilation cross section in the early Universe, there are some overlapping regions where the actual classification of a point is ambiguous, \emph{e.g.} the cut on $\sigma_\text{ann} v$ in step 3 above is to separate the overlapping $A^0$-pole annihilation and stau coannihilation regions in the second quadrant.  This will not have a qualitative impact on any of our conclusions below.

Figs~\ref{Fig: FullSpaceA}-\ref{Fig: FullSpaceD} can be utilized to navigate the viable parameter space.  Many disconnected continents are apparent.  The light $\chi$ and well-tempered regions are characterized by a narrow interval around $A_0/M_0 = 0$.  The $A^0$-pole annihilation region can be distinguished from well-tempered points by larger values of $\tan \beta$ or larger $|A_0/M_0|$.  Both the stau and stop coannihilation islands exist for $-15 \lsim A_0/M_0 \lsim 1$.  However, the former exists for smaller values of $M_\half$ and the later manifests for $2 \lsim A_0/M_0 \lsim 6$.  

 The next section discusses the specifics of the assumptions and the tools used to make these plots.    \refsec{Regions} discusses the observable consequences of each CMSSM continent including benchmarks which exemplify how one would search for these classes of models.

%\pagebreak

%\pagebreak

\vspace{30pt}

\section{Approach}
\label{Sec: Method}
This section discusses the assumptions made in this study.  The task of mapping the full parameter space required developing a novel scan strategy which will be discussed below.  This section also explains why unfolding the parameter space into $(\text{sign}(A_0),\, \text{sign}(\mu))$ quadrants leads to a clean presentation of the results.

\texttt{SoftSUSY v3.3.7} is used to evolve the CMSSM input parameters from the unification scale down to the weak scale \cite{Allanach:2001kg} using the Renormalization Group Equations (RGEs).  The 2-loop flags for MSSM RGEs and for the Higgs effective potential calculation are used since these provide accurate, validated calculations of the low energy CMSSM spectra.  For completeness, an example input card is given in the appendix.  In the conventions taken by \texttt{SoftSUSY}, an on-shell mass is computed using the two-loop massless $\overline{\text{DR}}$ scheme, including one-loop $\overline{\text{DR}}$ finite corrections \cite{Pierce:1996zz} which are added at the scale $M_S = \sqrt{m_{\tilde{t}_1} m_{\tilde{t}_2}}$.  

Given the low energy spectrum, the lightest CP even Higgs boson mass is constrained to lie between 
\begin{eqnarray}
122 \GeV < m_h < 128 \GeV,  
\end{eqnarray}
which is based on a $3\GeV$ uncertainty estimate \cite{Allanach:2004rh}.  One non-trival contribution to this range is the error bar on the measured value of the top quark Yukawa coupling.  In addition, there are 3-loop investigations \cite{Martin:2007pg, Harlander:2008ju, Kant:2010tf} which claim an $\OO(\!\GeV)$ shift in $m_h$ with respect to the 2-loop Higgs mass calculation.

Each low energy spectrum is then fed through \texttt{DarkSUSY v5.1.1} \cite{Gondolo:2004sc} which gives the the relic density, direct detection, and indirect detection cross sections for each point in parameter space.  A variety of cross checks with \texttt{MicrOmegas v.2.4.5} were performed \cite{Belanger:2010gh, Belanger:2004yn, Belanger:2001fz}.  There are a few differences between these programs.  \texttt{DarkSUSY} takes all the Standard Model parameters from the \texttt{SoftSUSY} output while \texttt{MicrOmegas} uses internal values.  This can cause an $\OO(10\%)$ variation for the light dark matter points since the annihilation to $b$ quarks through an $s$-channel $Z^0$ is sensitive to the $b$ Yukawa coupling (due to helicity suppression).  Another case where the two programs differ is when $A^0$-pole annihilation dominates.  This occurs because both programs compute the width of the $A^0$ internally and they tend to disagree on this value by $\OO(10\%)$.  Typically \texttt{MicrOmegas} matches the output from \texttt{SUSYHIT} \cite{Djouadi:2006bz} more closely than \texttt{SoftSUSY}.  None of these differences have a qualitative impact on the results and \texttt{DarkSUSY} is used for all relic density and direct detection results.

\subsection{Scan Strategy}
\label{Sec: Scan Strategy}

Building the maps presented in Figs.~\ref{Fig: FullSpaceA}-\ref{Fig: FullSpaceD} required a more targeted computational strategy than simply randomly scanning.  The results in this article began from an extensive random scan of the input parameter space.  This ``seed" scan was performed until a few points on all continents were discovered.  Many of the regions had partner disconnected components in the other quadrants and this helped discover several of the continents.  Ultimately, the discovery of any isolated region is limited by this original seed scan and there is no way to guarantee that every island was discovered.  Nevertheless, $\OO(10^7)$ random points were attempted with the following bounding box
\begin{eqnarray}
 0\le M_0 \le 10 \TeV; \quad  0 \le M_{\half} \le 10\TeV; \quad\quad\quad \nonumber \\
   -6 \le A_0/m_0 \le 6; \quad  1.5 \le \tan\beta \le 50; \quad \text{sign}(\mu) = \pm 1,
\end{eqnarray}
which limits the size of the undiscovered regions which limits the size of any islands not discovered at the 95\% confidence level to be smaller than
$$  \Delta M_0\times  \Delta M_\half \times \Delta \frac{A_0}{M_0} \times \Delta \tan \beta \le 0.036 \TeV^2.$$  
The smaller islands of parameter space are more susceptible to the implicit uncertainty in the numerical calculations used to identify the regions, which limits the potential relevance of searching for arbitrarily small regions.   This article focuses on finding the full extent of the large regions of parameter space and the methods below use the continuity of the parameter space to find the range in the parameter space.

 A variety of methods were used to extend the 4-dimensional parameter space after the random seed scans.  For example, a useful method of filling in sparsely populated parameter space is to take two valid points; draw lines connecting the values of each set of parameters for each of these points; and perform a scan which was restricted to these lines, both between the original points and extrapolated beyond them in either direction.  

Once this seed data was established, the remaining parameter space was filled in with a more efficient algorithm.  The key was to target a specific slice which would ultimately be scatter plotted.  Given a two-dimensional plane, a uniformly spaced grid can be applied to the plot and all viable CMSSM models then associated to a grid point.  All squares on the grid which contain at least a single viable point would be filled in.  Any empty point in the grid with two or more nearest neighbors would be attempted.  Keeping two of its coordinates fixed by their position in the grid, the other two directions were randomly scanned using the parameters of the filled neighbors to determine the range.  In order to ensure that the boundary was being appropriately sampled, this range was extended by $\OO(10\%)$ beyond the minimum and maximum value of its bounding neighbors.  The number of attempts to fill an empty square was proportional to the number of neighbors --- a point with more filled neighbors would be more likely to itself be valid.  Once a point was found, the grid was updated and the algorithm continued.  Besides filling in the bulk of the continents, this strategy allowed for a systematic test of all discovered boundaries.  

Recently, it was pointed out that public MSSM spectrum generators including \texttt{SoftSUSY} do not provide unique solutions to the RGEs \cite{Allanach:2013cda}.  This is a pathology of the algorithm used by these programs, namely that boundary conditions are imposed at three different scales, $m_Z,\,\sqrt{m_{\tilde{t}_1}\,m_{\tilde{t}_2}}, \,\text{ and } M_\mr{GUT}$.  In fact, this behavior was found in the scans in the light $\chi$ region; by imposing the 100 GeV bound on the charginos mass these points with spurious behavior were removed.\footnote{We thank Ben Allanach for discussion on this point.}  Given the extent of the scans, all physically relevant regions of the low energy parameter space have been explored.  However, the caveat remains that there could be additional regions which are not found using the default implementation of \texttt{SoftSUSY}.

\subsection{Quadrants}
When visualizing the allowed regions,  it instructive to unfold the parameter space by weighting the $x$ and $y$ axes by the signs of $A_0$ and $\mu$ respectively. The canonical convention is that the gaugino masses and Higgs bilinear soft mass $B_\mu$ are positive. This yields four distinct phase combinations that are specified in terms of $\!\!\sign(\mu)$ and $\!\!\sign(A_0)$.   It would be more economical to take $\mu>0$ and allow $M_\half$ to take either sign.  However, as this is just a $U(1)_R$ rotation of the standard choice, there would result a sign flip redefinition for $A_0$.  In order to avoid confusion with the existing extensive literature, the standard sign convention is maintained and the explicit signs of $\mu$ and $A_0$ are kept explicit throughout.

The 1-loop the renormalization group equations for the $A$-terms and the $B$-term take the following schematic form
\begin{eqnarray}
16\,\pi^2 \frac{d}{dt} A &\sim& A\,\left( |y|^2 - g^2 \right) + y\, g^2 \,M, \label{eq:ATermRGE}\\
 16\,\pi^2 \frac{d}{dt} B &\sim& B\,\left( |y|^2 - g^2\right) + \mu\, \left( A\, y^\dagger + g^2 \,M\,\right),
\end{eqnarray}
where $M$ is a gaugino mass, $y$ is a supersymmetric Yukawa coupling associated with the $A$-term, $g$ is a gauge coupling, and $t$ is the log of the renormalization scale \cite{Martin:1997ns}.  Clearly each choice of phase leads to an independent renormalization group trajectory.  The physics between quadrants are not simply related to each other. 

As an example, consider the case when $A \gg 0$ and $|y|^2 > g^2$.  Then Eq.~(\ref{eq:ATermRGE}) implies that the presence of a non-zero gaugino masses will suppress the $A$-term as it is evolved to lower energies --- the magnitude of the low energy $A$-term will be smaller than the unification scale input.  This should be contrasted with $A \ll 0$ where the magnitude grows as it is evolved to the weak scale.  

If $M_0$ is small and $\mu \,A>0$ $\left(\mu \,A<0\right)$, the $B$-term is suppressed (enhanced) at low energy.  Given a set of inputs which yield broken electroweak symmetry breaking, changing the sign of $\mu$ can result in low energy parameters which violate the requirement of a stable, non-zero electroweak vacuum expectation value.

These two considerations motivate designating quadrants.  In practice, one can see in Figs.~\ref{Fig: FullSpaceA} - \ref{Fig: FullSpaceD} that when the CMSSM is plotted this way the various regions with different dark matter properties have smooth boundaries.

\subsection{Metastability}
\label{Sec: Metastability}

$A$-terms can play a non-trivial role in the phenomenology of the CMSSM.  When trilinear scalar couplings are large, unstable directions open up in the scalar potential.  These new vacua are color and charge breaking and are therefore not phenomenologically viable. Working in the $D$-flat direction $\vev{H_u} = \vev{\widetilde{q}}= \vev{\widetilde{u}^c}$, it is straightforward to find a constraint on the $A$-term such that the color and charge breaking minima are not the absolute minima of the potential.  For the top squark this bound on $A_t$ is \cite{Casas:1997ze, Abel:1998wr}:
\begin{eqnarray}
|A_t|^2 < 3\,\left(m_{\tilde{q}_3}^2 +m_{\tilde{u}^c_3}^2 + (m_{H_u}^2 + |\mu|^2)\right).
\label{Eq: Absolute Stability}
\end{eqnarray}
Since $m_{H_u}^2 + |\mu|^2 \sim m^2_{Z}$, Eq.~(\ref{Eq: Absolute Stability}) implies $|A_t| \lsim 6 \sqrt{m_{\tilde{t}_1}\,m_{\tilde{t}_2}}$ to good approximation.  Notice that this allows for $A_t/m_{\tilde{t}} \simeq \sqrt{6}$ which is the condition for ``maximal mixing" in the top squark sector.  There is an analogous condition for the stau direction in the scalar potential.
  
The constraint in Eq.~(\ref{Eq: Absolute Stability}) is too restrictive.   Absolute stability is not a sufficient requirement.  As long as the tunneling rate from the standard model vacua to the color and charge breaking minima is longer than the age of the Universe, the theory is phenomenologically viable.  This more complicated condition does not yield a simple analytic constraint \cite{Falk:1995cq, Falk:1996zt, Kusenko:1996jn}.  However by performing a scan over a limited subset of the CMSSM parameter space, it has been argued that the metastability requirement relaxes the bound to \cite{Kusenko:1996jn}
\be
|A_t|^2 < \left(7.5\,m_{\tilde{q}_3}^2 + 7.5\,m_{\tilde{u}^c_3}^2 +3\, (m_{H_u}^2 + |\mu|^2)\right).
\label{Eq: Meta Stability}
\ee
This article will use this less restrictive requirement, though most regions satisfy the absolute stability bound.  

The charge/color breaking minima are typically close to the origin in field space.  Therefore it is appropriate to evaluate this condition at low energies.  In practice  the $\overline{DR}$ values from \texttt{SoftSUSY} are evaluated at the scale $M_S=\sqrt{m_{\tilde{t}_1}\,m_{\tilde{t}_2}}$ for checking the condition in Eq.~(\ref{Eq: Meta Stability}).

\renewcommand{\thefigure}{\thesubsection.\arabic{figure}}
\section{The Multiple Continents}
\label{Sec: Regions}

The CMSSM contains numerous disconnected regions of parameter space.  This article classifies each continent by dark matter annihilation mechanism and quadrant.  The purpose of this section is to present several benchmark models.  These are chosen to exhibit some of the distinctive signatures which are possible within each CMSSM region.  Additional, data files provided with the {\tt arXiv} submission give a set of CMSSM inputs which can be used to reproduce all of the points in the plots.  

The goal is to give a rough picture for how to discover any point within the entire CMSSM.  Given the scope of this task, only the roughest description of the phenomenology is presented.   Many constraints, which are traditionally used to explore the CMSSM, are neglected, such as $B\rightarrow s\,\gamma$, $(g-2)_\mu$, $B_s\rightarrow \mu^+\,\mu^-$, etc.   Furthermore, since the majority of the presented spectra are quite heavy, the CMSSM tends to match the Standard Model for these predictions.  The primary interest here is exploring the variety of discovery modes.

%For all the benchmarks discussed below the spectrum is computed with \texttt{SoftSUSY} v3.3.7~\cite{Allanach:2001kg} which performs two loop renormalization group running from the unification scale to the scale of soft masses.  The properties of dark matter are computed with \texttt{DarkSUSY} v5.1.1~\cite{Gondolo:2004sc} often with cross checks using \texttt{MicrOmegas} v.2.4.5~\cite{Belanger:2010gh, Belanger:2004yn, Belanger:2001fz}.

Many benchmark models presented below will have observable LHC signatures.  To demonstrate these claims quantitatively requires knowledge of the dominant production cross sections and most visible branching ratios.   \texttt{Prospino v2.1} \cite{Beenakker:1996ed, Beenakker:1996ch, Beenakker:1997ut, Beenakker:1999xh} is used to compute NLO cross sections at $\sqrt{s} = 13 \TeV$ (and $\sqrt{s} = 33 \TeV$ for the contours presented in the squark-gluino Simplified Model planes) and \texttt{SUSYHIT v1.3} \cite{Djouadi:2006bz} is used to compute the decay tables.  

Cascade decays have relevant mass scales beyond the parent particle and LSP masses.  For a one-step cascade decay, \emph{i.e.}, one which proceeds by emitting an additional particle, a useful dimensionless variable to describe the amount of phase space available is \cite{Alwall:2008ve,Alwall:2008va}
\be
r \equiv \frac{m_I - m_D}{m_P - m_D},
\ee
when the parent particle $P$ decays into an intermediate particle $I$ and a Standard Model state, followed by the decay of $I$ into another Standard Model state and the daughter particle $D$.  The range of $r$ is 
\be
0\le r\le 1,
\ee
where $r=0$ corresponds to the intermediate particle and daughter particle being degenerate and $r=1$ corresponds to the intermediate particle and parent particle being degenerate.  A canonical example is 
\be
\widetilde{q} \rightarrow \chi_1^+ \, q' \rightarrow \chi_1^0 \,W^+\,q' \quad\Longrightarrow \quad r = \frac{m_{\chi_1^+}-m_{\chi_1^0}}{m_{\tilde{q}} - m_{\chi^0_1}}.
\ee
In what follows, the values for $r$ are specified for all cascade decays.  This allows easy comparison of the benchmarks with Simplified Model results from the LHC collaborations.

One ton scale spin-independent direct detection are projected to reach \cite{Aprile:2002ef, Li:2012zs, Akerib:2012ys, Brink:2005ej, DMTools}
\begin{eqnarray}
\sigma_\text{SI}^{\text{1\,ton}} 
\sim 10^{-11}\pb 
\quad\quad \text{at} \quad\quad 
m_\chi = 300 \GeV.
\end{eqnarray}
In the following discussion we will use the projected limit obtainable for a one ton Xenon experiment from \cite{DMTools} to estimate the future reach of direct detection.

One caveat to consider when comparing direct detection limits to predictions is the range for the plausible size of the Higgs-nucleon effective Yukawa coupling.  This can imply up to an order of magnitude variation in the predictions for direct detection \cite{Ellis:2008hf, Giedt:2009mr}.  The main point of contention is the determination of the strange quark content of the nucleon.  There is a discrepancy when comparing lattice determinations with the value derived from a combination of chiral perturbation theory and measurements of the pion-nucleon scattering sigma term.  It is worth noting that there seems to be a consensus among the lattice community \cite{Junnarkar:2013ac}.  For concreteness, we take the default values in \texttt{DarkSUSY} --- if we had used the lattice values instead our predictions for the spin-independent cross section would be a factor of a few lower.  This should not have a qualitative impact on any of the statements we make below.

This article will not emphasize the fine-tuning associated with any benchmark because the tolerance of fine tuning is a subjective preference.  Since the entire CMSSM augmented with plausible theoretical constraints is bounded in all directions, it is not necessary to impose a bound on fine tuning.  However, since it is of general interest to the community values of the canonical Barbieri-Giudice tuning measure \cite{Barbieri:1987fn} 
\be
\label{EQ: BG tuning measure}
\Delta_v \equiv \text{max}\left(\frac{\partial \ln m_Z}{\partial \ln X}\right),
\ee
where $X \in \{M_0,\,M_\half,\,A_0,\,B_\mu\}$ will be provided.  We use the built in \texttt{SoftSUSY} routines to compute this value.  We find that $250 \lsim \Delta_v \lsim 60000$ for viable points (before applying any LHC or direct detection bounds).

We also present the fine-tuning associated with the relic density $\Delta_\Omega$.  This must be considered when discussing naturalness given that this is an ``orthogonal" tuning to $\Delta_v$.  Furthermore, some of the regions require a conspiracy in the spectrum to reproduce the observed value of $\Omega\, h^2$.  Analogous to Eq.~(\ref{EQ: BG tuning measure}) we define
\be
\Delta_\Omega \equiv \text{max}\left(\frac{\partial \ln \Omega\,h^2}{\partial \ln X}\right),
\ee
where again $X \in \{M_0,\,M_\half,\,A_0,\,B_\mu\}$.  We perform this derivative numerically by interfacing \texttt{SoftSUSY} and \texttt{MicrOmegas}.  Given that determining $\Delta_\Omega$ is computationally expensive, we have only explored this tuning for the benchmarks.  We find that $22 \lsim \Delta_v \lsim 1100$ for the models presented below.

Before discussing the detailed regions and benchmarks we need to quickly clarify our notation.  We will interchangeably use the terms LSP, lightest neutralino, $\chi$, and $\chi_1^0$.  The other neutralinos will be denoted with $\chi^0_i$ and we will call the charginos $\chi^\pm_i$.  All other superpartners will be demarcated with a tilde.  When they are nearly pure we will also refer to the electroweakino states as bino $\widetilde{B}$, wino $\widetilde{W}$, and Higgsino $\widetilde{H}$.    We will often refer to the light flavor squarks $\widetilde{q}$ which includes the superpartners of the $u$, $d$, $c$, and $s$ quarks.

%\draftnote{$c\,\tau$ conversion factor is $1 \GeV^{-1} = 1.9733\times 10^{-14} \text{ cm}$ or inverted this is $1 = 5.068\times 10^{13} /(\text{cm GeV})$.}

%%%%%%%%%%%%%%%%%%%%%%%%%%%%%%%%%%%%%%%%%%%%%%%%%%%%%%
%%%%%%%%%%%%%%%%%%%%%%%%%%%%%%%%%%%%%%%%%%%%%%%%%%%%%%
%%%%%%%%%%%%%%%%%%%%%%%%%%%%%%%%%%%%%%%%%%%%%%%%%%%%%%

\clearpage

\subsection{{\color{LightChi}Light $\chi$}}

\stepcounter{region}
\setcounter{model}{0}
\setcounter{table}{0}
\setcounter{figure}{0}

\begin{figure}[h!!]
\centering
\includegraphics[width=0.35\textwidth]{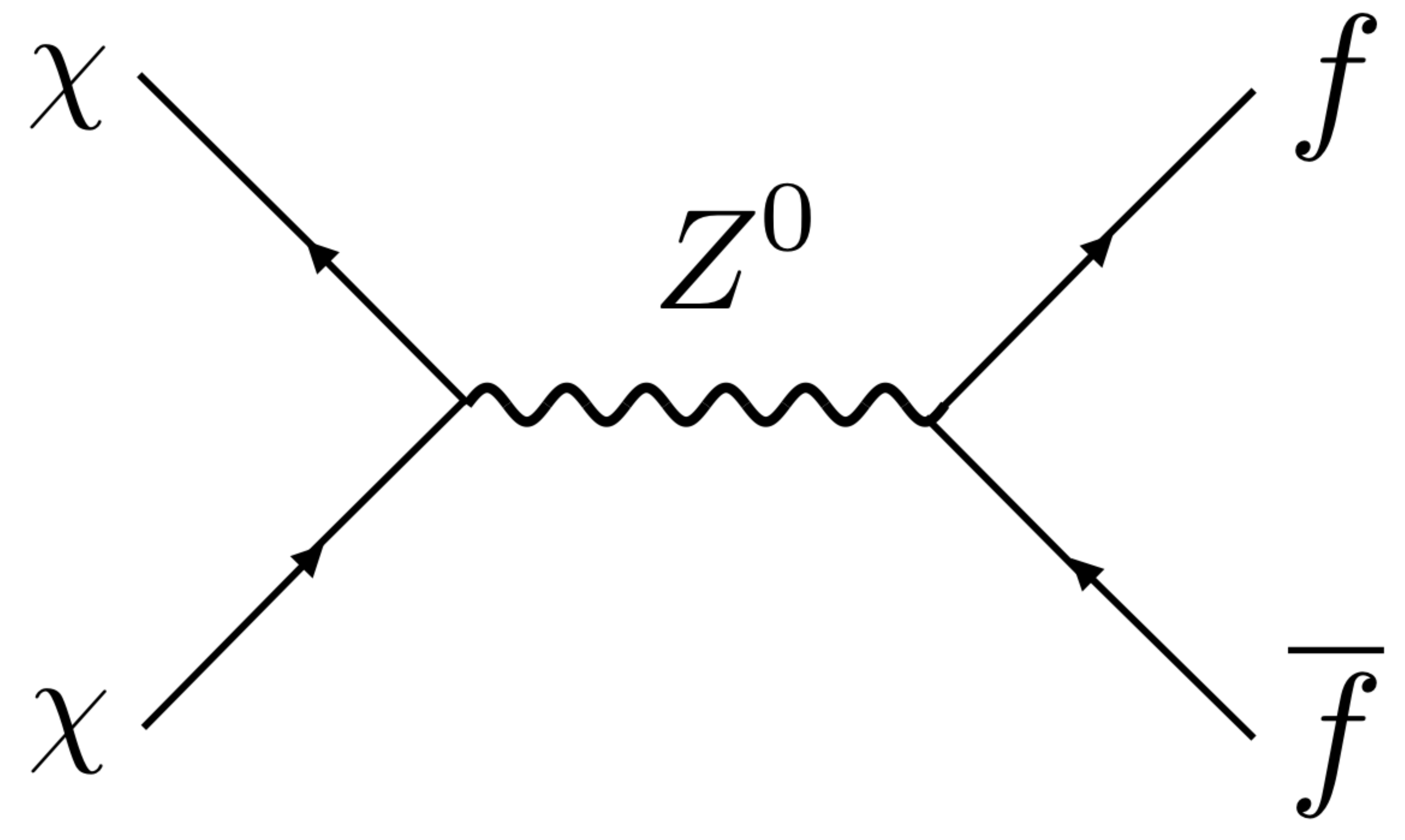}
\caption{A typical diagram which contributes to the dark matter annihilation cross section for the light $\chi$ region.  The final state consists of light standard model fermions.}
\label{Fig: Diagram LightChi}
\end{figure}

\label{Sec: ZhFunnel}
The light $\chi$ region is distinguished by requiring the LSP mass be less than 70~GeV.  Since $2\times m_\chi \sim m_Z$ and/or $m_h$, $Z^0-$ and $h-$ pole annihilation determines the relic density (the diagram in Fig.~\ref{Fig: Diagram LightChi} provides an example process).  One of the characteristic features of this region that sets it apart from the well-tempered dark matter of \refsec{WellTempered} is the small mixing between the bino-like LSP and the Higgsinos.

The bound on $m_\chi$ corresponds to an upper bound on $M_{\half}\lsim 160\GeV$.   The region exists in all four quadrants for $|A_0/M_0| \lsim 1$.  In order to make the Higgs sufficiently heavy, the masses of the scalars must be heavy; $2 \TeV \lsim M_0 \lsim 12\TeV$.  This region can be thought of as version of ``Mini-Split Supersymmetry'' \cite{Arvanitaki:2012ps}.  The scalars will play little roll for phenomenology.  

\begin{figure}[h!]
\begin{tabular}{c}
{\bf{\color{LightChi}$\quad\quad$Light $\chi$}}\\
\includegraphics[width=0.4\textwidth]{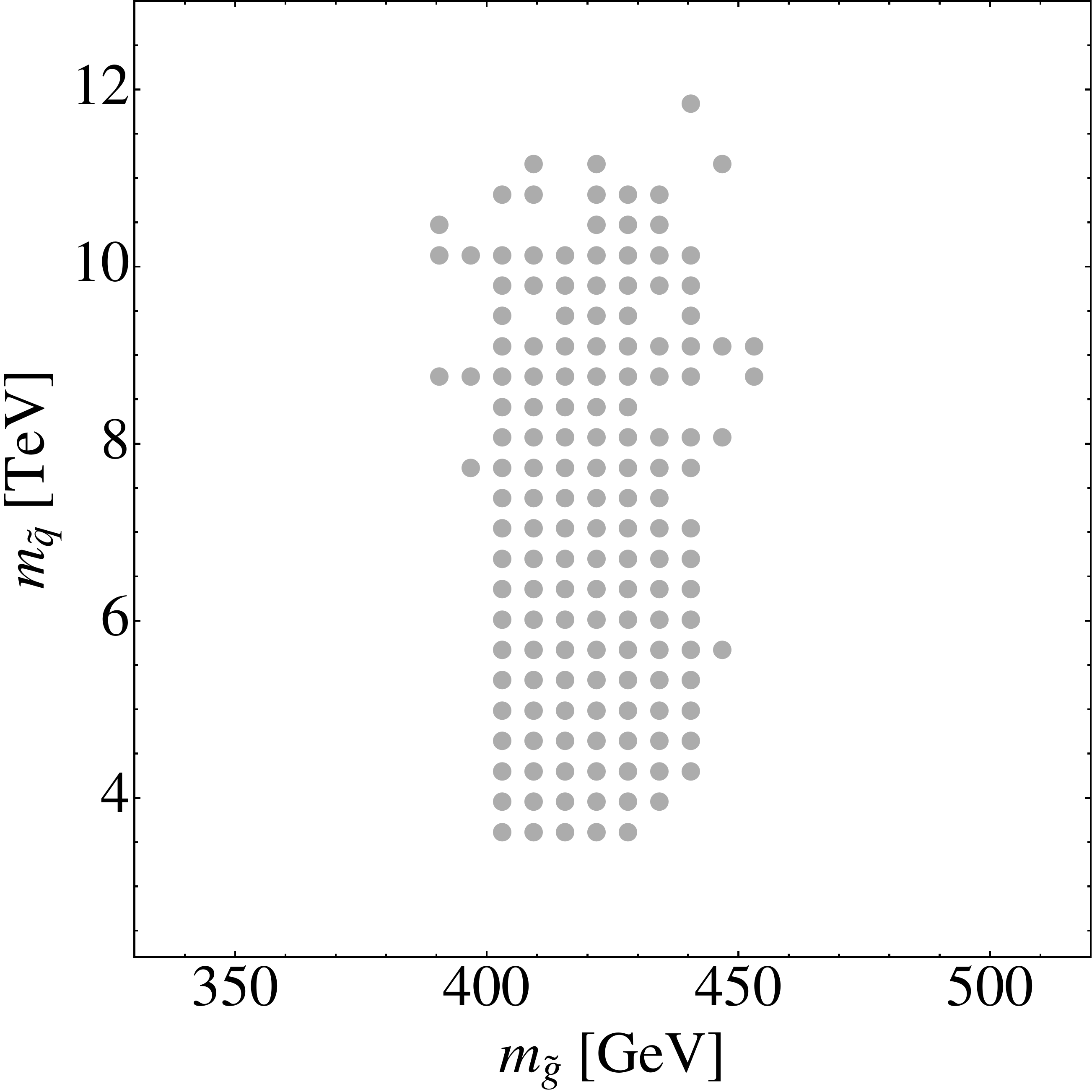}
\end{tabular}
\caption{The squark mass versus gluino mass plane for points in the light $\chi$ region.  The 7~TeV and 8~TeV LHC data sets can be used to constrain all of these models.}
\label{Fig: LightChi Gluino-Squark Plane}
\end{figure}

The range of the gluino and light flavor squark masses in the light $\chi$ region are shown in Fig.~\ref{Fig: LightChi Gluino-Squark Plane}.  The squark masses lie in the range $4\TeV \lsim m_{\tilde{q}} \lsim 11\TeV$.    Due to the presence of a light gluino, $m_{\tilde{g}}\lsim450\GeV$, this region is excluded by direct searches for this state.  However, as is demonstrated by the following benchmark, this is non-trivial to show as the gluino tends to have many competing decay modes.  Direct electroweakino production can also be constraining.  As the scalar superpartners decouple, the electroweak tuning can also be large.  This region spans the range $260 \lsim \Delta_v \lsim 9200$.

\begin{figure}[h!]
\begin{tabular}{c}
{\bf{\color{LightChi}$\quad\quad$Light $\chi$}}\\
\includegraphics[width=0.4\textwidth]{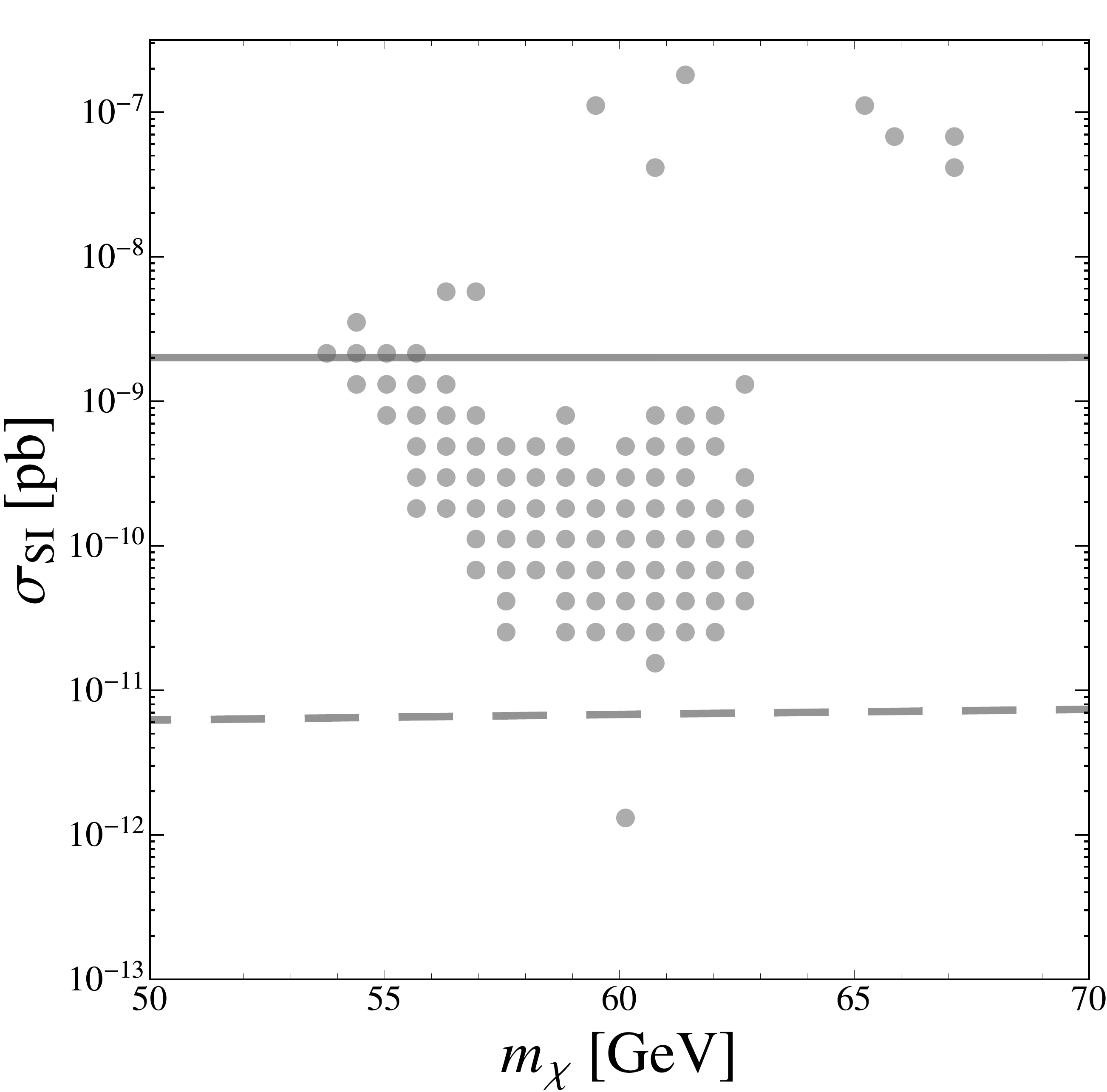}
\end{tabular}
\caption{The spin independent direct detection cross section versus LSP mass for points in the light $\chi$ region. The solid line gives the current bound from XENON100 \cite{Aprile:2012nq} and the dashed line is a projected limit for a ton scale Xenon experiment \cite{DMTools}.}
\label{Fig: LightChi DD}
\end{figure}

The second impact of having a bino LSP is that spin-independent direct detection tends to be small.  This is shown explicitly in Fig.~\ref{Fig: LightChi DD}.  Since many of these models rely on coupling to the $Z^0$, it is possible that spin-dependent direct detection could be important --- we find that these cross sections are outside the reach of near term experiments.   None of this is relevant for phenomenology as these models have already been excluded by the LHC.

\clearpage

\stepcounter{model}
\tocless\subsubsection{ \Cmodel{LightChi}}
\begin{table}[h!]
\begin{centering}
\renewcommand{\arraystretch}{1.1}
\setlength{\tabcolsep}{4pt}
\begin{tabular}{|c|c|c|c|c||c|c|}
\hline
 \multicolumn{7}{|c|}{Input parameters} \\
 \hline
 \hline
$M_0$ & $M_\half$ & $A_0$ & $\tan\beta$ & $\text{sign}(\mu)$ & $|\mu|$ & $\sqrt{B_\mu}$ \\
\hline
5455.8 & 132.315 & -3480.24 & 15.5977 & 1 & 301.773 & $14204.3$
\\
\hline
\end{tabular}\\
\vspace{5pt}
\begin{tabular}{|c|c|c|c|c|c|c|c||c|c||c|c|}
\hline
 \multicolumn{12}{|c|}{Low energy spectrum} \\
 \hline
 \hline
$m_{\tilde{g}}$ & $m_{\tilde{q}}$ & $m_{\tilde{t}_1}$& $m_{\tilde{\tau}_1}$ & $m_\chi$ & $m_{\chi_1^\pm}$ &  $m_h$ & $m_A$  & $\Omega\,h^2$ & $\sigma_\text{SI} \text{ [pb]}$ & $\Delta_{v} $ & $\Delta_{\Omega}$\\
\hline
409 & 5390 & 3100 & 5330 & 57.1 & 111 & 124 & 5210 & 0.105 & $3.92\times 10^{-10}$ & 1200  & 35  \\ 
\hline
\end{tabular}
\caption{\label{Tab: BM LightChi} Light $m_\chi$ benchmark.  This model is excluded by LHC7 searches for the gluino.  Dimensionful values are in GeV unless otherwise stated.}
\end{centering}
\end{table}

This benchmark provides a concrete example of the issues that arise when attempting to exclude a model with a variety of competing decay modes.  Both $M_\half$ and $\mu$ are a factor of 15 to 30 smaller than $M_0$; this is a CMSSM realization of ``Mini-Split Supersymmetry'' \cite{Arvanitaki:2012ps}.  While this benchmark does not have a small $A$ term, there exist points that do. Table~\ref{Tab: BM LightChi} shows a typical light bino dark matter candidate that is dominantly annihilating through the $h$ and $Z^0$ resonances.  In this case, $m_{h} - 2\,m_{\chi}=10.8\GeV$.    Even though this spectrum is relatively light, the fine tuning is $\Delta_v=1200$.  This point demonstrates that models with small $\mu$ can still be highly fine-tuned.    

Given the low energy spectrum shown in Table \ref{Tab: BM LightChi}, it is clear that the squarks and sleptons are well out of range of colliders and play no role in determining the signatures of this model. The most promising avenue for discovery is via gluino pair production.  The gluino mass is $\simeq 409\GeV$ which implies that the  7~TeV gluino pair production cross section is $\sigma(p\,p\rightarrow \widetilde{g}\,\widetilde{g})\simeq 9.0\pb$. Hence, it is likely that this model has already been excluded by the LHC.  In order to determine if this is true, we need the full neutralino spectrum 
\begin{eqnarray}
\renewcommand{\arraystretch}{1.1}
\begin{array}{|c||c|c|c|c|c|c|}
\hline
&\chi_1^0&\chi_2^0&\chi_3^0&\chi_4^0&\chi_1^\pm&\chi_2^\pm\\
\hline
m\, [\text{GeV}] & 57.1 & 111 & 326 & 338 & 111 & 340 \\ 
\hline
\end{array}
\end{eqnarray}
and mixings matrices
\begin{eqnarray}
\renewcommand{\arraystretch}{1.1}
\begin{array}{|c||c|c|c|c|}
\hline
&\chi^0_1& \chi^0_2& \chi^0_3&\chi^0_4\\
\hline\hline
\widetilde{B} & 97.0\% & 0.4\% & 2.5\% & 0.1\% \\
\hline
\widetilde{W} & 1.1\% & 90.9\% & 6.8\% & 1.1 \\
\hline
\widetilde{H}_d & 0.6\% & 1.4\% & 47.5 & 50.5 \\
\hline
\widetilde{H}_u & 1.4\% & 7.3\% & 43.1\% & 48.2 \\
\hline
\end{array}
\qquad
\begin{array}{|c||c|c|c|c|}
\hline
&\chi^-_1& \chi^-_2& \chi^+_1&\chi^+_2\\
\hline\hline
\widetilde{W}^+ &  &  & 97.6\% & 2.4\% \\
\hline
\widetilde{H}^+_u  &  &  & 2.4\% & 97.6\% \\
\hline
\widetilde{W}^- & 85.6\% & 14.4\% &  & \\
\hline
\widetilde{H}^-_d & 14.4 & 85.6 &  & \\
\hline
\end{array}
\end{eqnarray}
%%%%

This allows the determination of the gluino branching ratios; the rates involving Higgsinos are suppressed.  However, there are a variety of gluino cascades involving the winos which make limit setting non-trivial for this benchmark.   The signatures of this benchmark are well-approximated by three of the standard Simplified Models: 
\begin{equation}
  \widetilde{g}\rightarrow  
  \begin{cases}
  \widetilde{B} \, q\,\overline{q} \quad& 1.9\% \\
\widetilde{\chi}_1^\pm\, q\,\bar{q}\rightarrow  \widetilde{B}\, W^\pm \, q\,\overline{q}' \quad&\, 45\% \quad[r=0.181]\\
\widetilde{\chi}_2^0\, q\,\overline{q}\rightarrow  \widetilde{B} \,Z^0\, q\,\overline{q} \quad& \,34\% \quad[r=0.181]
 \end{cases}
\end{equation}

So far the only applicable LHC results which provide the limits on Simplified Models with $\text{BR} < 1$ have been released using 7~TeV data.  There are many searches with sensitivity to this model \cite{Aad:2012fqa, Chatrchyan:2012qka, CMS:2012yua}.  The most relevant of these is an ATLAS search for jets, $\MET\!\!$, and no high $p_T$ electrons or muons using $4.7 \ifb$ of data \cite{Aad:2012fqa}.  The collaboration has recast this search for the final state 
\be
\widetilde{g}\,\widetilde{g}\rightarrow W^\pm \,W^\pm\, q \,\bar{q} \, q\,\bar{q}\, \chi\,\chi,
\ee
providing a limit of $\sigma\times\text{BR} \lsim 1 \pb$.  The corresponding prediction for this benchmark is $1.8 \pb$ when all combinations for the sign of the $W^\pm$ bosons are included.  Furthermore, the decay involving $Z^0$ bosons will have a very similar efficiency for this search \cite{Alwall:2008va, Alwall:2008ve, Alves:2011sq}.   Finally, we note that a direct search for wino-like charginos and neutralinos decaying to electroweak gauge bosons and bino-like neutralinos using the full 8 TeV data set is also sensitive to this model \cite{ATLAS:2013rla}.  These considerations exclude this benchmark.

%%%%%%%%%%%%%%%%%%%%%%%%%%%%%%%%%%%%%%%%%%%%%%%%%%%%%%
%%%%%%%%%%%%%%%%%%%%%%%%%%%%%%%%%%%%%%%%%%%%%%%%%%%%%%
%%%%%%%%%%%%%%%%%%%%%%%%%%%%%%%%%%%%%%%%%%%%%%%%%%%%%%

\newpage

\subsection{{\color{WT}Well-tempered}}
\label{Sec: WellTempered}

\stepcounter{region}
\setcounter{model}{0}
\setcounter{table}{0}
\setcounter{figure}{0}

The well-tempered region of the CMSSM is characterized by a non-trivially mixed LSP.  Specifically, the lightest neutralino has a non-trivial wino and/or Higgsino component.  This implies that the dominant process which determines the relic density is the annihilation channel $\chi\,\chi\rightarrow W^+\,W^-$ as shown in Fig.~\ref{Fig: Diagram Well Tempered}.  The well-tempered region encompasses the so-called focus point region \cite{Feng:1999mn,Feng:2000gh,Feng:2011aa,Feng:2012jfa}.   The electroweak tuning spans the range $270 \lsim \Delta_v \lsim 33000$ for this region.  The focusing effect lowers the upper bound on tuning (as defined in \Eref{EQ: BG tuning measure}) by about a factor of 10 as compared to points with similar values of $M_0$ in other regions.
\begin{figure}[h!!]
\centering
\includegraphics[width=0.4\textwidth]{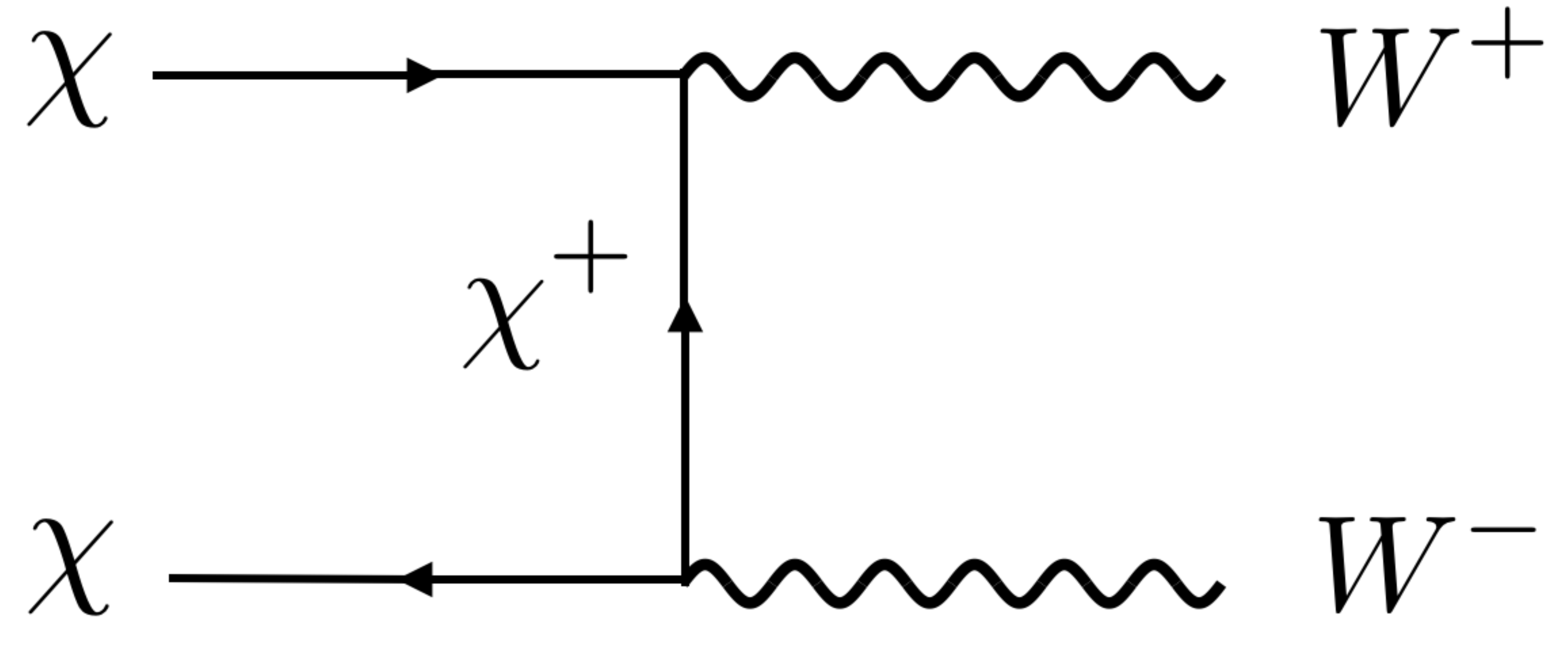}
\caption{A typical diagram which contributes to the dark matter annihilation cross section for the well-tempered region.}
\label{Fig: Diagram Well Tempered}
\end{figure}

This region also contains the pure Higgsino limit for the LSP with $m_\chi \simeq 1.1 \TeV$ and the scalars and gauginos arbitrarily heavy.  Imposing $m_h < 128 \GeV$ bounds $M_0$.  An approximate pure Higgsino limit {\em is possible} with $\OO(10^{-3})$ bino fraction.   These correspond to the points with the smallest direct detection cross sections in Fig.~\ref{Fig: WellTempered DD}.

Quantitatively, there is a huge range of allowed values for both $M_0$ and $M_\half$ in this region: $3 \TeV \lsim M_0 \lsim 20 \TeV$ and $160 \GeV \lsim M_\half \lsim 20 \TeV$.  The lower bound on $M_\half$ is determined by the designation of the cut-off between this and the light $\chi$ regions.  Note that similar to the light $\chi$ region, the well-tempered region only exists for a limited range of $|A_0/M_0|\lsim 1$. 
   
Fig.~\ref{Fig: WellTempered Squark Gluino plane} shows the range of squark and gluino masses which result.  The lower bound on the gluino mass is $m_{\tilde{g}}\simeq 600 \GeV$.  This simply results from the fact that any points which had smaller gluino masses would be classified as being in the light $\chi$ region since at the low scale $M_3/M_1 \simeq 7$.  Many of these points with light gluinos $ m_{\tilde{g}} \lsim 1\TeV$ will already be excluded by some combination of 7~TeV and 8~TeV LHC results.  The maximum gluino masses occur as the LSP approaches the pure Higgsino limit.  Fig.~\ref{Fig: WellTempered Squark Gluino plane} demonstrates that gluino masses as high as 13~TeV are possible.

\begin{figure}[t!]
\begin{tabular}{c}
{\bf{\color{WT}Well-tempered}}\\
\includegraphics[width=0.85\textwidth]{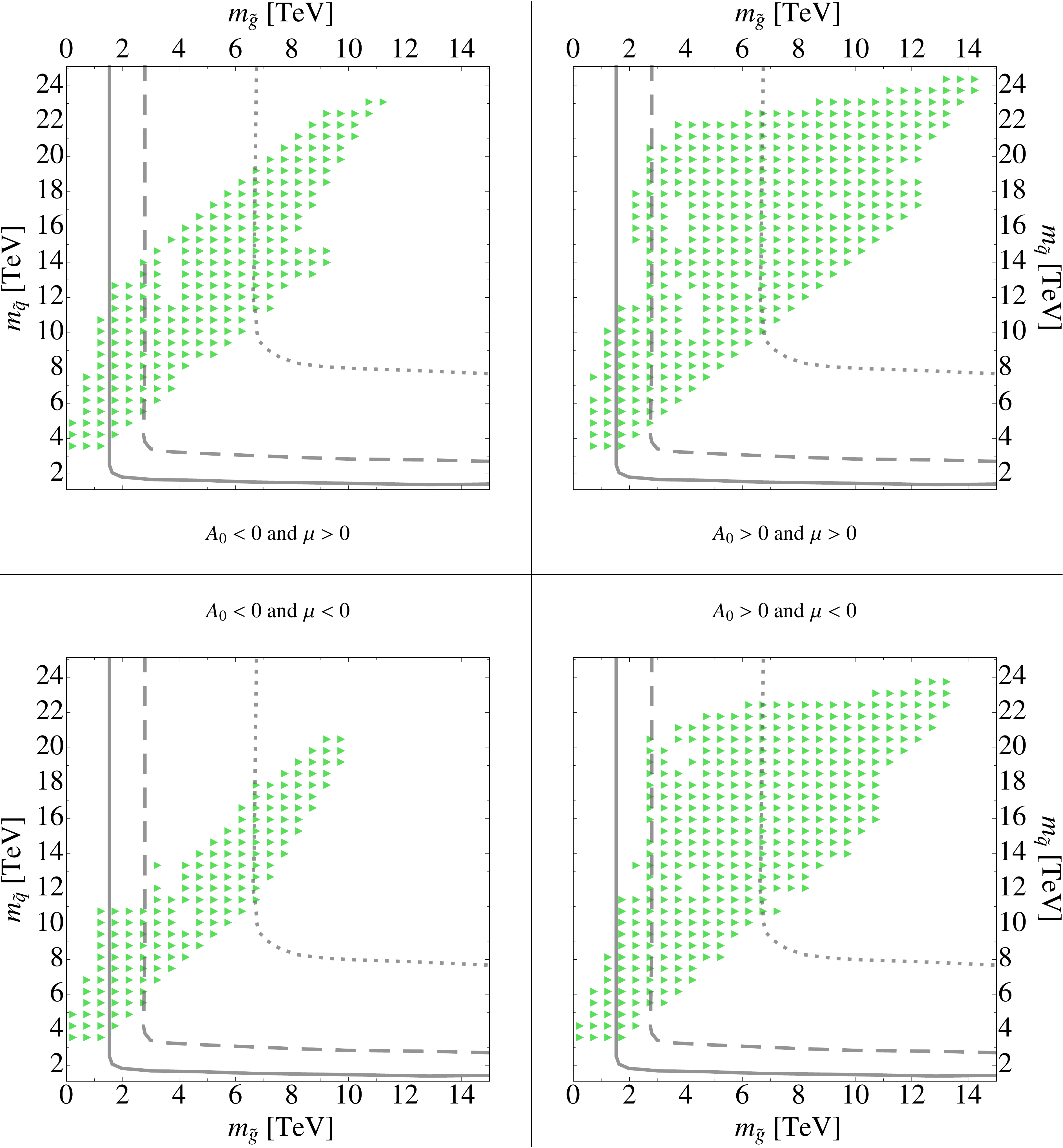}
\end{tabular}
\caption{The squark mass versus gluino mass plane for points in the well-tempered region.  Each plot only includes points from the corresponding quadrant.  Also plotted are contours corresponding to 10 squark and/or gluino events for $30 \ifb$ integrated luminosity at $\sqrt{s}=8\TeV$ [solid], $300 \ifb$ integrated luminosity at $\sqrt{s}=13\TeV$ [dashed], and $3000 \ifb$ integrated luminosity at $\sqrt{s}=33\TeV$ [dotted].}
\label{Fig: WellTempered Squark Gluino plane}
\end{figure}

\begin{figure}[t!]
\begin{tabular}{c}
{\bf{\color{WT}Well-tempered}}\\
\includegraphics[width=0.85\textwidth]{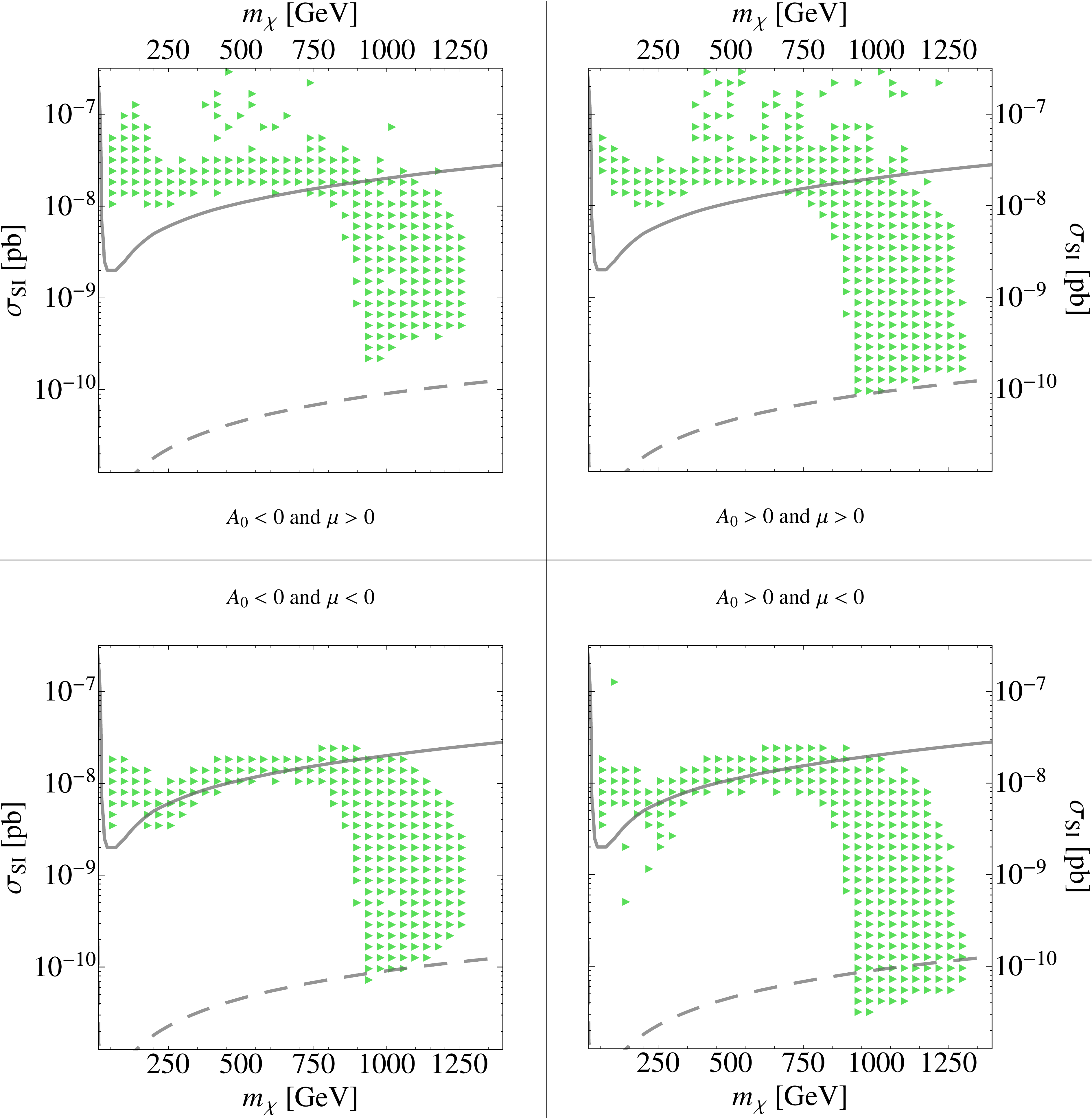}
\end{tabular}
\caption{The spin independent direct detection cross section versus LSP mass for points in the well-tempered region.  Each plot only includes points from the corresponding quadrant.  The solid line gives the current bound from XENON100 \cite{Aprile:2012nq} and the dashed line is a projected limit for a ton scale Xenon experiment \cite{DMTools}.}
\label{Fig: WellTempered DD}
\end{figure}

The squark masses lie in the range $3.5 \TeV \lsim m_{\tilde{q}} \lsim 24 \TeV$ as shown in Fig.~\ref{Fig: WellTempered Squark Gluino plane}.  The squark masses tend to be larger than the gluino mass because larger values of $M_0$ are required in order to achieve a Higgs boson mass of 125~GeV.  The squarks in these models will lie outside the range of the 13~TeV LHC.  Only a small range of these models will be testable at colliders through the direct production of gluinos.  

The most effective way to discover or exclude this region is through direct detection.   As shown in Fig.~\ref{Fig: WellTempered DD}, the ton scale limits on spin independent scattering will cover most well-tempered models.  Multi-ton scale direct detection should have sensitivity to this remaining sliver of parameter space.  Even accounting for the uncertainty in the nucleon form factor, the well-tempered region is probable utilizing near-term direct detection proposed experiments.

\clearpage

\stepcounter{model}
\tocless\subsubsection{\Cmodel{WT}}
\begin{table}[h!]
\begin{centering}
\renewcommand{\arraystretch}{1.1}
\setlength{\tabcolsep}{4pt}
\begin{tabular}{|c|c|c|c|c||c|c|}
\hline
 \multicolumn{7}{|c|}{Input parameters} \\
 \hline
 \hline
$M_0$ & $M_\half$ & $A_0$ & $\tan\beta$ & $\text{sign}(\mu)$ & $|\mu|$ & $\sqrt{B_\mu}$ \\
\hline
4103.76 & 525.385 & 905.88 & 13.6663 & -1 & 292.034 & $10805.$
\\
\hline
\end{tabular}\\
\vspace{5pt}
\begin{tabular}{|c|c|c|c|c|c|c|c||c|c||c|c|}
\hline
 \multicolumn{12}{|c|}{Low energy spectrum} \\
 \hline
 \hline
$m_{\tilde{g}}$ & $m_{\tilde{q}}$ & $m_{\tilde{t}_1}$& $m_{\tilde{\tau}_1}$ & $m_\chi$ & $m_{\chi_1^\pm}$ &  $m_h$ & $m_A$  & $\Omega\,h^2$ & $\sigma_\text{SI} \text{ [pb]}$ & $\Delta_{v} $ & $\Delta_{\Omega}$\\
\hline 
1330 & 4180 & 2510 & 4040 & 218 & 292 & 122 & 4000 & 0.139 & $5.15\times 10^{-9}$ & 400  & 37  \\ 
\hline
\end{tabular}
\caption{\label{Table: BM WT1} Well-tempered benchmark.  Dimensionful values are in GeV unless otherwise stated.}
\end{centering}
\end{table}

This well-tempered benchmark model is in the focus point supersymmetry region.  It can be probed using both the 13 TeV LHC and direct detection.  The electroweakino sector of the theory has masses   
\begin{eqnarray}
\renewcommand{\arraystretch}{1.1}
\begin{array}{|c||c|c|c|c|c|c|}
\hline
&\chi_1^0&\chi_2^0&\chi_3^0&\chi_4^0&\chi_1^\pm&\chi_2^\pm\\
\hline
m\, [\text{GeV}] & 218 & 297 & 310 & 466 & 292 & 465 \\ 
\hline
\end{array}
\end{eqnarray}
and mixings
\begin{eqnarray}
\begin{array}{|c||c|c|c|c|}
\hline
&\chi^0_1& \chi^0_2& \chi^0_3&\chi^0_4\\
\hline\hline
\widetilde{B} & 86.6\% & 0.4\% & 8.9\% & 4.1\% \\
\hline
\widetilde{W} & 12.8\% & 9.3\% & 39.9\% & 38.0 \%\\
\hline
\widetilde{H}_d & 0.4\% & 0.7\% & 48.5 & 50.4 \%\\
\hline
\widetilde{H}_u & 0.2\% & 89.6\% & 2.7\% & 7.4 \% \\
\hline
\end{array}
\qquad
\begin{array}{|c||c|c|c|c|}
\hline
&\chi^-_1& \chi^-_2& \chi^+_1&\chi^+_2\\
\hline\hline
\widetilde{W}^+ &  &  & 14.7\% & 85.3\% \\
\hline
\widetilde{H}^+_u  &  &  & 85.3\% & 14.7\% \\
\hline
\widetilde{W}^- & 5.3\% & 94.7\% &  & \\
\hline
\widetilde{H}^-_d & 94.7 \% & 5.3\% &  & \\
\hline
\end{array}
\end{eqnarray}
%%%%
The LSP is dominantly bino dark matter with a non-trivial wino ad-mixture --- this is clearly a well-tempered neutralino.  This model a sizable annihilation cross section to $W^+\,W^-$ in the early Universe so that the computed relic abundance can match the observation.

The gluino has a mass of $m_{\widetilde{g}}=1333\GeV$ and currently is too heavy to have been directly produced.  Its cross section at the 13~TeV LHC is 
\be
\sigma(p\,p\rightarrow \widetilde{g}\,\widetilde{g}) = 30 \fb.
\ee
The most important gluino decays for phenomenology are cascades involving the electroweakinos that have a large Higgsino fraction, $\chi_1^\pm\,\chi_2^0,\text{ and} \,\chi_3^0$.

The majority of the gluinos decay into heavy flavor.  The cumulative light flavor branching ratio is 
\be
\widetilde{g}\,\rightarrow q\, \overline{q}'\, X \quad 11.5\%
\ee
where $X$ is a neutralino or chargino.  This pattern of branching ratios can be understood from the pattern of the squark masses,
\begin{eqnarray}
\renewcommand{\arraystretch}{1.1}
\begin{array}{|c||c|c|c|c|}
\hline
& \widetilde{q} & \widetilde{q}_3 & \widetilde{d}_3^c & \widetilde{u}_3^c  \\
\hline
m\, [\text{TeV}] & 4.2 & 3.4 & 4.1 & 2.5  \\  
\hline
\end{array}
\end{eqnarray}
Since the gluino decays are mediated through off-shell squarks, the branching ratios are proportional to $1/m^4_{\widetilde{q}_i}$.  The process with an off-shell right handed stop is enhanced by a factor of 7.7 over light flavored squarks.  All of the properties of the spectra will have to be inferred from these decay widths since the direct production of these squarks is beyond the reach of the 13~TeV LHC.  They should be accessible at the 33~TeV LHC.

The LHC phenomenology of this model is dominated by the following simplified models involving the lightest Higgsinos
\be
 \widetilde{g} \rightarrow 
 \begin{cases}
  t\,\overline{b}\, \chi_1^- + \cc \rightarrow   t\, \overline{b}\, \left(W^{-}\right)^*\,\chi_1^0 &\quad33\%\quad[r=0.09] \\
t\,\overline{t}\, \chi_2^0 \rightarrow   t\, \overline{t}\, \left(Z^0\right)^*\,\chi_1^0 &\quad 15\%\quad[r=0.08]\\
 t\,\overline{t}\, \chi_3^0 \rightarrow   t\, \overline{t}\, Z^0\,\chi_1^0 &\quad 15\%\quad[r=0.07] 
 \end{cases}
\ee

The Higgsino decays to the bino via an off-shell $W^\pm$ or $Z^0$.  
%The wino dominantly decays to the Higgsino via on-shell $W^\pm$, $Z^0$ and $h$ bosons.  
%These decay products are not particularly boosted with $p_T\sim 60\GeV$.  
The Higgsino cross sections are sufficiently large such that they will be discoverable at the 13~TeV LHC; for example 
\be
\sigma(p\,p\rightarrow \chi^+_1\, \chi^0_2) = 73\fb 
\ee
at $\sqrt{s} = 13\TeV$.

Direct detection is a good way to discover this model with a spin independent cross section of 
\be
\sigma_\mr{SI} = 5.2\times 10^{-9}\pb.
\ee
This is within a factor of two of current sensitivity.  Therefore, this benchmark will likely be discovered using direct detection instead of at the LHC.  Given a direct detection signal the LHC signatures will provide a crucial compliment for understanding the properties of this model.

%The fine tuning to get the right dark matter density is moderate, having a sensitivity of 49, which is not particularly severe relative to other regions.

%%
%\begin{eqnarray}
%\Br(\widetilde{g}\rightarrow \chi_1^+ \bar{t}b +\cc) = 35.0\% \qquad
%\Br(\widetilde{g}\rightarrow \chi_{2,3}^0 \bar{t}t ) = 33.1\% \qquad
%\Br(\widetilde{g}\rightarrow \chi_2^+ \bar{t}b+\cc) = 10.3\% \qquad
%\Br(\widetilde{g} \rightarrow \chi_1^0 \bar{t}t) = 4.3\% .
%\end{eqnarray}
%%

\begin{comment}
\begin{itemize}
\item $\sigma_{gg} = 452 \fb$; 
$\sigma_{qq^*} = XX \pb$; 
$\sigma_{gq} = XX\pb$; 
$\sigma_{t t} = \pb$; 
$\sigma_{n n} = \pb$; 
\end{itemize}
\end{comment}

\clearpage
\stepcounter{model}
\tocless\subsubsection{\Cmodel{WT}}
\begin{table}[h!]
\renewcommand{\arraystretch}{1.1}
\setlength{\tabcolsep}{4pt}
\begin{centering}
\begin{tabular}{|c|c|c|c|c||c|c|}
\hline
 \multicolumn{7}{|c|}{Input parameters} \\
 \hline
 \hline
$M_0$ & $M_\half$ & $A_0$ & $\tan\beta$ & $\text{sign}(\mu)$ & $|\mu|$ & $\sqrt{B_\mu}$ \\
\hline
7250. & 2123.36 & 3559.09 & 24.078 & 1 & 897.284 & $32815.5$
\\
\hline
\end{tabular}\\
\vspace{5pt}
\begin{tabular}{|c|c|c|c|c|c|c|c||c|c||c|c|}
\hline
 \multicolumn{12}{|c|}{Low energy spectrum} \\
 \hline
 \hline
$m_{\tilde{g}}$ & $m_{\tilde{q}}$ & $m_{\tilde{t}_1}$& $m_{\tilde{\tau}_1}$ & $m_\chi$ & $m_{\chi_1^\pm}$ &  $m_h$ & $m_A$  & $\Omega\,h^2$ & $\sigma_\text{SI} \,\,\left[\text{pb}\right]$ & $\Delta_{v} $ & $\Delta_{\Omega}$\\
\hline
4700 & 8120 & 5390 & 6920 & 888 & 906 & 126 & 6660 & 0.106 & $1.72\times 10^{-8}$ & 2100  & 30  \\ 
\hline
\end{tabular}
\caption{\label{Table: BM WT2} Well tempered benchmark.  Dimensionful values are in GeV unless otherwise stated.}
\end{centering}
\end{table}

This benchmark provides an example of a well-tempered neutralino with a very heavy squark and gluino spectrum.  Clearly from Table~\ref{Table: BM WT2}, these states are outside the reach of the 13~TeV LHC.  The electroweakinos are somewhat lighter with masses given by
\begin{eqnarray}
\renewcommand{\arraystretch}{1.1}
\begin{array}{|c||c|c|c|c|c|c|}
\hline
&\chi_1^0&\chi_2^0&\chi_3^0&\chi_4^0&\chi_1^\pm&\chi_2^\pm\\
\hline
m\, [\text{GeV}] & 888 & 910 & 968 & 1800 & 906 & 1800 \\ 
\hline
\end{array}
\end{eqnarray}
and mixings
\begin{eqnarray}
\begin{array}{|c||c|c|c|c|}
\hline
&\chi^0_1& \chi^0_2& \chi^0_3&\chi^0_4 \\
\hline\hline
\widetilde{B} & 20.5\% & 0.3\% & 40.4\% & 38.8\% \\
\hline
\widetilde{W} & 0.\% & 0.\% & 49.9\% & 50.1 \% \\
\hline
\widetilde{H}_d & 79.5\% & 0.1\% & 9.6 & 10.8\% \\
\hline
\widetilde{H}_u & 0.\% & 99.5\% & 0.1\% & 0.4 \% \\
\hline
\end{array}
\qquad
\begin{array}{|c||c|c|c|c|}
\hline
&\chi^-_1& \chi^-_2& \chi^+_1&\chi^+_2\\
\hline\hline
\widetilde{W}^+ &  &  & 0.7\% & 99.3\% \\
\hline
\widetilde{H}^+_u  &  &  & 99.3\% & 0.7\% \\
\hline
\widetilde{W}^- & 0.2\% & 99.8\% &  & \\
\hline
\widetilde{H}^-_d & 99.8\% & 0.2\% &  & \\
\hline
\end{array}
\end{eqnarray}

The decays of the lighter electroweakinos occur via off-shell decays mediated by the $W^\pm$ and $Z^0$.   The cross sections for producing these states directly at the 13~TeV LHC are below $1\fb$.  The wino decays yield boosted $W^\pm$, $Z^0$ and $h$ with $p_T \sim 600\GeV$.  However, given the large diboson background, these states may not be observable at the 13~TeV LHC.  

The direct detection cross section is at the edge of the current XENON100 exclusion:
\be
\sigma_\text{SI} = 1.72 \times 10^{-8}\pb.
\ee
This benchmark will be probed by direct detection using existing technology.

\clearpage
\stepcounter{model}
\tocless\subsubsection{\Cmodel{WT}}
\begin{table}[h!]
\renewcommand{\arraystretch}{1.1}
\setlength{\tabcolsep}{4pt}
\begin{centering}
\begin{tabular}{|c|c|c|c|c||c|c|}
\hline
 \multicolumn{7}{|c|}{Input parameters} \\
 \hline
 \hline
$M_0$ & $M_\half$ & $A_0$ & $\tan\beta$ & $\text{sign}(\mu)$ & $|\mu|$ & $\sqrt{B_\mu}$ \\
\hline
13927.9 & 5700. & 6837.31 & 51.1892 & 1 & 1170.51 & $96009.4$
\\
\hline
\end{tabular}\\
\vspace{5pt}
\begin{tabular}{|c|c|c|c|c|c|c|c||c|c||c|c|}
\hline
 \multicolumn{12}{|c|}{Low energy spectrum} \\
 \hline
 \hline
$m_{\tilde{g}}$ & $m_{\tilde{q}}$ & $m_{\tilde{t}_1}$& $m_{\tilde{\tau}_1}$ & $m_\chi$ & $m_{\chi_1^\pm}$ &  $m_h$ & $m_A$  & $\Omega\,h^2$ & $\sigma_\text{SI} \text{ [pb]}$ & $\Delta_{v} $ & $\Delta_{\Omega}$\\
\hline
11700 & 16900 & 11900 & 10200 & 990 & 991 & 128 & 3910 & 0.0901 & $1.45\times 10^{-10}$ & 12000  & 34  \\ 
\hline
\end{tabular}
\caption{\label{Table: BM WT3} Well tempered benchmark in the ``pure Higgsino" limit.  Dimensionful values are in GeV unless otherwise stated.}
\end{centering}
\end{table}

This benchmark is provided as an example of a model in the ``pure Higgsino" limit of the CMSSM.   From Table~\ref{Table: BM WT3}, it is clear that the superpartners lie far beyond the reach of the 13~TeV LHC.  The electroweakino masses are
\begin{eqnarray}
\renewcommand{\arraystretch}{1.1}
\begin{array}{|c||c|c|c|c|c|c|}
\hline
&\chi_1^0&\chi_2^0&\chi_3^0&\chi_4^0&\chi_1^\pm&\chi_2^\pm\\
\hline
m\, [\text{GeV}] & 990 & 992 & 2640 & 4840 & 991 & 4840 \\ 
\hline
\end{array}
\end{eqnarray}
and mixings
\begin{eqnarray}
\begin{array}{|c||c|c|c|c|}
\hline
&\chi^0_1& \chi^0_2& \chi^0_3&\chi^0_4 \\
\hline\hline
\widetilde{B} & 0.038\% & 0.021\% & 50.0\% & 49.9\% \\
\hline
\widetilde{W} & 0.007\% & 0.009\% & 49.95\% & 50.0\% \\
\hline
\widetilde{H}_d & 99.96\% & 0.00006\% & 0.006 & 0.04\% \\
\hline
\widetilde{H}_u & 0.00002\% & 99.97\% & 0.001\% & 0.03\% \\
\hline
\end{array}
\qquad
\begin{array}{|c||c|c|c|c|}
\hline
&\chi^-_1& \chi^-_2& \chi^+_1&\chi^+_2\\
\hline\hline
\widetilde{W}^+ &  &  & 0.06\% & 99.9\% \\
\hline
\widetilde{H}^+_u  &  &  & 99.9\% & 0.06\% \\
\hline
\widetilde{W}^- & 0.003\% & 100.0\% &  & \\
\hline
\widetilde{H}^-_d & 100.0 & 0.003 &  & \\
\hline
\end{array}
\end{eqnarray}

The LSP is very nearly Higgsino.  Given 
\be
\sigma_\text{SI} = 1.44 \times 10^{-10}\pb,
\ee
this model will require a multi-ton scale direct detection experiment to be convinced it has been probed.  This model demonstrates that even the most difficult limit of the well-tempered region can eventually be explored. 

%%%%%%%%%%%%%%%%%%%%%%%%%%%%%%%%%%%%%%%%%%%%%%%%%%%%%%
%%%%%%%%%%%%%%%%%%%%%%%%%%%%%%%%%%%%%%%%%%%%%%%%%%%%%%
%%%%%%%%%%%%%%%%%%%%%%%%%%%%%%%%%%%%%%%%%%%%%%%%%%%%%%
\clearpage
\subsection{{\color{A}$A^0$-pole Annihilation}}
\stepcounter{region}
\setcounter{model}{0}
\setcounter{table}{0}
\setcounter{figure}{0}

\label{Sec: A}

For models in this region, the dark matter relic density is dominated by $s$-channel annihilation through the pseudo-scalar Higgs $A^0$ into $b\,\overline{b}$.  Because this channel experiences a resonant enhancement, the LSP can be nearly pure bino.  The mass of the $A^0$ is largely controlled by $M_0$ while the bino mass $M_1$ is determined by $M_\half$.  Naively, points should fall in this region with $M_0 > 2 \times M_\half$.  The electroweak tuning spans the range $400 \lsim \Delta_v \lsim 15000$ for this region.

\begin{figure}[h!!]
\centering
\includegraphics[width=0.35\textwidth]{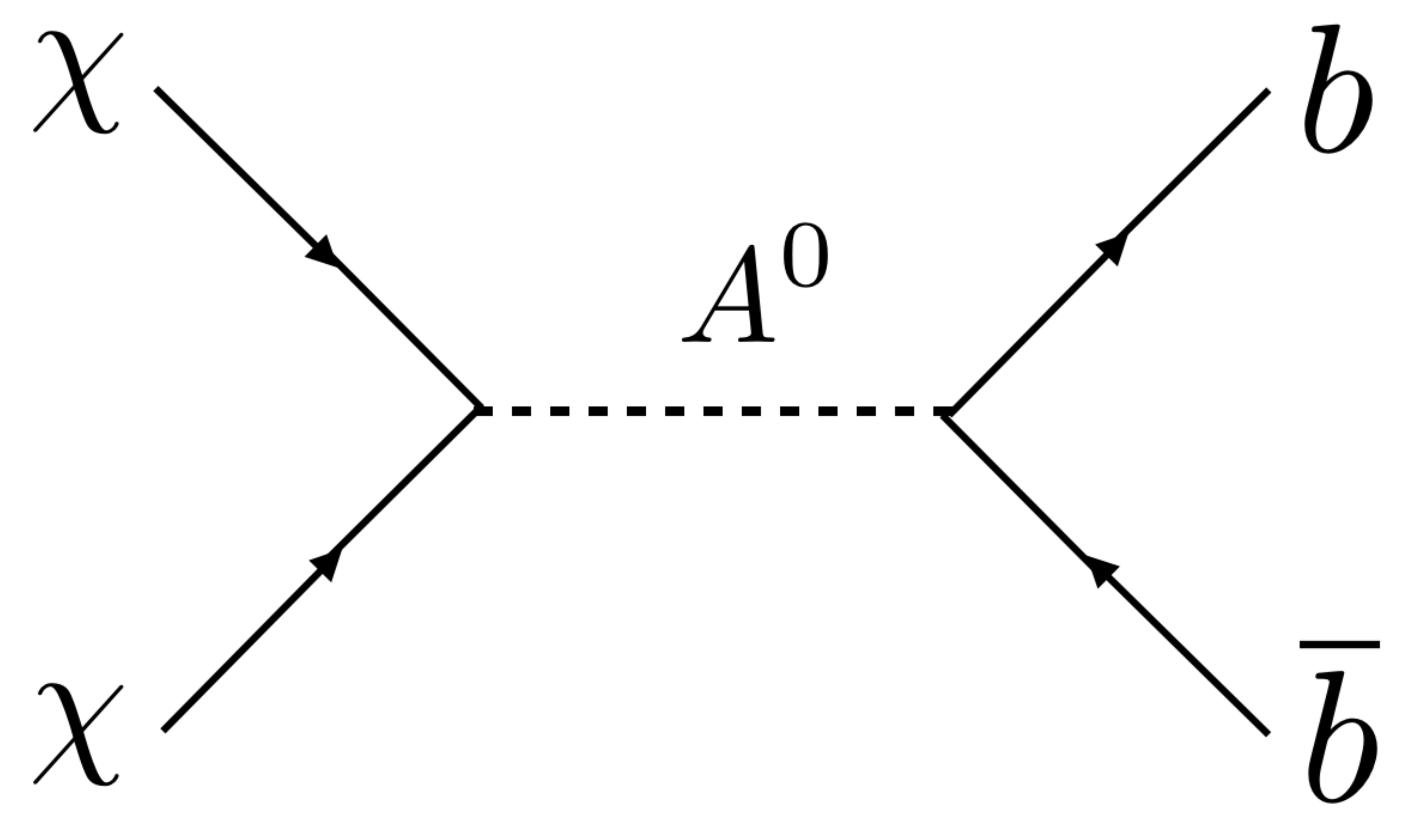}
\caption{A typical diagram which contributes to the dark matter annihilation cross section for the $A^0$-pole annihilation region.}
\label{Fig: Diagram A Pole}
\end{figure}

The $A^0$-pole annihilation regions exist in the $1^\text{st}$, $2^\text{nd}$, and $4^\text{th}$ quadrants.  This region has significant overlap with the well-tempered region, and can be distinguished by either larger values of $\tan \beta$ or a larger magnitude of $A_0/M_0$.  It also extends to lower values of $M_0$, $0 < \text{sign}(\mu) \times M_0 \lsim 4000 \GeV$ where no well-tempered points exist.

Fig.~\ref{Fig: FullSpaceD} shows that the $2^\text{nd}$ quadrant $A^0$-pole region transitions into the stau coannihilation region.  There is some overlap where both stau coannihilation and $A^0$-pole resonant annihilation are active.  The phenomenology of these points is more stau coannihilation-like.  The detailed discussions of these signatures are contained in Sec.~\ref{Sec: Stau}.

Fig.~\ref{Fig: APole Gluino Squark Plane} shows the squark mass versus gluino mass plane for the three quadrants of $A^0$-pole annihilation.  There is a huge range of allowed masses extending up to 13 TeV for the gluino and 19 TeV for the squarks.  All three quadrants contain models which would be visible at the 13~TeV LHC.  An example benchmark of one such spectrum is provided below.

\begin{figure}[t!]
\begin{tabular}{c}
{\bf{\color{A}$A^0$-pole Annihilation}}\\
\includegraphics[width=0.85\textwidth]{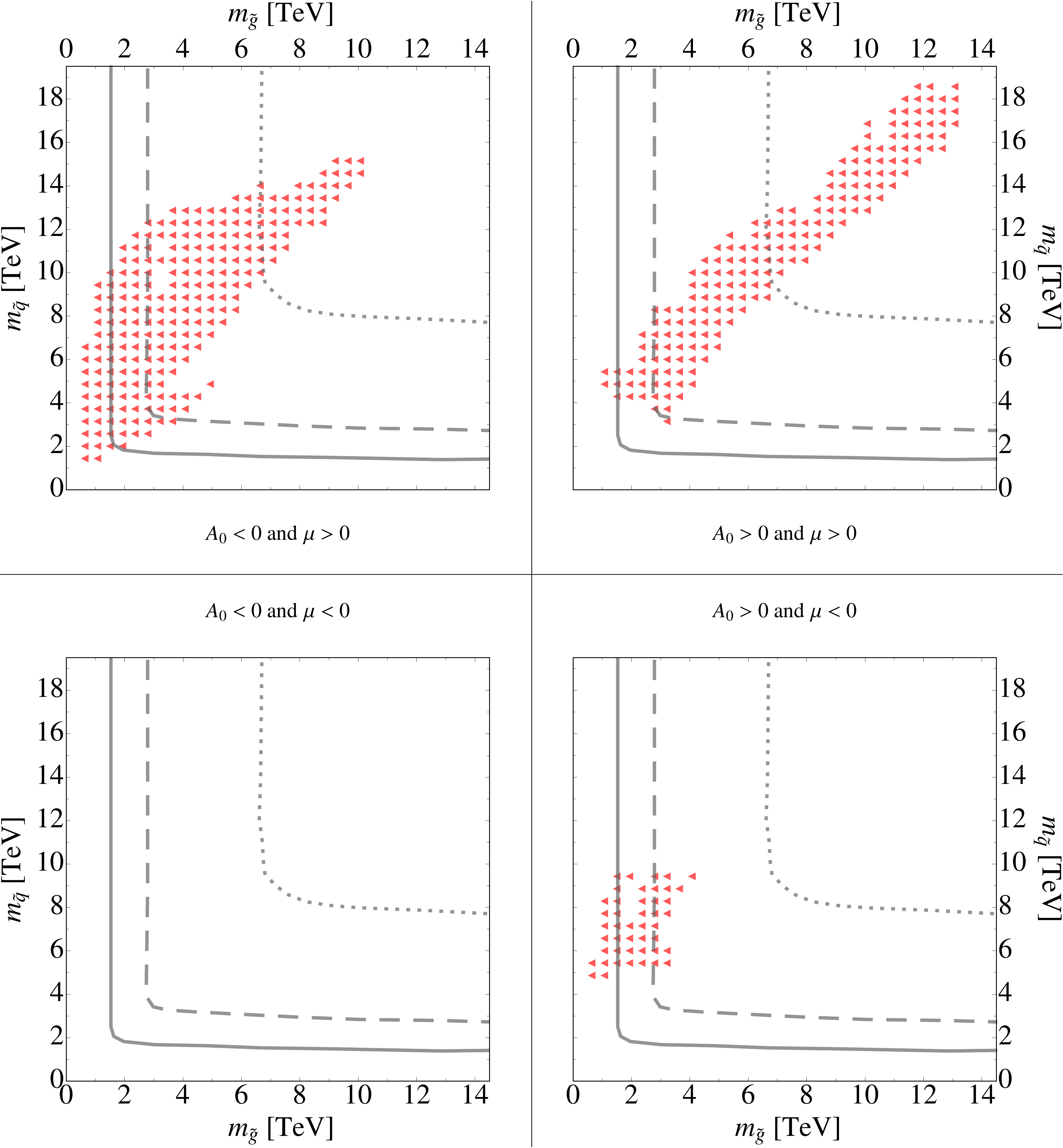}
\end{tabular}
\caption{The squark mass versus gluino mass plane for points in the $A^0$-pole annihilation region.  Each plot only includes points from the corresponding quadrant.   Also plotted are contours corresponding to 10 squark and/or gluino events for $30 \ifb$ integrated luminosity at $\sqrt{s}=8\TeV$ [solid], $300 \ifb$ integrated luminosity at $\sqrt{s}=13\TeV$ [dashed], and $3000 \ifb$ integrated luminosity at $\sqrt{s}=33\TeV$ [dotted].}
\label{Fig: APole Gluino Squark Plane}
\end{figure}

The story for direct detection is more favorable.   Fig.~\ref{Fig: APole DD} shows that some of these models are already in tension with the XENON100 limit.  The $1^\text{st}$ and $2^\text{nd}$ quadrants can be almost entirely covered using ton scale direct detection.  However, the $4^\text{th}$ quadrant will remain outside the capabilities of these experiments.  We note that the shape of the region which extends to low values of $\sigma_\text{SI}$ in the 2$^\text{nd}$ quadrant of Fig.~\ref{Fig: APole DD} is determined by the exact classification scheme employed in this paper (see Sec.~\ref{Sec: Cartography} for details).

\begin{figure}[t!]
\begin{tabular}{c}
{\bf{\color{A}$A^0$-pole Annihilation}}\\
\includegraphics[width=0.85\textwidth]{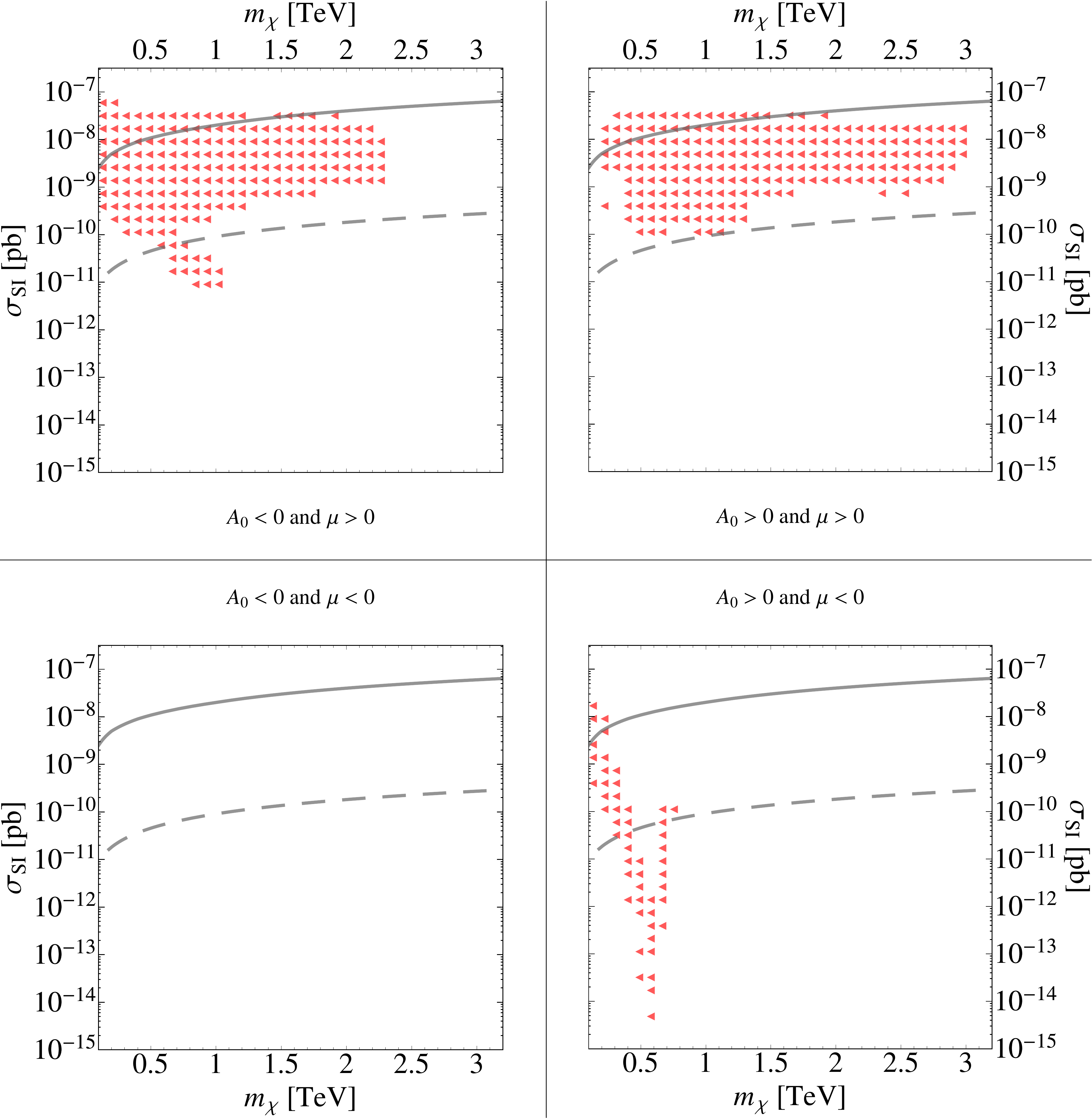}
\end{tabular}
\caption{The spin independent direct detection cross section versus LSP mass for points in the $A^0$-pole region.  Each plot only includes points from the corresponding quadrant.  The solid line gives the current bound from XENON100 \cite{Aprile:2012nq} and the dashed line is a projected limit for a ton scale Xenon experiment \cite{DMTools}.}
\label{Fig: APole DD}
\end{figure}

In order to probe the remaining models, another class of experiment is necessary.  One promising avenue is indirect direct detection.  In particular, searches for continuum $\gamma$-rays from dark matter annihilations could be sensitive to these models in the future \cite{Scott:2009jn, Ripken:2010ja}.  The relic density in this region is determined by annihilation to bottom quarks.  Annihilations today through this channel could be observable.  The range of annihilation cross sections which result are plotted in Fig.~\ref{Fig: APole sigmaAnn}. Once the $b\,\overline{b}$ pairs are produced, they decay into a hadronic shower which produces a continuum of photons.  These photons can be searched for by experiments, \emph{e.g.} the Fermi LAT.  In particular, a limit derived by stacking the results from a survey of 10 dwarf galaxies results in the solid line plotted in Fig.~\ref{Fig: APole sigmaAnn} \cite{Ackermann:2011wa}.  One can also derive complimentary limits using the Fermi LAT galactic center data \cite{Hooper:2012sr}.  This appears to be the most promising way to test these $4^\text{th}$ quadrant points.  Future experiments such as the proposed Cherenkov Telescope Array (CTA) will be relevant \cite{Doro:2012xx, Ripken:2012db}.

\begin{figure}[t!]
\begin{tabular}{c}
{\bf{\color{A}$A^0$-pole Annihilation}}\\
\includegraphics[width=0.85\textwidth]{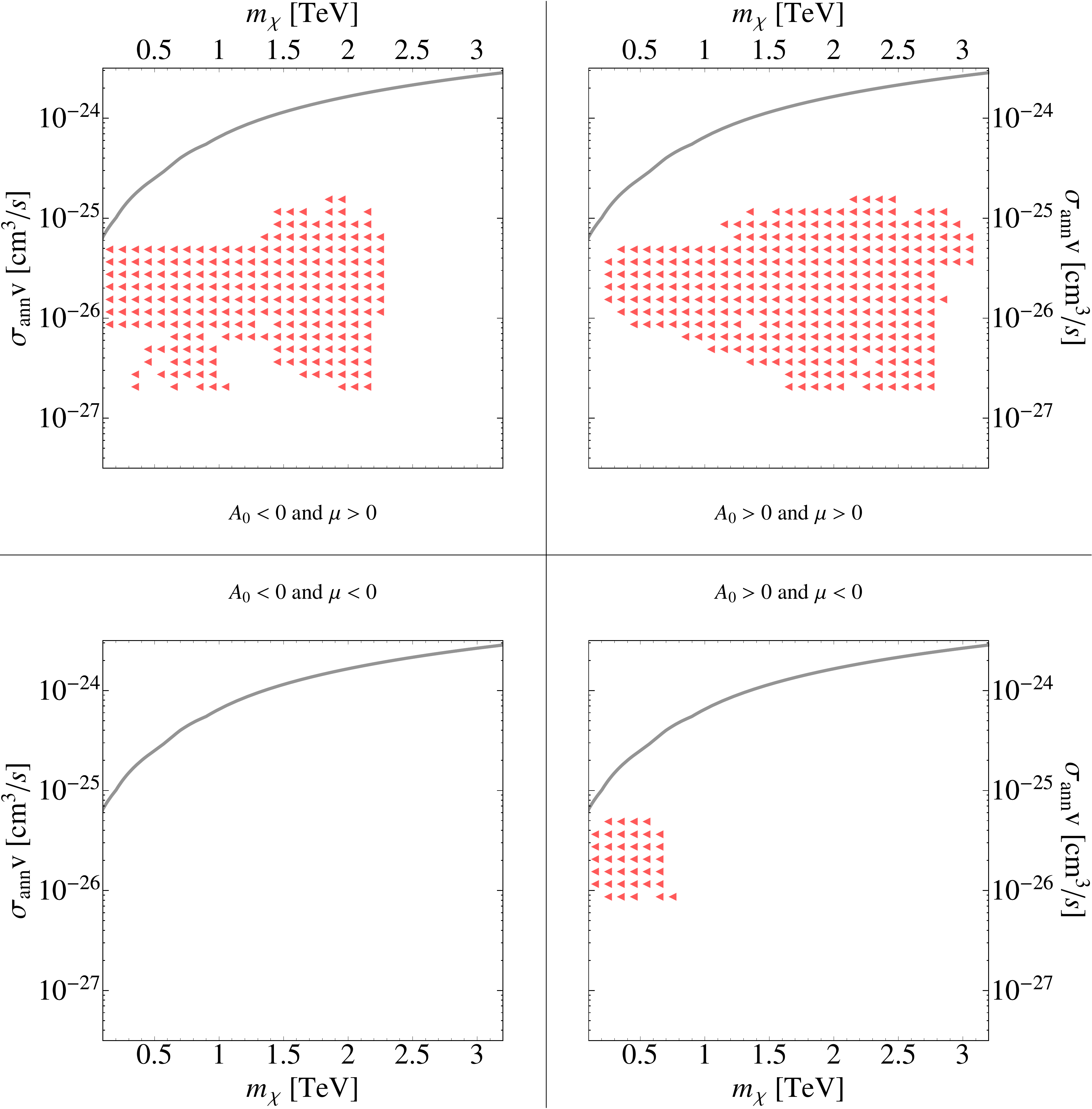}
\end{tabular}
\caption{The annihilation cross section for the $A^0$-pole region versus the LSP mass.  Each plot only includes points from the corresponding quadrant.  The solid line gives the current bound from the Fermi-LAT  \cite{Ackermann:2011wa}}
\label{Fig: APole sigmaAnn}
\end{figure}

As discussed above, \texttt{DarkSUSY} and \texttt{MicrOmegas} tend to disagree by roughly $30\%$ in this region.  All of our benchmarks match the observed relic density according to both calculations; we provide both values below.

%%%%%%%%%%%%%%%%%%%
\clearpage
\stepcounter{model}
\tocless\subsubsection{\Cmodel{A}}
\begin{table}[h!]
\renewcommand{\arraystretch}{1.1}
\setlength{\tabcolsep}{4pt}
\begin{centering}
\begin{tabular}{|c|c|c|c|c||c|c|}
\hline
 \multicolumn{7}{|c|}{Input parameters} \\
 \hline
 \hline
$M_0$ & $M_\half$ & $A_0$ & $\tan\beta$ & $\text{sign}(\mu)$ & $|\mu|$ & $\text{sign}(B_\mu)\sqrt{|B_\mu|}$ \\
\hline
2311.11 & 666.667 & -3021.77 & 55.8605 & 1 & 1708.6 & $- 99290.9$
\\
\hline
\end{tabular}\\
\vspace{5pt}
\begin{tabular}{|c|c|c|c|c|c|c|c||c|c||c|c|}
\hline
 \multicolumn{12}{|c|}{Low energy spectrum} \\
 \hline
 \hline
$m_{\tilde{g}}$ & $m_{\tilde{q}}$ & $m_{\tilde{t}_1}$& $m_{\tilde{\tau}_1}$ & $m_\chi$ & $m_{\chi_1^\pm}$ &  $m_h$ & $m_A$  & $\Omega\,h^2$ & $\sigma_\text{SI} \text{ [pb]}$ & $\Delta_{v} $ & $\Delta_{\Omega}$\\
\hline
1610 & 2640 & 1430 & 1110 & 292 & 564 & 122 & 564 & 0.138 & $6.11\times 10^{-10}$ & 870  & 91  \\ 
\hline
\end{tabular}
\caption{\label{Table: BM A} $A^0$-pole annihilation benchmark.  \texttt{MicrOmegas} yields $\Omega h^2 = 0.106$ for this model. Dimensionful values are in GeV unless otherwise stated.}
\end{centering}
\end{table}

This $A^0$-pole annihilation benchmark was chosen as an example that it is discoverable at the 13~TeV LHC.  The gluino mass is 1.6 TeV and
\be
\sigma(p\,p\rightarrow \widetilde{g}\,\widetilde{g}) = 8.0 \fb
\ee
at 13~TeV.  The squark masses are 
\begin{eqnarray}
\renewcommand{\arraystretch}{1.1}
\begin{array}{|c||c|c|c|c|c|}
\hline
& \widetilde{q} & \widetilde{d}_3^c & \widetilde{q}_3 & \widetilde{u}_3^c \\
\hline
m\, [\text{TeV}] & 2.6 & 1.7 & 1.9 & 1.4 \\ 
\hline
\end{array}
\end{eqnarray}

It is clear that the gluino decays will involve heavy flavor.  In fact 
\be
\widetilde{g} \rightarrow \overline{t}\,\widetilde{t}_1+\cc \quad 100\%.
\ee
The electroweakinos are very pure with mixings in the $10^{-3}$ range.  Their masses and orderings are
\begin{eqnarray}
\renewcommand{\arraystretch}{1.1}
\begin{array}{|c||c|c|c|c||c|c|}
\hline
&\widetilde{B}&\widetilde{W}&\multicolumn{2}{|c||}{\widetilde{H}}&\widetilde{W}&\widetilde{H}\\ 
\hline
& \chi_1^0 & \chi_2^0 & \chi_3^0  & \chi_4^0 & \chi_1^\pm& \chi_2^\pm \\
\hline
m\, [\text{GeV}] & 292 & 563 & 1370 & 1370 & 564 & 1370 \\ 
\hline
\end{array}
\end{eqnarray}
The lightest stop decays will dominantly involve the bino and winos:
\begin{eqnarray}
\widetilde{t}_1\rightarrow 
\begin{cases}
t\, \chi_1^0& \quad18\% \\
t\,\chi_2^0 & \quad 25\% \\
b\,\chi_1^+ & \quad 53\%  
\end{cases}
\end{eqnarray}
The neutral wino decays to $h\,\chi^0_1$ over 90\% of the time while the charged wino decays to $W^+\,\chi^0_1$ with a 100\% branching ratio.  Therefore, the dominant simplified models to describe the first LHC signals of this benchmark are 
\be
 \widetilde{g} \rightarrow
 % \widetilde{t}\,\,\overline{t} +\cc\rightarrow
\begin{cases} 
  t\,\overline{t}\,\chi_1^0  \quad& 18\%\\
 t\,\overline{t}\, \chi_2^0 \rightarrow   t\, \overline{t}\, h \,\chi_1^0 \quad& 22.5\%\quad[r=0.26]\\
\overline{t}\, \overline{b}\, \chi_1^+ + \cc \rightarrow   \overline{t}\, \overline{b}\, W^+\,\chi_1^0 + \cc \quad& 53\%\quad\,\,\,\,\,[r=0.26]
\end{cases}
\ee
The decay involving Higgs bosons is an interesting feature of this model.

Since they are pure winos, there is a large 13~TeV cross section for 
\be
\sigma(p\,p\rightarrow \chi_2^0\,\chi_1^+) = 14\fb.
\ee  
The $\chi^0_2$ decays will involve boosted Higgs bosons.  There may be a possibility of distinguishing this signal from the electroweak backgrounds \cite{Howe:2012xe, Arbey:2012fa, Baer:2012ts}.

It should also be possible to test this model using ton scale direct detection experiments:
\be
\sigma_\mr{SI} = 6.1\times 10^{-10} \pb.
\ee
Finally, there is a complimentary signal for indirect detection.  The annihilation cross section to bottom quarks is 
\be
\sigma_\mr{ann}\,v = 2.75 \times 10^{-26} \cmcs.
\ee
It is possible that CTA would have sensitivity to this model \cite{Doro:2012xx, Ripken:2012db}.  However, given the uncertainty associated with the dark matter profile it is unlikely that CTA can conclusively exclude a cross section of this size.  

All together, this benchmark involves many interesting signatures with a high degree of complimentarily between many experiments.

%%%%%%%%%%%%%%%%%%%

\clearpage
\stepcounter{model}
\tocless\subsubsection{\Cmodel{A}}
\begin{table}[h!]
\renewcommand{\arraystretch}{1.1}
\setlength{\tabcolsep}{4pt}
\begin{centering}
\begin{tabular}{|c|c|c|c|c||c|c|}
\hline
 \multicolumn{7}{|c|}{Input parameters} \\
 \hline
 \hline
$M_0$ & $M_\half$ & $A_0$ & $\tan\beta$ & $\text{sign}(\mu)$ & $|\mu|$ & $\sqrt{B_\mu}$ \\
\hline
5559.87 & 1900. & 909.796 & 52.0675 & 1 & 1458.4 & $18254.3$
\\
\hline
\end{tabular}\\
\vspace{5pt}
\begin{tabular}{|c|c|c|c|c|c|c|c||c|c||c|c|}
\hline
 \multicolumn{12}{|c|}{Low energy spectrum} \\
 \hline
 \hline
$m_{\tilde{g}}$ & $m_{\tilde{q}}$ & $m_{\tilde{t}_1}$& $m_{\tilde{\tau}_1}$ & $m_\chi$ & $m_{\chi_1^\pm}$ &  $m_h$ & $m_A$  & $\Omega\,h^2$ & $\sigma_\text{SI} \text{ [pb]}$ & $\Delta_{v} $ & $\Delta_{\Omega}$\\
\hline
4210 & 6510 & 4410 & 3930 & 847 & 1190 & 124 & 1660 & 0.105 & $6.7\times 10^{-10}$ & 1500  & 36  \\ 
\hline
\end{tabular}
\caption{\label{Table: BM A2} $A^0$-pole annihilation benchmark.  \texttt{MicrOmegas} yields $\Omega\,h^2 = 0.0915$ for this model.  Dimensionful values are in GeV unless otherwise stated.}
\end{centering}
\end{table}

Table~\ref{Table: BM A2} gives an example of an $A^0$-pole annihilation benchmark that will not yield LHC signatures.  This model has all light squarks above 6 TeV and heavy flavor squarks above 4.4 TeV.  The second lightest electroweakino is a Higgsino at 1.2 TeV which will also be difficult to explore at the LHC.   However, the model should be testable with direct detection:
\be
\sigma_\mr{SI} = 6.7\times 10^{-10} \pb.
\ee
Note that it also has an annihilation cross section of 
\be
\sigma_\mr{ann} v = 4.1\times 10^{-26} \cmcs,
\ee
which, given optimistic assumptions about the dark matter profile, would be also possible to explore with CTA \cite{Doro:2012xx, Ripken:2012db}.

\begin{comment}

\begin{eqnarray}
\begin{array}{|c||c|c|c|c|}
\hline
&\chi^0_1& \chi^0_2& \chi^0_3&\chi^0_4\\
\hline\hline
\widetilde{B} & 99.1\% & 0.\% & 0.6\% & 0.3\% \\
\hline
\widetilde{W} & 0.9\% & 1.9\% & 48.8\% & 48.4 \\
\hline
\widetilde{H}_d & 0.\% & 0.\% & 49.9 & 50. \\
\hline
\widetilde{H}_u & 0.\% & 98.1\% & 0.7\% & 1.2 \\
\hline
\end{array}
\qquad
\begin{array}{|c||c|c|c|c|}
\hline
&\chi^-_1& \chi^-_2& \chi^+_1&\chi^+_2\\
\hline\hline
\widetilde{W}^+ &  &  & 2.4\% & 97.6\% \\
\hline
\widetilde{H}^+_u  &  &  & 97.6\% & 2.4\% \\
\hline
\widetilde{W}^- & 1.4\% & 98.6\% &  & \\
\hline
\widetilde{H}^-_d & 98.6 & 1.4 &  & \\
\hline
\end{array}
\end{eqnarray}

\begin{eqnarray}
\begin{array}{|c||c|c|c|c|c|c|}
\hline
%& $\chi_1^0$ & $\chi_2^0$ & $\chi_3^0$ & $\chi_4^0$ & $\chi_1^\pm$ & $\chi_2^\pm$\\
\hline
m\, [\text{GeV}] & 846.95 & 1192.1 & -1198.4 & 1606.2 & 1190.5 & 1606.2 \\ 
\hline
\end{array}
\end{eqnarray}

\begin{eqnarray}
\begin{array}{|c||c|c|c|c|c|}
\hline
%&$ \widetilde{q}$ & $\widetilde{d}_3^c $& $\widetilde{q}_3$ & $\widetilde{u}_3^c$ \\
\hline
m\, [\text{TeV}] & 6.5 & 4.9 & 5.1 & 4.4 & 4.9 \\ 
\hline
\end{array}
\end{eqnarray}

\end{comment}

%%%%%%%%%%%%%%%%%%%

\clearpage
\stepcounter{model}
\tocless\subsubsection{\Cmodel{A}}
\begin{table}[h!]
\renewcommand{\arraystretch}{1.1}
\setlength{\tabcolsep}{4pt}
\begin{centering}
\begin{tabular}{|c|c|c|c|c||c|c|}
\hline
 \multicolumn{7}{|c|}{Input parameters} \\
 \hline
 \hline
$M_0$ & $M_\half$ & $A_0$ & $\tan\beta$ & $\text{sign}(\mu)$ & $|\mu|$ & $\text{sign}(B_\mu)\sqrt{|B_\mu|}$ \\
\hline
7457.53 & 1300. & 8542.26 & 48.3871 & -1 & 2053.96 & $-143680.$
\\
\hline
\end{tabular}\\
\vspace{5pt}
\begin{tabular}{|c|c|c|c|c|c|c|c||c|c||c|c|}
\hline
 \multicolumn{12}{|c|}{Low energy spectrum} \\
 \hline
 \hline
$m_{\tilde{g}}$ & $m_{\tilde{q}}$ & $m_{\tilde{t}_1}$& $m_{\tilde{\tau}_1}$ & $m_\chi$ & $m_{\chi_1^\pm}$ &  $m_h$ & $m_A$  & $\Omega\,h^2$ & $\sigma_\text{SI} \text{ [pb]}$ & $\Delta_{v} $ & $\Delta_{\Omega}$\\
\hline
3010 & 7750 & 4580 & 5330 & 570 & 1080 & 124 & 1150 & 0.116 & $1.23\times 10^{-13}$ & 2200  & 22  \\ 
\hline
\end{tabular}
\caption{\label{Table: BM A3} $A^0$-pole annihilation benchmark.   \texttt{MicrOmegas} yields $\Omega\,h^2 = 0.133$ for this model.  Dimensionful values are in GeV unless otherwise stated.}
\end{centering}
\end{table}
Table~\ref{Table: BM A3} presents an $A^0$-pole annihilation benchmark which will be unobservable at the LHC and outside the reach of ton scale direct detection.  The squarks are far beyond the reach of the 13~TeV LHC.  The gluino is 3 TeV which will also be difficult for the LHC to discover since the large $\widetilde{g}\,\widetilde{q}$ production channel will not be active.  The next to lightest electroweakino is a pure wino at 1.1 TeV.  The spin-independent direct detection cross section is also beyond the reach of a 1 ton scale experiment.  The annihilation cross section is 
\be
\sigma_\mr{ann}\, v = 2.1\times10^{-26} \cmcs,
\ee  
and is dominated by the $b\,\overline{b}$ channel.   Therefore the only possibility for probing this model will be indirect detection with the CTA experiment \cite{Doro:2012xx, Ripken:2012db}.  

This is an example of a model within the CMSSM which will be very difficult to test.

\begin{comment}
\begin{eqnarray}
\begin{array}{|c||c|c|c|c|}
\hline
&\chi^0_1& \chi^0_2& \chi^0_3&\chi^0_4\\
\hline\hline
\widetilde{B} & 99.9\% & 0.\% & 0.1\% & 0.\% \\
\hline
\widetilde{W} & 0.\% & 99.1\% & 0.6\% & 0.3 \\
\hline
\widetilde{H}_d & 0.\% & 0.\% & 49.9 & 50. \\
\hline
\widetilde{H}_u & 0.1\% & 0.9\% & 49.3\% & 49.7 \\
\hline
\end{array}
\qquad
\begin{array}{|c||c|c|c|c|}
\hline
&\chi^-_1& \chi^-_2& \chi^+_1&\chi^+_2\\
\hline\hline
\widetilde{W}^+ &  &  & 99.5\% & 0.5\% \\
\hline
\widetilde{H}^+_u  &  &  & 0.5\% & 99.5\% \\
\hline
\widetilde{W}^- & 98.7\% & 1.3\% &  & \\
\hline
\widetilde{H}^-_d & 1.3 & 98.7 &  & \\
\hline
\end{array}
\end{eqnarray}

\begin{eqnarray}
\begin{array}{|c||c|c|c|c|c|c|}
\hline
%& $\chi_1^0$ & $\chi_2^0$ & $\chi_3^0$ & $\chi_4^0$ & $\chi_1^\pm$ & $\chi_2^\pm$\\
\hline
m\, [\text{GeV}] & 570.31 & 1081.1 & -1649.1 & 1653.5 & 1081.5 & 1653.8 \\ 
\hline
\end{array}
\end{eqnarray}

\begin{eqnarray}
\begin{array}{|c||c|c|c|c|c|}
\hline
%&$ \widetilde{q}$ & $\widetilde{d}_3^c $& $\widetilde{q}_3$ & $\widetilde{u}_3^c$ \\
\hline
m\, [\text{TeV}] & 7.7 & 5.1 & 5.6 & 4.6 & 5.1 \\ 
\hline
\end{array}
\end{eqnarray}
\end{comment}

%%%%%%%%%%%%%%%%%%%%%%%%%%%%%%%%%%%%%%%%%%%%%%%%%%%%%%
%%%%%%%%%%%%%%%%%%%%%%%%%%%%%%%%%%%%%%%%%%%%%%%%%%%%%%
%%%%%%%%%%%%%%%%%%%%%%%%%%%%%%%%%%%%%%%%%%%%%%%%%%%%%%
\clearpage
\subsection{{\color{STAU}Stau Coannihilation}}

\stepcounter{region}
\setcounter{model}{0}
\setcounter{table}{0}
\setcounter{figure}{0}

\label{Sec: Stau}

Stau co-annihilation is a commonly studied mechanism for setting the dark matter abundance within the CMSSM~\cite{Ellis:1998kh, Arnowitt:2006jq}.  If the stau mass is  
\be
m_{\chi} \le m_{\tilde{\tau}_1} \lsim m_{\chi} + T_{\text{f.o}} \simeq m_{\chi}+ m_\chi/20,
\ee
where $T_{\text{f.o}}$ is the LSP freeze-out temperature, the staus may annihilate with the otherwise inert LSP.   For a range of input parameters, the appropriate rate to achieve the measured relic abundance can be found.   Figs.~\ref{Fig: FullSpaceB} and \ref{Fig: FullSpaceC} show that stau coannihilation regions are characterized by small values of both $M_0$ and $M_\half$ along with $A_0 < 0$.  The region in the $2^\mr{nd}$ quadrant extends up to $A_0/M_0$ approaching zero.  This is where the so-called ``coannihilation strip" resides.  Note that the region extends to large negative values of $A_0/M_0$ and that there exists a second disconnected island with similar phenomenology.  The electroweak tuning spans the range $490 \lsim \Delta_v \lsim 5000$ for this region.

\begin{figure}[h!!]
\centering
\includegraphics[width=0.35\textwidth]{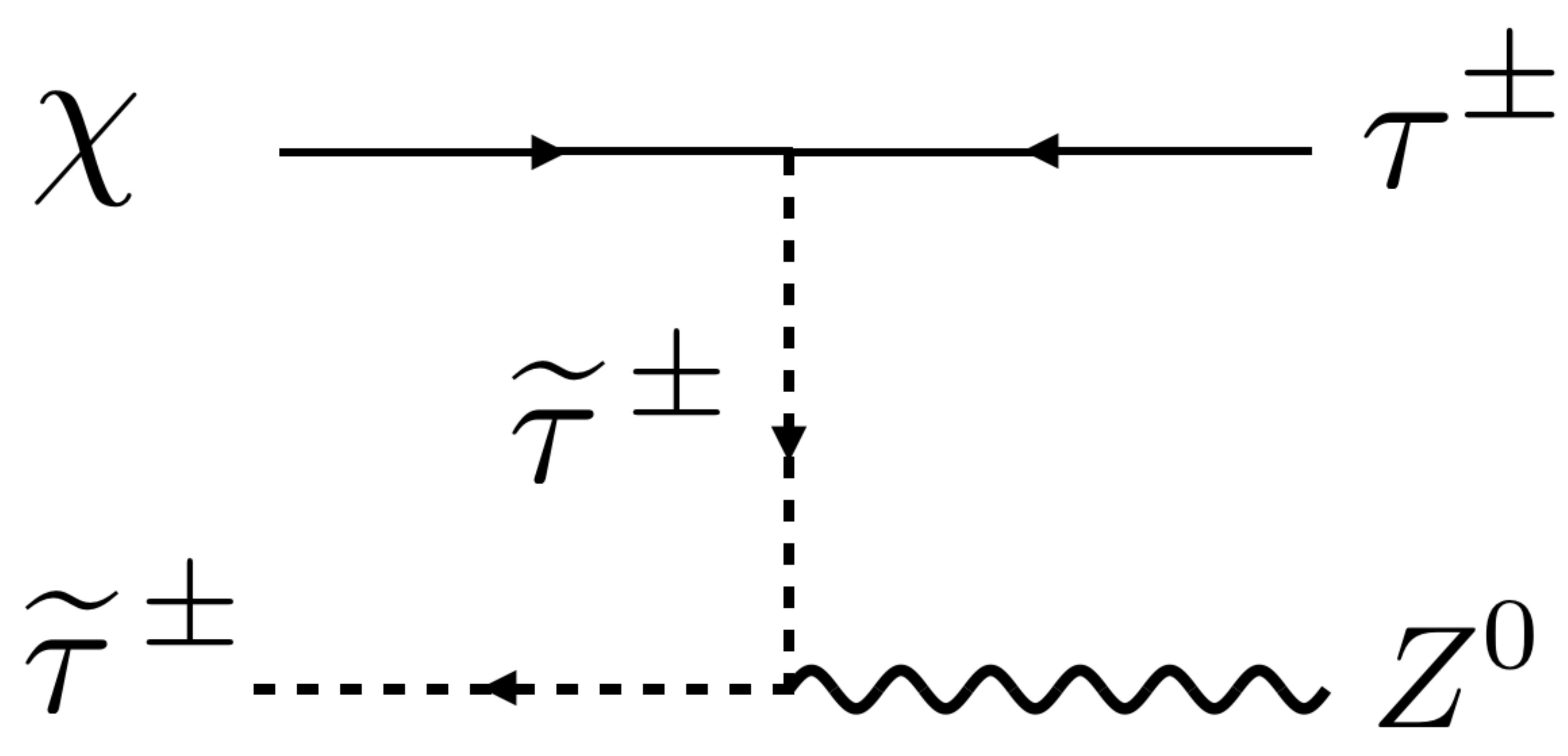}
\caption{A typical diagram which contributes to the dark matter annihilation cross section for the stau coannihilation region.}
\label{Fig: Diagram Stau Coann}
\end{figure}

The LSP mass is less than 800 GeV within the stau coannihilation regions.  Therefore, the stau must be light and the scalar masses must be low.  This in turn forces a fairly light supersymmetric spectrum.  It is clear from the plot of the squark mass versus gluino mass plane shown in Fig.~\ref{Fig: stau Gluino Squark Plane} that nearly all of this region will be observable at the LHC.  

\begin{figure}[t!]
\begin{tabular}{c}
{\bf{\color{STAU}Stau Coannihilation}}\\
\includegraphics[width=0.85\textwidth]{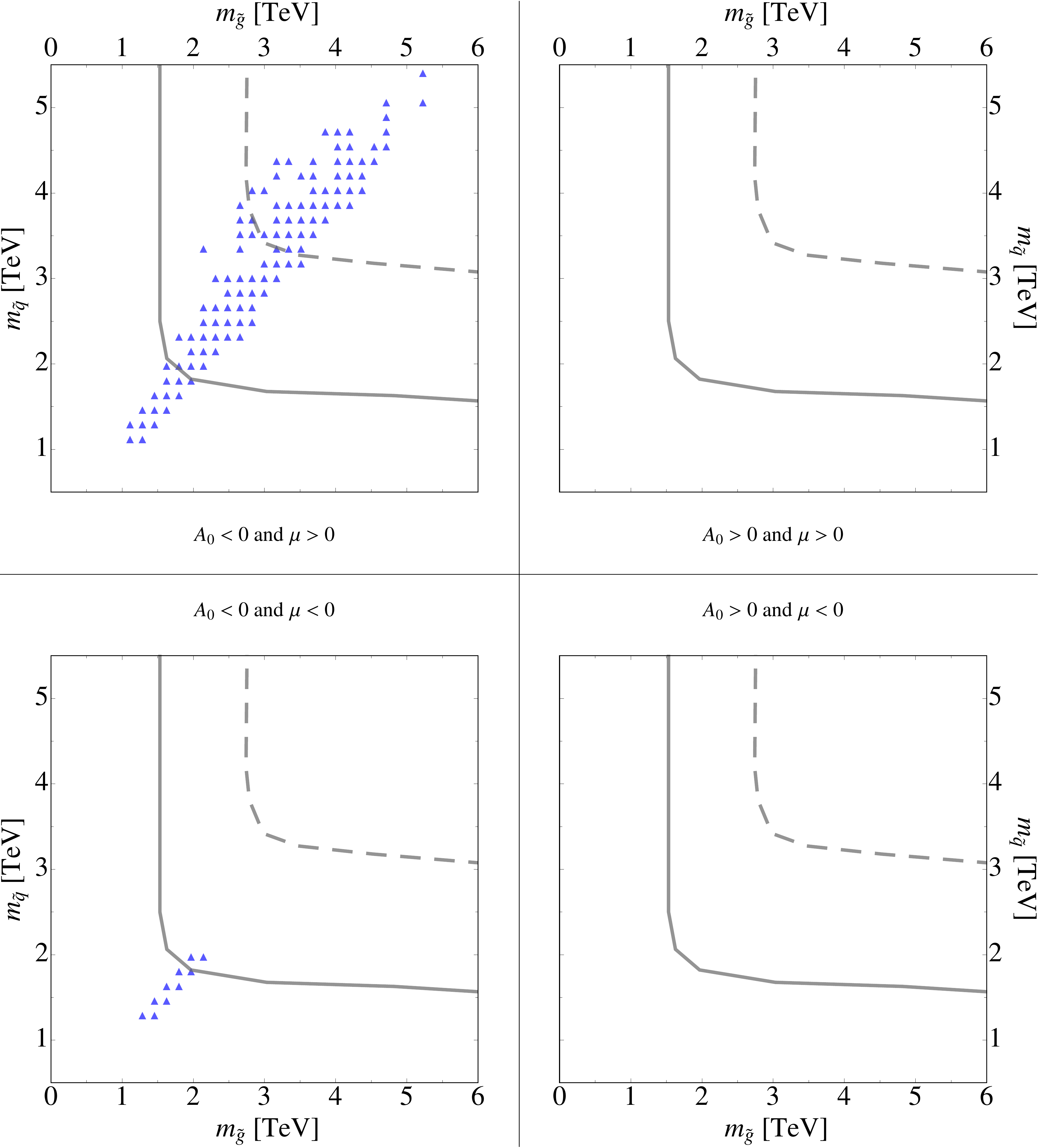}
\end{tabular}
\caption{The squark mass versus gluino mass plane for points in the stau coannihilation region.  Each plot only includes points from the corresponding quadrant.   Also plotted are contours corresponding to 10 squark and/or gluino events for $30 \ifb$ integrated luminosity at $\sqrt{s}=8\TeV$ [solid] and $300 \ifb$ integrated luminosity at $\sqrt{s}=13\TeV$ [dashed].}
\label{Fig: stau Gluino Squark Plane}
\end{figure}

As shown in Fig.~\ref{Fig: stau DD}, the entire $2^\mr{nd}$ quadrant region should be visible to ton scale direct detection experiments.  This implies that the few points in the $2^\mr{nd}$ quadrant with squark mass above 3.5 TeV which may remain unprobed by the 13~TeV LHC will be tested other ways (see Table \ref{Table: BM Stau3} for a benchmark).

\begin{figure}[t!]
\begin{tabular}{c}
{\bf{\color{STAU}Stau Coannihilation}}\\
\includegraphics[width=0.85\textwidth]{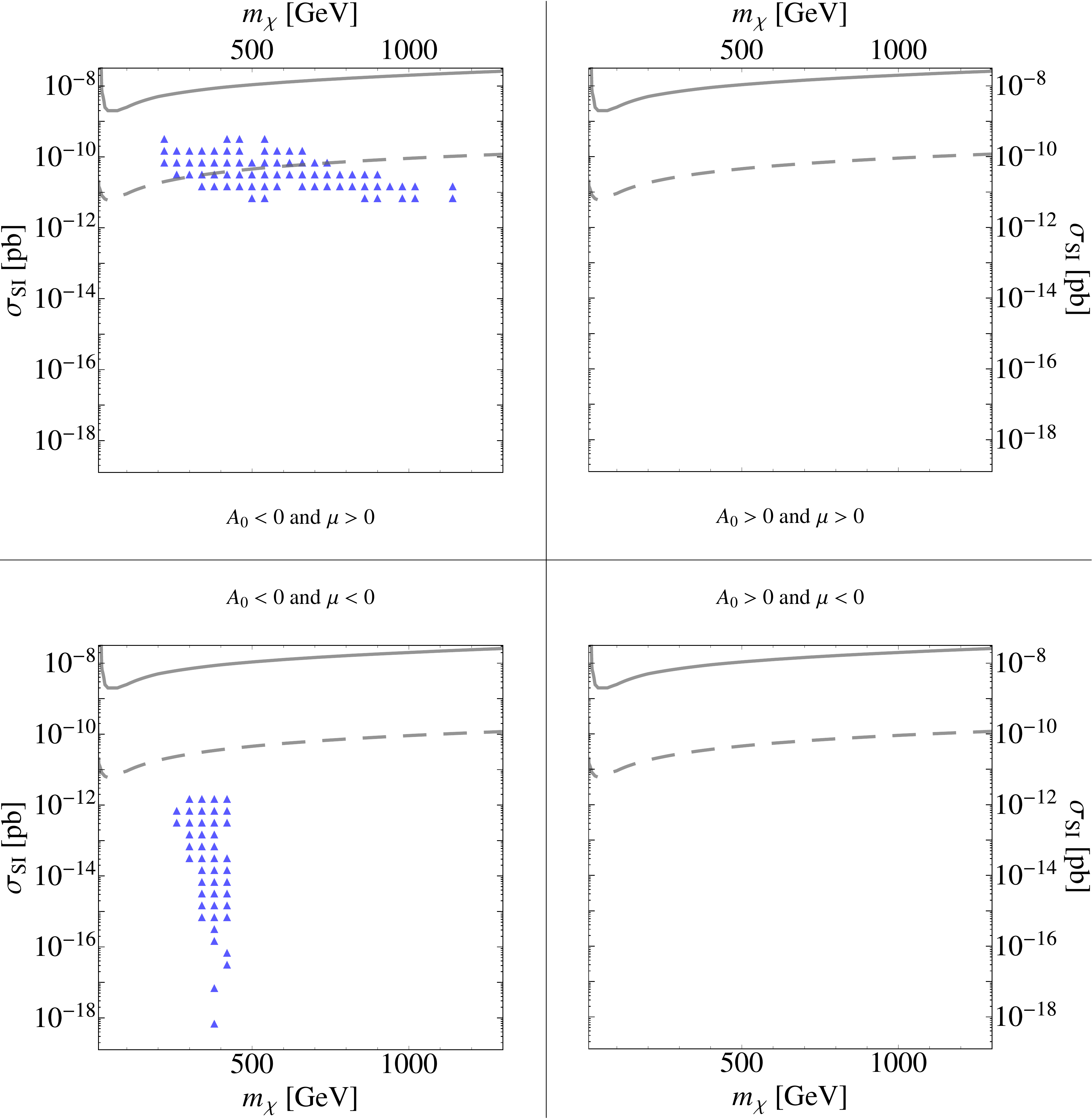}
\end{tabular}
\caption{The spin independent direct detection cross section versus LSP mass for points in the stau coannihilation region.  Each plot only includes points from the corresponding quadrant.  The solid line gives the current bound from XENON100 \cite{Aprile:2012nq} and the dashed line is a projected limit for a ton scale Xenon experiment \cite{DMTools}.}
\label{Fig: stau DD}
\end{figure}

One final characteristic of this region is that the stau is the NLSP.  Depending on the mass splitting between the stau and the LSP, the stau will decay promptly, inside the detector, or long after it has passed through \cite{Citron:2012fg}.  An example of the first and last possibilities  are given in the following discussion and the variety of LHC phenomenology that can result is presented.

%%%%%%%%%%%%%%%%
%%%%%%%%%%%%%%%%
%%%%%%%%%%%%%%%%

\clearpage
\stepcounter{model}
\tocless\subsubsection{\Cmodel{STAU}}
\label{Sec: STAU 1}
\begin{table}[h!]
\renewcommand{\arraystretch}{1.1}
\setlength{\tabcolsep}{4pt}
\begin{centering}
\begin{tabular}{|c|c|c|c|c||c|c|}
\hline
 \multicolumn{7}{|c|}{Input parameters} \\
 \hline
 \hline
$M_0$ & $M_\half$ & $A_0$ & $\tan\beta$ & $\text{sign}(\mu)$ & $|\mu|$ & $\text{sign}(B_\mu)\sqrt{|B_\mu|}$ \\
\hline
765.97 & 900. & -2882.83 & 28.3588 & 1 & 1736.46 & $31794.6$\\
\hline
\end{tabular}\\
\vspace{10pt}
\begin{tabular}{|c|c|c|c|c|c|c|c||c|c||c|c|}
\hline
 \multicolumn{12}{|c|}{Low energy spectrum} \\
 \hline
 \hline
$m_{\tilde{g}}$ & $m_{\tilde{q}}$ & $m_{\tilde{t}_1}$& $m_{\tilde{\tau}_1}$ & $m_\chi$ & $m_{\chi_1^\pm}$ &  $m_h$ & $m_A$  & $\Omega\,h^2$ & $\sigma_\text{SI} \text{ [pb]}$ & $\Delta_{v} $ & $\Delta_{\Omega}$\\
\hline
1990 & 1950 & 988 & 389 & 386 & 736 & 125 & 1580 & 0.103 & $2.21\times 10^{-11}$ & 1400  & 160  \\ 
\hline
\end{tabular}
\caption{\label{Table: BM Stau1} Stau coannihilation benchmark with a promptly decaying stau.  Dimensionful values are in GeV unless otherwise stated.}
\end{centering}
\end{table}

The first stau coannihilation benchmark features a promptly decaying stau; the mass splitting between the lightest stau and the LSP is $m_{\tilde{\tau}_1} - m_\chi = 3.36 \GeV$ so that the decay $\widetilde{\tau} \rightarrow \tau\,\chi$ can proceed on-shell.  However, given that this splitting is small, the resulting $\tau$ will be very soft and essentially invisible at the LHC.  In practice, any $\widetilde{\tau}_1$ which is produced results in $\MET\!\!$.

In principle one would like a direct confirmation that the stau and LSP were degenerate in order to determine that the stau coannihilation was the dominant process for determining the relic density.  The cleanest channel would be through direct stau production.  At the 13~TeV this process has a 0.55 fb cross section.  However, since the stau decays only produce $\MET\!\!$, this would be essentially impossible to distinguish from the background.

Another avenue would be to produce staus in the decays of neutralinos and charginos.  The electroweakinos are nearly pure with the following spectrum
\begin{eqnarray}
\renewcommand{\arraystretch}{1.1}
\begin{array}{|c||c|c|c|c||c|c|}
\hline
&\widetilde{B}&\widetilde{W}&\multicolumn{2}{|c||}{\widetilde{H}}&\widetilde{W}&\widetilde{H}\\ 
\hline
& \chi_1^0 & \chi_2^0 & \chi_3^0  & \chi_4^0 & \chi_1^\pm& \chi_2^\pm \\
\hline
m\, [\text{GeV}] & 386 & 736 & 1690 & 1700 & 736 & 1700 \\ 
\hline
\end{array}
\end{eqnarray}
The degeneracy between $\widetilde{\tau}_1$ and the LSP is driven by the presence of a large $A$-term for the stau at low energies.  Therefore, the stau is the only slepton with a mass below the lightest chargino; the other sleptons have masses between 800-900 GeV.  This impacts the decay pattern for the lightest chargino and second lightest neutralino.  Specifically 
\bea
\chi^+_1 &\rightarrow& \widetilde{\tau}_1^+ \,\nu_\tau\quad \,97\%\\
\chi^0_2 &\rightarrow& \widetilde{\tau}_1^\pm \,\tau^\mp \quad 98\%
\eea
The $\chi^+_1$ decay effectively results in $\MET\!\!$ while the $\chi^0_2$ decay yields a $\tau$ and $\MET\!\!$.  The best channel for observing these states at the 13~TeV LHC would be though $\chi_2^0 \, \chi^\pm_1$ production
\be
\sigma(p\,p\rightarrow \chi_2^0 \, \chi^\pm_1) = 2.3\fb.
\ee
Given that the final state is a $\tau$ and $\MET\!\!$ this will be very challenging to observe.  However, it should be possible to discover these states at a TeV $e^+\,e^-$ collider.  

The largest production cross section at the 13~TeV LHC is light squark direct production:
\bea
\sigma(p\,p\rightarrow \widetilde{q}\,\widetilde{q}) &=& 6.0 \fb;\\
\sigma(p\,p\rightarrow \widetilde{q}\,\widetilde{q}^{\,*}) &=& 1.0\fb.
\eea
These two contributions will both provide similar signatures so for the purpose of arguing the first signals of this model, these can be combined. The right handed squarks decay to $q + \chi$ while left handed squarks decay to $q' + \chi^+_1$.  As described above, $\chi^+_1$ yields $\MET\!\!$ so the signature will jets + $\MET\!\!$ with the caveat that the phase space distribution for the left handed squark decay will be distorted by the cascade.

The next largest colored production channel is squark-gluino associated production followed by gluino pair production
\bea
\sigma(p\,p\rightarrow \widetilde{g}\,\widetilde{q}) &=& 4.6 \fb;\\
\sigma(p\,p\rightarrow \widetilde{g}\,\widetilde{g}) &=& 0.48 \fb.
\eea
The stop is the lightest squark due to the large $A$-term.  
\begin{eqnarray}
\begin{array}{|c||c|c|c|c|c|}
\hline
& \widetilde{q} & \widetilde{b}_1 & \widetilde{b}_2 & \widetilde{t}_1 & \widetilde{t}_2 \\
\hline
m\, [\text{TeV}] & 1.9 & 1.5 & 1.7 & 0.99 & 1.5 \\ 
\hline
\end{array}
\end{eqnarray}
Therefore, the gluino decay patters are dominated by heavy flavor:
\begin{eqnarray}
\widetilde{g} \rightarrow \widetilde{q}_3 \,\overline{q}_3+\cc\quad 
94\%\times
\begin{cases} 
\,\,\widetilde{t}_1\,\overline{t} + \cc & 45\%\\
\,\,\widetilde{t}_2\,\overline{t} + \cc & 22\%\\
\,\,\widetilde{b}_1\,\overline{b}+ \cc & 20\%\\
\,\,\widetilde{b}_2\,\overline{b}+ \cc & 8.0\%
\end{cases}
\end{eqnarray}
The rest of the gluinos decay to light flavor quarks and squarks.  

The large $A$-term impacts the decays of the stops and sbottoms since it gives these states a large coupling to the Higgs boson.  Given
\bea
\widetilde{t}_1 &\rightarrow& 
\begin{cases}
\chi^0_1\,t &\quad 45\% \\
\chi^+_1\,b &\quad 42\% 
\end{cases}\\
\widetilde{t}_2 &\rightarrow& \widetilde{t}_1\, h\,\, \quad\quad \quad74\%\\
\widetilde{b}_1 &\rightarrow& \widetilde{t}_1 \, W^- \,\,\quad\quad71\%\\
\widetilde{b}_2 &\rightarrow& \widetilde{b}_1\, h \,\,\quad\quad\quad 27\%
\eea
a gluino produces heavy flavor and a boosted Higgs in its decays 18\% of the time.  Relying on the gluino-squark associated production channel results in a production cross section of
\be
\sigma(p\,p\rightarrow h\,t\,\MET\! X +\cc) = 0.82\fb
\ee
which is $\OO(100)$ events at $100\ifb$.  This is a distinctive signal to search for at the 13~TeV LHC that currently has not been targeted.

%%%%%%%%%%%
%%%%%%%%%%%
%%%%%%%%%%%
\clearpage
\stepcounter{model}
\tocless\subsubsection{\Cmodel{STAU}}

\begin{table}[h!]
\renewcommand{\arraystretch}{1.1}
\setlength{\tabcolsep}{4pt}
\begin{centering}
\begin{tabular}{|c|c|c|c|c||c|c|}
\hline
 \multicolumn{7}{|c|}{Input parameters} \\
 \hline
 \hline
$M_0$ & $M_\half$ & $A_0$ & $\tan\beta$ & $\text{sign}(\mu)$ & $|\mu|$ & $\sqrt{B_\mu}$ \\
\hline
259.515 & 900.862 & -2296.71 & 9.23077 & -1 & 1555.68 & $8702.87$
\\
\hline
\end{tabular}\\
\vspace{5pt}
\begin{tabular}{|c|c|c|c|c|c|c|c||c|c||c|c|}
\hline
 \multicolumn{12}{|c|}{Low energy spectrum} \\
 \hline
 \hline
$m_{\tilde{g}}$ & $m_{\tilde{q}}$ & $m_{\tilde{t}_1}$& $m_{\tilde{\tau}_1}$ & $m_\chi$ & $m_{\chi_1^\pm}$ &  $m_h$ & $m_A$  & $\Omega\,h^2$ & $\sigma_\text{SI} \text{ [pb]}$ & $\Delta_{v} $ & $\Delta_{\Omega}$\\
\hline
1980 & 1820 & 1070 & 384 & 384 & 732 & 122 & 1680 & 0.116 & $1.52\times 10^{-14}$ & 1300  & 33  \\ 
\hline
\end{tabular}
\caption{\label{Table: BM Stau2} Stau coannihilation benchmark with a long lived stau.  Dimensionful values are in GeV unless otherwise stated.}
\end{centering}
\end{table}

This stau coannihilation benchmark features a long lived stau; the mass splitting between the lightest stau and the LSP is $m_{\tilde{\tau}_1} - m_\chi = 0.28 \GeV$.  The stau lifetime will be $\OO(10^{-2} \text{ s})$ \cite{Citron:2012fg}.  It is stable on detector time scales and will manifest as a CHArged Massive Particle (CHAMP).  

Using 7~TeV data, ATLAS has already placed bound of $m_{\tilde{\tau}_1} \lsim 280 \GeV$ on the direct production of long lived staus \cite{Aad:2012pra}.  At the 13~TeV LHC, 
\be
\sigma(p\,p\rightarrow \widetilde{\tau}_1^+\,\widetilde{\tau}_1^-) = 0.59 \fb.
\ee
Since these particles are CHAMPs it should only require a handful of event to discover them.  

Decays involving the staus and the tau-sneutrinos are particularly relevant for LHC searches.  Their masses are
\begin{eqnarray}
\begin{array}{|c||c|c|c|}
\hline
& \widetilde{\tau}_1^\pm & \widetilde{\tau}_2^\pm & \widetilde{\nu}_\tau \\
\hline
m\, [\text{GeV}] & 384 & 644 & 638  \\ 
\hline
\end{array}
\end{eqnarray}

The electroweakinos are very pure:
\begin{eqnarray}
\renewcommand{\arraystretch}{1.1}
\begin{array}{|c||c|c|c|c||c|c|}
\hline
&\widetilde{B}&\widetilde{W}&\multicolumn{2}{|c||}{\widetilde{H}}&\widetilde{W}&\widetilde{H}\\ 
\hline
& \chi_1^0 & \chi_2^0 & \chi_3^0  & \chi_4^0 & \chi_1^\pm& \chi_2^\pm \\\hline
m\, [\text{GeV}] & 384 & 732 & 1580 & 1580 & 732 & 1580 \\ 
\hline
\end{array}
\end{eqnarray}
Given that the LSP is very nearly pure bino, direct detection is too small to be observed.  The winos will play an interesting role in the potential collider signatures.

The gluino and squarks are observable at the 13~TeV LHC.  The patterns of squark masses determines the gluino branching ratios into different flavors:
\begin{eqnarray}
\begin{array}{|c||c|c|c|c|c|}
\hline
& \widetilde{q} & \widetilde{b}_1 & \widetilde{b}_2 & \widetilde{t}_1 & \widetilde{t}_2 \\
\hline
m\, [\text{TeV}] & 1.8 & 1.5 & 1.7 & 1.1 & 1.6 \\ 
\hline
\end{array}
\end{eqnarray}
The gluino has a similar decay pattern to the previous benchmark with decays involving stops and sbottoms 73\% of the time. 

The most interesting signature of this model is the presence of a CHAMP in the gluino and squark decays.  In fact, the discovery mode seems likely to be squark pair production:
\bea
\sigma(p\,p \rightarrow \widetilde{q}\,\widetilde{q}) &\simeq& 10\fb;\\
\sigma(p\,p \rightarrow \widetilde{q}\,\widetilde{q}^*) &\simeq& 1.9\fb.
\eea
To understand the light flavor squark decay patterns, take the up squark as an example:
\be
\widetilde{u}_R \rightarrow u \, \chi_1^0 \quad 100\%.
\ee
The decay pattern of $\widetilde{u}_L$ is more interesting since this can result in CHAMPs.  To determine the fraction of $\widetilde{u}_L$ decay modes that have CHAMPs requires knowledge of the following decays
\bea
\widetilde{u}_L \rightarrow
\begin{cases}
u\,\chi^0_2\quad\, 33\% \\
d \, \chi_1^+ \quad 66\%
\end{cases}
&\quad&
\widetilde{\chi}^0_2 \rightarrow
\begin{cases} 
\,\,\widetilde{\nu}_\tau \, \overline{\nu}_\tau + \cc& 21\%\\
\,\,\widetilde{\tau}^+_1\,\tau^- + \cc & 1.4\%\\
\,\,\widetilde{\tau}^+_2\,\tau^- + \cc &18\%
\end{cases}
\qquad
\widetilde{\chi}^+_1 \rightarrow 
\begin{cases} 
\,\,\widetilde{\nu}_\tau \, \tau^+ & 21\%\\
\,\,\widetilde{\tau}^+_1\,\nu_\tau  & 1.4\%\\
\,\,\widetilde{\tau}^+_2\,\nu_\tau  &18\%
\end{cases}\nonumber\\
\widetilde{\tau}^+_2 &\rightarrow&
\begin{cases} 
Z^0 \, \widetilde{\tau}^+_1&\quad 18\%\\
h \, \widetilde{\tau}^+_1 &\quad 19\%\\
\end{cases}
\qquad\quad
\widetilde{\nu}_\tau \rightarrow \widetilde{\tau}^+_1 \, W^- \quad 36\% 
\eea
Noting that squark pair production leads to at least one left handed squark roughly $75\%$ of the time, all of this information can be combined together to give
\be
\sigma(p\,p \rightarrow \text{CHAMP} + j + X) \simeq 1.4\fb.
\ee
This cross section is larger than that for direct stau production by more than a factor of two.  This benchmark would provide an an early discovery for the LHC.\footnote{The search efficiency depends in a non-trivial way on the velocity of the CHAMP and the properties of the rest of the event \cite{Aad:2012pra}.  A careful consideration of these effects is required to be sure which signal would be observed first at the 14 TeV LHC.}

The $A$-term is not as large in this model as it was in the previous benchmark.  This implies that there are many light sleptons.  It would be complimentary to search for this model via squark production with cascade decays involving the other sleptons.  This would result in a classic jets, $\MET\!\!$, and same-sign leptons signature.

%%%%%%%%%%%
%%%%%%%%%%%
%%%%%%%%%%%
%\clearpage
\stepcounter{model}
\tocless\subsubsection{\Cmodel{STAU}}
\begin{table}[h!]
\renewcommand{\arraystretch}{1.1}
\setlength{\tabcolsep}{4pt}
\begin{centering}
\begin{tabular}{|c|c|c|c|c||c|c|}
\hline
 \multicolumn{7}{|c|}{Input parameters} \\
 \hline
 \hline
$M_0$ & $M_\half$ & $A_0$ & $\tan\beta$ & $\text{sign}(\mu)$ & $|\mu|$ & $\text{sign}(B_\mu)\sqrt{|B_\mu|}$\\
\hline
3389.47 & 1733.33 & -5503.95 & 59.1701 & 1 & 3660.9 & $-224661$
\\
\hline
\end{tabular}\\
\vspace{5pt}
\begin{tabular}{|c|c|c|c|c|c|c|c||c|c||c|c|}
\hline
 \multicolumn{12}{|c|}{Low energy spectrum} \\
 \hline
 \hline
$m_{\tilde{g}}$ & $m_{\tilde{q}}$ & $m_{\tilde{t}_1}$ & $m_{\tilde{\tau}_1}$ & $m_\chi$ & $m_{\chi_1^\pm}$ &  $m_h$ & $m_A$  & $\Omega\,h^2$ & $\sigma_\text{SI} \text{ [pb]}$ & $\Delta_{v} $ & $\Delta_{\Omega}$\\
\hline
3790 & 4670 & 2730 & 779 & 779 & 1470 & 126 & 1040 & 0.134 & $2.78\times 10^{-11}$ & 3800  & 670  \\ 
\hline
\end{tabular}
\caption{Stau coannihilation benchmark.  Dimensionful values are in GeV unless otherwise stated.}
\label{Table: BM Stau3}
\end{centering}
\end{table}

This benchmark provides an example which would likely be observed first in direct detection.  
\be
\sigma_\mr{SI} = 2.8\times 10^{-11}\pb.
\ee
Given $m_{\tilde{\tau}_1} - m_\chi = 0.13 \GeV$, the $\widetilde{\tau}_1$ is a CHAMP.  The stau pair production cross section is
\be
\sigma(p\,p\rightarrow \widetilde{\tau}_1^+\,\widetilde{\tau}_1^-) = 0.019 \fb 
\ee
at the 13~TeV LHC.  Given $100\ifb$ of data, there would be roughly 2 events.  Assuming a decent efficiency for CHAMP searches, it is likely that this model could also be probed at the LHC.

%%%%%%%%%%%%%%%%%%%%%%%%%%%%%%%%%%%%%%%%%%%%%%%%%%%%%%
%%%%%%%%%%%%%%%%%%%%%%%%%%%%%%%%%%%%%%%%%%%%%%%%%%%%%%
%%%%%%%%%%%%%%%%%%%%%%%%%%%%%%%%%%%%%%%%%%%%%%%%%%%%%% 
\clearpage
\subsection{{\color{STOP}Stop Coannihilation}}

\stepcounter{region}
\setcounter{model}{0}
\setcounter{table}{0}
\setcounter{figure}{0}

\label{Sec: Stop}
The remaining regions are characterized by stop coannihilation.\footnote{Although the dominant loop corrections to the relic abundance for stop coannihilaiton have been computed \cite{Harz:2012fz}, we use the tree results as implimented in \texttt{DarkSUSY} in this paper.}  If the stop mass is within
\be
m_{\chi} \le m_{\tilde{t}_1} \lsim m_{\chi} + T_{\text{f.o}} \simeq m_{\chi}+ m_\chi/20,
\ee
coannihilations can be important.  Stop coannihilation regions exist for large swaths of the 1\st, 3\rd, and 4\nth quadrants.  The Higgs mass constraint is incompatible with potential stop coannihilation points in the 2\nd quadrant.  The electroweak tuning spans the range $2300 \lsim \Delta_v \lsim 57000$ for this region.

\begin{figure}[h!!]
\centering
\includegraphics[width=0.35\textwidth]{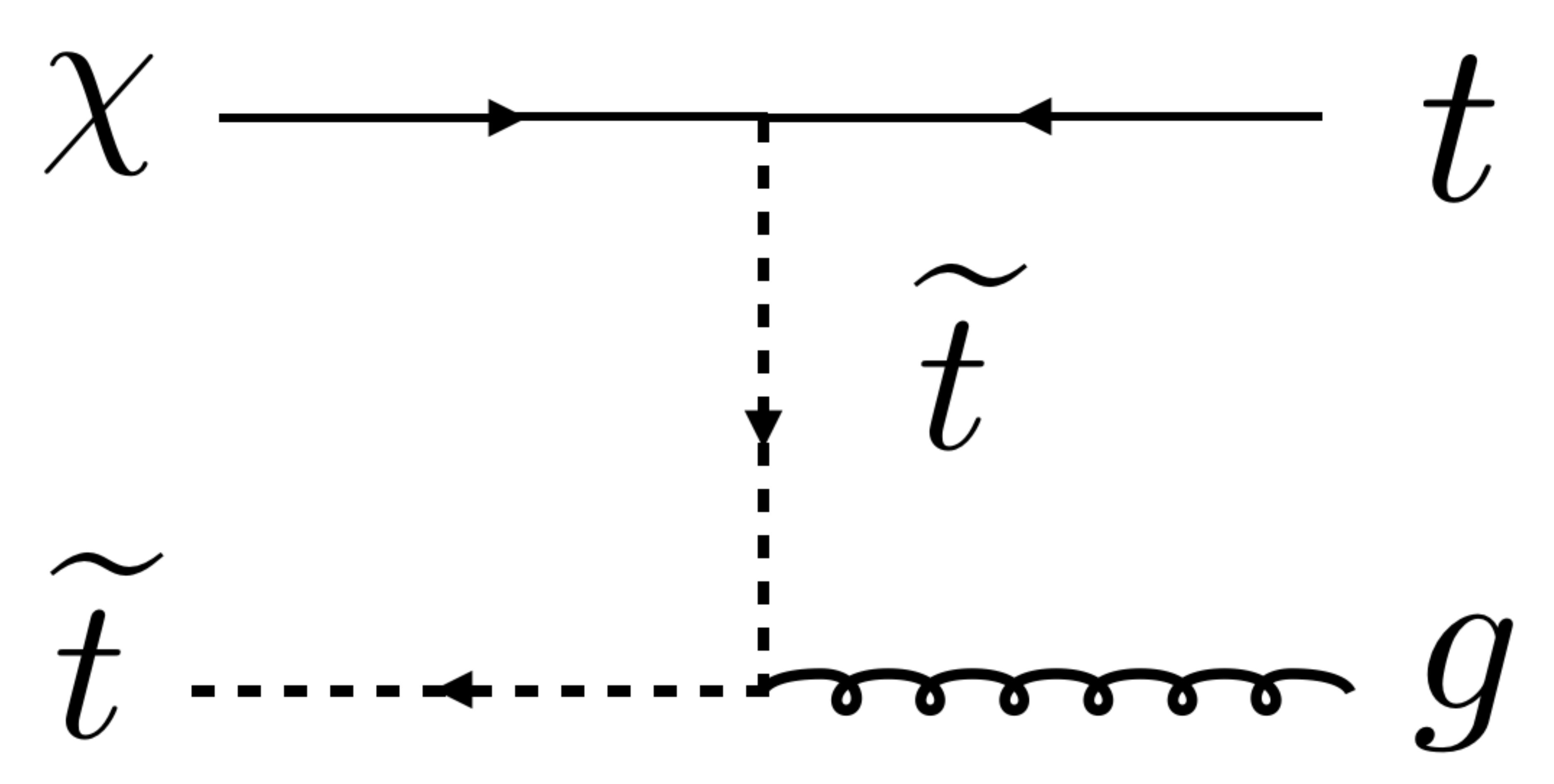}
\caption{A typical diagram which contributes to the dark matter annihilation cross section for the stop coannihilation region.}
\label{Fig: Diagram Stop Coann}
\end{figure}

In order to realize stop coannihilation within the CMSSM requires large $A$-terms at the low scale.  These off-diagonal parameters in the stop mass matrix lead to the suppression of one eigenvalue due to ``level-repulsion."  Quantitatively, the stop coannihilation regions are characterized by $3\lsim A_0/M_0 \lsim 6$ and  $-15\lsim A_0/M_0 \lsim -3$.  These ranges can be understood by considering how the signs of the $A$-terms enter the RGEs (see Eq.~(\ref{eq:ATermRGE})); for $A_0 > 0$, the low energy value does not change dramatically while for $A_0 < 0$, the RGE tends to drive the magnitude of the $A$-term to smaller values so that a larger GUT scale value is required to push the lightest stop down towards the LSP.

This region is characterized by larger $M_\half$ than the corresponding values which occur for the stau coannihilation regions.  Since the stop coannihilation annihilation cross section involves factors of the strong gauge coupling, the measured relic abundance can be achieved for larger values of the LSP mass.  This also implies that the majority of this region will lie outside the reach of the 13~TeV LHC.   The range of squark and gluino masses are shown in Fig.~\ref{Fig: stop Gluino Squark Plane}.  

Since the relic abundance is dominated by coannihilation, the LSP can be very bino-like.  Hence, the tree-level direct detection cross sections can be very small, as shown in Fig.~\ref{Fig: stop DD}.  However, since the low energy $A$-terms are large, there is a 1-loop diagram which can possibly bring many of these points into reach.  Loop corrections to direct detection are discussed in more detail in \refsec{Stop1LoopDD}.

\begin{figure}[t!]
\begin{tabular}{c}
{\bf{\color{STOP}Stop Coannihilation}}\\
\includegraphics[width=0.85\textwidth]{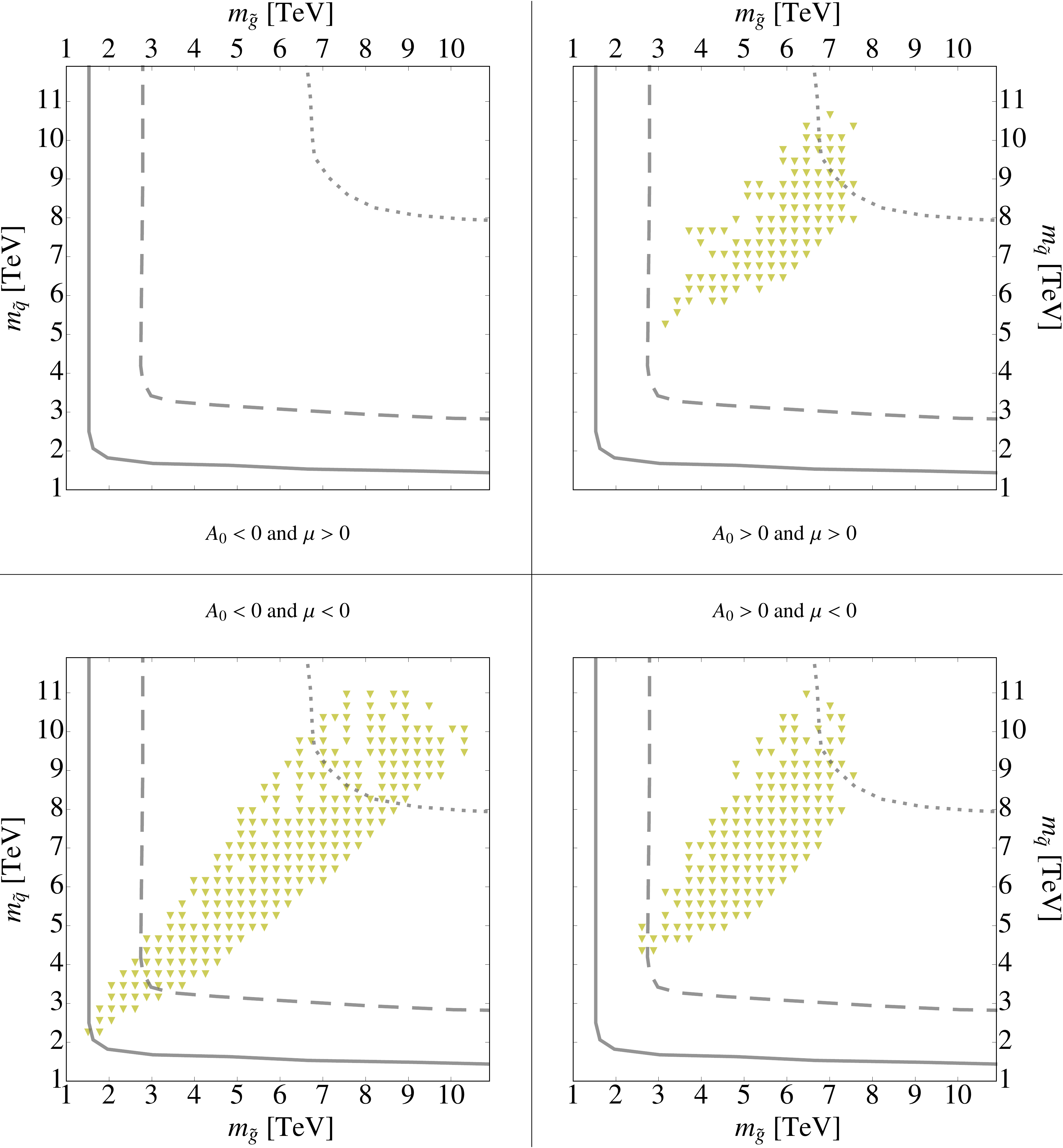}
\end{tabular}
\caption{The squark mass versus gluino mass plane for points in the stop coannihilation region.   Also plotted are contours corresponding to 10 squark and/or gluino events for $30 \ifb$ integrated luminosity at $\sqrt{s}=8\TeV$ [solid], $300 \ifb$ integrated luminosity at $\sqrt{s}=13\TeV$ [dashed], and $3000 \ifb$ integrated luminosity at $\sqrt{s}=33\TeV$ [dotted].}
\label{Fig: stop Gluino Squark Plane}
\end{figure}

\begin{figure}[t!]
\begin{tabular}{c}
{\bf{\color{STOP}Stop Coannihilation}}\\
\includegraphics[width=0.85\textwidth]{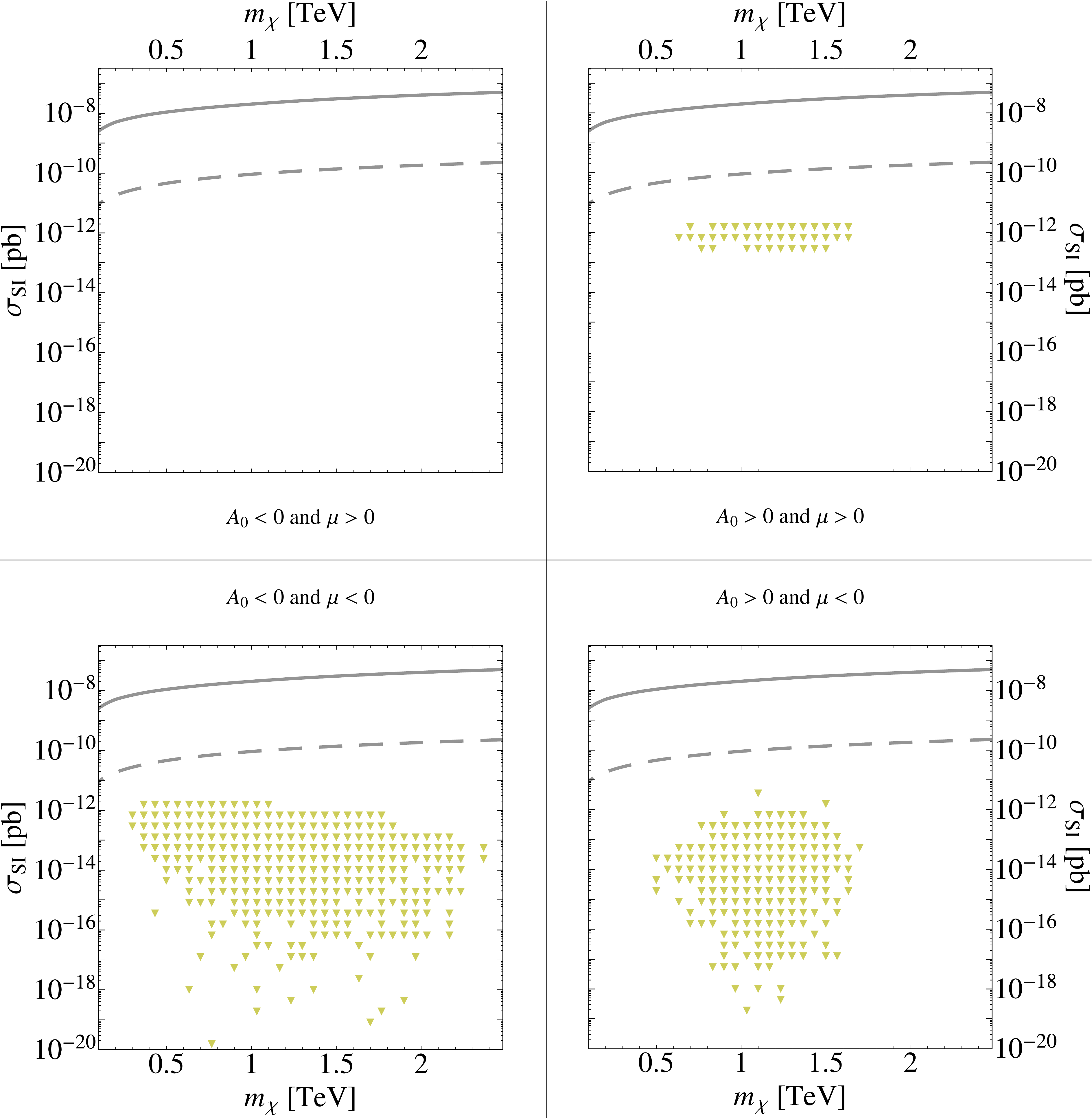}
\end{tabular}
\caption{The  leading-order spin independent direct detection cross section versus LSP mass for points in the stop coannihilation region.  Each plot only includes points from the corresponding quadrant.  The solid line gives the current bound from XENON100 \cite{Aprile:2012nq} and the dashed line is a projected limit for a ton scale Xenon experiment \cite{DMTools}.  Loop corrections could significantly alter these results.}
\label{Fig: stop DD}
\end{figure}

\tocless\subsubsection{Large 1-Loop Contributions to Direct Detection}
\label{Sec: Stop1LoopDD}
Figure \ref{Fig: stop DD} contains many examples of models with direct detection cross sections which are far too low to ever be discovered.  This behavior results because the LSP is approaching the limit of pure bino.  In this limit, the $\mu$ term is becoming heavy which implies that the scalar masses are also becoming large.  Therefore, the low energy $A$-term for the stop must also become large in this region to result in a $\widetilde{t}_1$ eigenvalue which is nearly degenerate with the LSP.  

These large $A$-terms have another important physical consequence --- they can contribute to direct detection at 1 loop via the diagram in Fig.~\ref{Fig: Diagram 1LoopDD}.  The appropriate 1-loop calculation has been performed \cite{Drees:1992rr, Drees:1993bu, Djouadi:2001kba}.  However, the region of parameter space resulting in stop coannihilation has not been fully explored.  While a full reevaluation of the 1-loop diagrams are beyond the scope of this work, it is possible to estimate the size of these contributions.

\begin{figure}[h!!]
\centering
\includegraphics[width=0.35\textwidth]{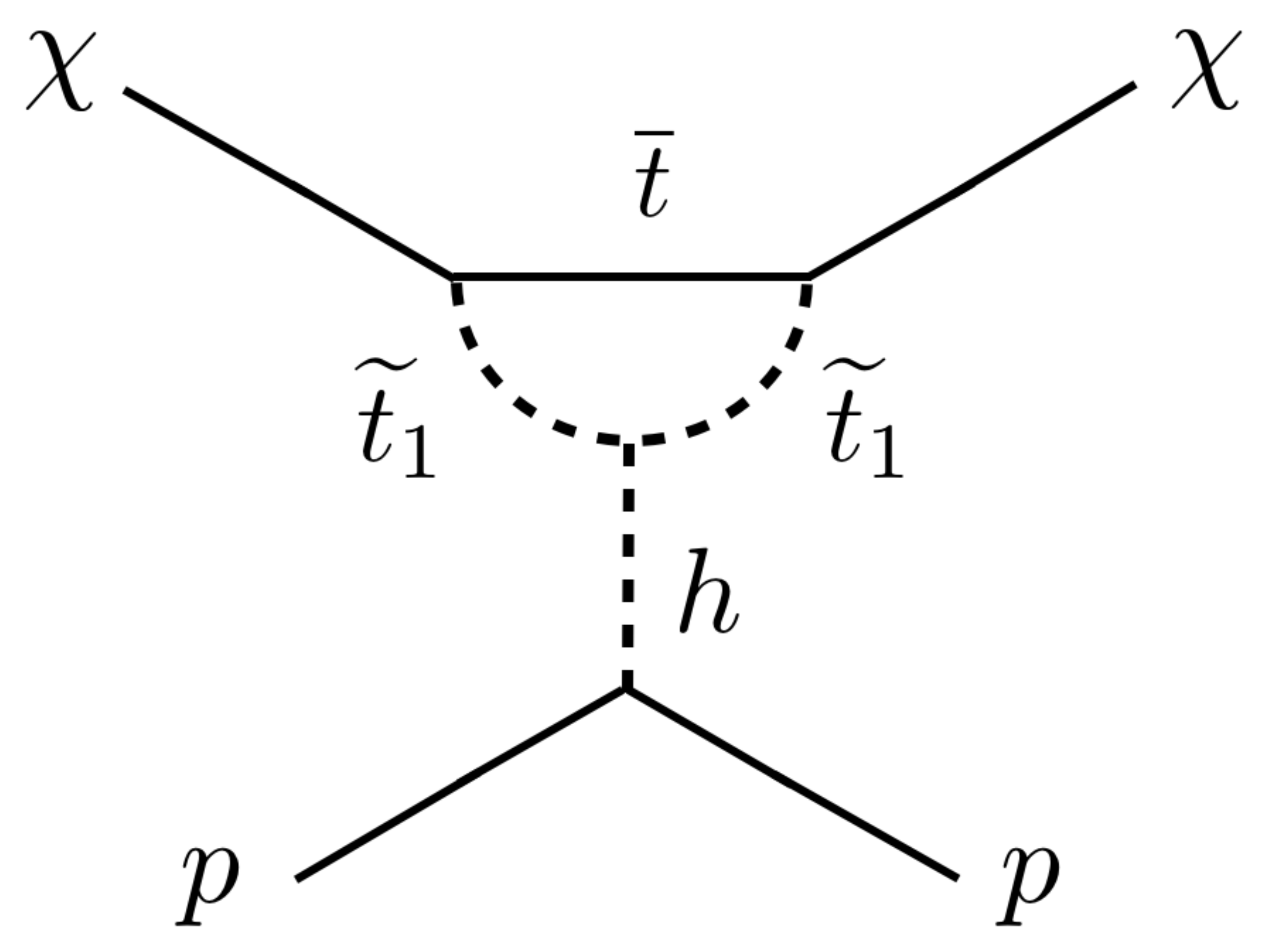}
\caption{One of the 1-loop direct detection diagrams which can dominate over the tree level contribution in the stop coannihilation region.}
\label{Fig: Diagram 1LoopDD}
\end{figure}

Consider the effective operator for Higgs mediated spin-independent direct detection after the Higgs boson has been integrated out:
\begin{eqnarray}
\OO_\mr{SI} = \left(\frac{y_q\,y_\chi}{m_h^2}\right)\,\overline{\chi}\, \chi\,\overline{q}\,q,
\end{eqnarray}
where $y_\chi$ is the effective coupling between the dark matter and the Higgs and $y_q$ is the quark Yukawa coupling.  The estimate for the size of the correction from Fig.~\ref{Fig: Diagram 1LoopDD} is
\begin{eqnarray}
y_\chi \sim \frac{g'^2\,N_c}{16\, \pi^2} \left(Q_Y^{q}\, Q_Y^{u^c}\right) \frac{ A_t (m_{\tilde{t}_1}+C\,m_t)}{m_{\tilde{t}_1}^2}
\end{eqnarray}
where $g'$ is the hypercharge gauge coupling, $N_c = 3$ is the number of colors, $Q_Y^i$ is the hypercharge of the particle $i$, and $C$ is a numerical constant that has not been computed.  Note that for this region $m_{\tilde{t}_1} \gg m_t$ so that it is safe to ignore the contribution to the estimate that is proportional to $m_t$.

Following the conventions of \cite{Cohen:2010gj}, this yields
\begin{eqnarray}
\sigma_\mr{SI}^\mr{1-loop} \sim 3\times 10^{-13}\pb\times\left(\frac{A_t}{m_{\tilde{t}_1}}\right)^2
\end{eqnarray}
The value of $A_t/m_{\tilde{t}_1}$ can range from 
\be
2 \lsim A_t/m_{\tilde{t}_1} \lsim  14.
\ee
In the absence of a accidental suppression, there is a class of models for which this process would be observable using ton scale technology.  This will be discussed for both benchmarks below.

\tocless\subsubsection{Possible Issue for Stop Coannihilation at Large Mass}
\label{Sec: StopFailure}
Before moving on to the benchmarks, there is one more relevant subtlety for some of the stop coannihilation models that is worth discussing.  The stop mass can reach $\OO(2\TeV)$ in the stop coannihilation region.  Naively this implies that the freeze-out of the stop-neutralino system will occur at $m_\chi/20 \sim O(100\GeV)$.  The electroweak phase transition is also occurring around this temperature.   Extrapolating the results of \cite{D'Onofrio:2012ni}, the Standard Model with a Higgs boson mass of 125~GeV begins the second order phase transition starting from $\vev{H} = 0$ at around $T \simeq 170\GeV$ and finally ending in the broken phase at $T\simeq 80-90 \GeV$. 

The lightest stop mass in this region is driven by the non-zero contribution from $A_t \times \vev{H}$.  If the Higgs vev is changing during freeze-out, the stop mass can be drastically different and the naive freeze-out calculation breaks down.  There are a variety of effects to consider.  For example, the neutralino-stop system can undergo a second period of annihilation after the Higgs field settles into its final value --- this is reminiscent of ``changing dark matter" models \cite{Cohen:2008nb}.

The largest freeze-out temperature in the stop coannihilation region of the CMSSM is $T_{\text{f.o.}} \simeq 87\GeV$ which means that this is not an important effect except at the highest mass boundaries of the stop coannihilation regions.  Include these modifications to the relic density calculation is beyond the scope of this work.

%%%%%%%%%%%%%%
%%%%%%%%%%%%%%
\clearpage
\stepcounter{model}
\tocless\subsubsection{\Cmodel{STOP}}
\begin{table}[h!]
\renewcommand{\arraystretch}{1.1}
\setlength{\tabcolsep}{4pt}
\begin{centering}
\begin{tabular}{|c|c|c|c|c||c|c|}
\hline
 \multicolumn{7}{|c|}{Input parameters} \\
 \hline
 \hline
$M_0$ & $M_\half$ & $A_0$ & $\tan\beta$ & $\text{sign}(\mu)$ & $|\mu|$ & $\sqrt{B_\mu}$ \\
\hline
2666.67 & 933.333 & -6444. & 8.52015 & -1 & 2794.86 & $18094.8$
\\
\hline
\end{tabular}\\
\vspace{5pt}
\begin{tabular}{|c|c|c|c|c|c|c|c||c|c||c|c|}
\hline
 \multicolumn{12}{|c|}{Low energy spectrum} \\
 \hline
 \hline
$m_{\tilde{g}}$ & $m_{\tilde{q}}$ & $m_{\tilde{t}_1}$& $m_{\tilde{\tau}_1}$ & $m_\chi$ & $m_{\chi_1^\pm}$ &  $m_h$ & $m_A$  & $\Omega\,h^2$ & $\sigma_\text{SI} \text{ [pb]}$ & $\Delta_{v} $ & $\Delta_{\Omega}$\\
\hline
2170 & 3200 & 446 & 2640 & 411 & 791 & 124 & 3880 & 0.116 & $2.06\times 10^{-13}$ & 4500  & 800  \\ 
\hline
\end{tabular}
\caption{Stop coannihilation benchmark.  Dimensionful values are in GeV unless otherwise stated.}
\label{Table: BM Stop 1}
\end{centering}
\end{table}

This benchmark provides an example of a point in the stop coannihilation region that is observable at the 13~TeV LHC.  The electroweakinos are nearly pure with masses and orderings given by
\begin{eqnarray}
\renewcommand{\arraystretch}{1.1}
\begin{array}{|c||c|c|c|c||c|c|}
\hline
&\widetilde{B}&\widetilde{W}&\multicolumn{2}{|c||}{\widetilde{H}}&\widetilde{W}&\widetilde{H}\\ 
\hline
& \chi_1^0 & \chi_2^0 & \chi_3^0  & \chi_4^0 & \chi_1^\pm& \chi_2^\pm \\
\hline
m\, [\text{GeV}] & 411 & 791 & 2830 & 2830 & 791 & 2840  \\ 
\hline
\end{array}
\end{eqnarray}

The squark spectrum is
\begin{eqnarray}
\begin{array}{|c||c|c|c|c|c|}
\hline
& \widetilde{q} & \widetilde{d}_3^c & \widetilde{q}_3 & \widetilde{u}_3^c \\
\hline
m\, [\text{TeV}] & 3.2 & 3.1 & 0.45 & 2.3 \\ 
\hline
\end{array}
\label{Eq: squarks Benchmark 5.2}
\end{eqnarray}

The lightest stop decays are
\begin{eqnarray}
\widetilde{t}_1 \rightarrow 
\begin{cases} 
\,\,c \, \chi^0_1& 69\%\\
\,\, b\,\left(W^+\right)^* \, \chi^0_1 &31\%
\end{cases}
\end{eqnarray}

The gluinos decay to one final state:
\be
\widetilde{g} \rightarrow \overline{t} \,\widetilde{t}_1 +\cc \quad 100\%
\ee

The production largest cross section is for $\widetilde{t}_1$ pair production.  At 13~TeV 
\be
\sigma(p\,p \rightarrow \widetilde{t}_1\,\widetilde{t}_1) = 0.96 \pb.
\ee
However, given that the stop and LSP are incredibly degenerate, the decay products would be extremely soft.  This process will be unobservable.  Furthermore, direct squark production is $\OO(10^{-2} \fb)$, and would therefore be difficult to observe.

The discovery mode will be either gluino pair production or gluino squark associated production:
\begin{eqnarray}
\sigma(p\,p \rightarrow \widetilde{g}\,\widetilde{g}) &=& 0.23 \fb\\
\sigma(p\,p \rightarrow \widetilde{g}\,\widetilde{q}) &=& 0.22 \fb.
\end{eqnarray}
Given the squark spectrum in Eq.~(\ref{Eq: squarks Benchmark 5.2}), it is clear that associated production will dominantly include the right handed up-type squarks.  

The squark decays are
\bea
\widetilde{q}_R&\rightarrow& q\,\widetilde{g}\quad \quad \,\,97\%\\
\widetilde{q}_L&\rightarrow& 
\begin{cases} 
q\,\widetilde{g}\quad \,\,\,88\%\\
q'\,\chi^+_1 \quad 8\%\\
q\,\chi^0_2\quad \,\,\,4\%\\
\end{cases}
\eea
Due to the stop-neutralino degeneracy, any $\widetilde{t}_1$ in the final state will manifest as missing energy.   The relevant electroweakino decays are
\bea
\chi^+_1 &\rightarrow& \widetilde{t}_1\,b\quad\quad\quad\quad 100\% \label{Eq: CharginoDecayStopBenchmark5.1}\\
\chi^0_2 &\rightarrow& \widetilde{t}_1\,\overline{t} +\cc \quad 100 \% \label{Eq: NeutralinoDecayStopBenchmark5.1}
\eea
Putting all of this together gives
\bea
\sigma(p\,p\rightarrow t\,t\, \MET\! X) = 0.22\fb\\
\sigma(p\,p\rightarrow t\,\overline{t}\,\MET\! X) = 0.22\fb
\eea
This benchmark motivates the study of a ``like-sign tops plus $\MET\!\!\!$" simplified model \cite{Kraml:2005kb} which is not currently being searched for at the LHC.

The direct production cross sections for $\chi_1^+\,\chi_1^-$ and $\chi_1^+\,\chi_2^0$ are $1.6 \fb$ and $2.6 \fb$ respectively.  The charged winos decay to $b+\MET\!\!$ and the neutral wino gives $t$+$\MET\!\!$.  The only electroweakino production signature which would be potentially observable is single top + $\MET\!\!$ with cross section of a few fb.  This is a very challenging signal to observe.

Tree level direct detection is very small
\be
\sigma_\mr{SI}^\mr{tree} = 2.1\times 10^{-13} \pb.
\ee

For this benchmark $A_t = -3795.85 \GeV$, $m_{\tilde{q}_3} = 2281.81 \GeV$, $m_{\tilde{u}_3} = 360.906 \GeV$ at $M_S = \sqrt{m_{\tilde{t}_1}\,m_{\tilde{t}_2}}$.  This implies that $A_t/m_{\tilde{u}_3} = 10.5$ and the 1-loop direct detection process described in Sec.~\ref{Sec: Stop1LoopDD} could be observable
\be
\sigma_\mr{SI}^{\mr{1-loop}} \sim 3 \times 10^{-11} \pb.
\ee
While the LHC will likely probe this model before direct detection experiments become sensitive to this cross section, $\OO(1)$ factors which could result from performing this calculation carefully could push this rate higher.  This motivates studying the contribution to direct detection at 1-loop in detail.

%%%%%%%%%%%%%%
%%%%%%%%%%%%%%
\clearpage
\stepcounter{model}
\tocless\subsubsection{\Cmodel{STOP}}
\begin{table}[h!]
\renewcommand{\arraystretch}{1.1}
\setlength{\tabcolsep}{4pt}
\begin{centering}
\begin{tabular}{|c|c|c|c|c||c|c|}
\hline
 \multicolumn{7}{|c|}{Input parameters} \\
 \hline
 \hline
$M_0$ & $M_\half$ & $A_0$ & $\tan\beta$ & $\text{sign}(\mu)$ & $|\mu|$ & $\text{sign}(B_\mu)\sqrt{|B_\mu|}$ \\
\hline
6250. & 2347.25 & 21477.3 & 17.1261 & -1 & 6512.64 & $-87608.2$\\
\hline
\end{tabular}\\
\vspace{5pt}
\begin{tabular}{|c|c|c|c|c|c|c|c||c|c||c|c|}
\hline
 \multicolumn{12}{|c|}{Low energy spectrum} \\
 \hline
 \hline
$m_{\tilde{g}}$ & $m_{\tilde{q}}$ & $m_{\tilde{t}_1}$& $m_{\tilde{\tau}_1}$ & $m_\chi$ & $m_{\chi_1^\pm}$ &  $m_h$ & $m_A$  & $\Omega\,h^2$ & $\sigma_\text{SI} \text{ [pb]}$ & $\Delta_{v} $ & $\Delta_{\Omega}$\\
\hline
5010 & 7480 & 1060 & 5660 & 1040 & 1950 & 126 & 8010 & 0.105 & $7.73\times 10^{-15}$ & 24000  & 1100  \\ 
\hline
\end{tabular}
\caption{Stop coannihilation benchmark.  Dimensionful values are in GeV unless otherwise stated.}
\label{Table: BM Stop2} 
\end{centering}
\end{table}

This benchmark serves as an example of something that is impossible to see at the 13~TeV LHC.  The gluino mass is 5 TeV and the squark masses are
\begin{eqnarray}
\begin{array}{|c||c|c|c|c|c|}
\hline
& \widetilde{q} & \widetilde{d}_3^c & \widetilde{q}_3 & \widetilde{u}_3^c \\
\hline
m\, [\text{TeV}] & 7.4  & 6.7 & 1.1 & 5.1 \\ 
\hline
\end{array}
\end{eqnarray}

Since this is a stop coannihilation point, the LSP has a similar mass to the lightest stop
\begin{eqnarray}
\renewcommand{\arraystretch}{1.1}
\begin{array}{|c||c|c|c|c||c|c|}
\hline
&\widetilde{B}&\widetilde{W}&\multicolumn{2}{|c||}{\widetilde{H}}&\widetilde{W}&\widetilde{H}\\ 
\hline
& \chi_1^0 & \chi_2^0 & \chi_3^0  & \chi_4^0 & \chi_1^\pm& \chi_2^\pm \\
\hline
m\, [\text{GeV}] & 1040 & 1950 & 6360 & 6360 & 1950 & 6360 \\ 
\hline
\end{array}
\end{eqnarray}

Tree level direct detection is beyond the reach of ton scale experiments:
\be
\sigma_\mr{SI}^\mr{tree} = 7.7\times 10^{-14} \pb.
\ee

Using $A_t =  3673.19 \GeV$, $m_{\tilde{q}_3} = 4981.14 \GeV$, $m_{\tilde{u}^c_3} =   1050.88 \GeV$ implies that $A_t/m_{\tilde{u}^c_3} = 3.5$, where all these parameters are evaluated at $\sqrt{m_{\tilde{t}_1}\,m_{\tilde{t}_2}}$.  An estimate of the 1-loop direct detection (see Sec.~\ref{Sec: Stop1LoopDD}) gives
\be
\sigma_\mr{SI}^{\mr{1-loop}} \sim 4 \times 10^{-12} \pb.
\ee
Given that $m_\chi \simeq 1 \TeV$, this will likely require multi-ton scale experiments to be observed.  However, further study is warranted to determine the precise value of this cross section.  

Taken together, this point provides an example of a CMSSM benchmark that will be extremely difficult to probe without an energy upgrade for the LHC.

%%%%%%%%%%%%%%%%%%%%%%%%%%%%%%%%%%%%%%%%%%%%%%%%%%%%%%
%%%%%%%%%%%%%%%%%%%%%%%%%%%%%%%%%%%%%%%%%%%%%%%%%%%%%%
%%%%%%%%%%%%%%%%%%%%%%%%%%%%%%%%%%%%%%%%%%%%%%%%%%%%%%
\clearpage
\section{Discussion}
\label{Sec: Discussion}
This article has mapped out the entire parameter space of the CMSSM {\em ansatz} in the post-Higgs discovery era.  
The constructed maps of the regions that are consistent with the measured values of the Higgs mass and dark matter relic density demonstrate that the CMSSM is compact.  The inputs can range from $\OO(100 \GeV)$ to $\OO(10 \TeV)$.   While the Giudice-Barbieri definition of fine-tuning indicates that the CMSSM is at least tuned to a part in 200, this quantity is bounded to be less than a part in 60,000.  

The near-term discovery or exclusion of the CMSSM shows an interesting interplay between the three common searches for physics beyond the Standard Model: direct production of superpartners, direct detection of the LSP, and indirect detection of the LSP annihilation products.  While it is not possible to fully exclude this {\em ansatz} by the end of the decade, a large portion of the CMSSM will be discovered or excluded.
Going through each of the five regions
\begin{itemize}
\item {\color{LightChi}Light $\chi$}: LHC~7 and LHC~8 --- completely excluded;
\item {\color{WT}Well-tempered}: multi-ton scale direct detection --- most likely discover or exclude;
\item {\color{A}$A^0$-pole annihilation}:  LHC~13, ton scale direct detection, and indirect detection --- some region will remain;
\item {\color{STAU}Stau coannihilation}:  LHC~13 and multi-ton direct detection --- most likely discover or exclude;
\item {\color{STOP}Stop coannihilation}:  LHC~13 and direct detection --- some region will remain.
\end{itemize}
After the full run of the 13~TeV LHC with $300\ifb$ and ton-scale direct detection only portions of the $A^0$-pole annihilation and stop coannihilation regions will go untouched.   

This article provided benchmarks and discussed a wide variety of the Simplified Models which can result from the CMSSM including some with the following features:
\begin{itemize}
\item gluino cascade decays involving heavy flavor and electroweak gauge bosons;
\item gluino cascade decays to heavy flavor and Higgs bosons;
\item electroweakino production resulting in boosted Higgs bosons;
\item colored production with stable charged particles in the final state;
\item same sign top production with missing energy.
\end{itemize}

This paper demonstrates a general philosophy that can be taken when attempting to exclude the entire parameter space of any restrictive slice of a model such as the MSSM.  The CMSSM serves as a nice example for demonstrating the importance of complimentary experiments --- it is our job to be sure we are looking under every possible rock as we search for the signs of beyond the Standard Model physics.

Going into the future, due to the compactness of the CMSSM parameter space, the masses of the heaviest particles are all beneath $30\TeV$ and the heaviest particles are colored.   Since all $R$-odd superparticles can be made through the decays of colored particles, it is possible to discover all of these states at a ``human-buildable," \emph{e.g.} $\sqrt{s}=100\TeV$, hadron collider in the foreseeable future.  Fortunately, for most of the parameter space, discoveries of physics beyond the Standard Model should have occurred beforehand.

\vspace{20pt}
\begin{center}
{\bf Acknowledgments}
\end{center}

We thank Ben Allanach,  Asimina Arvanitaki, Patrick Draper, Paolo Gondolo, Kiel Howe, Junwu Huang, Mariangela Lisanti, Aaron Pierce, Michael Peskin, David Sanford, and Hai Bo Yu for many helpful discussions, and Kiel Howe for the data to make the 10 event lines in the figures.
TC and JGW are supported by the US Department of Energy under contract number DE-AC02-76SF00515.
We thank the Stanford FarmShare cluster and especially Jason Bishop for providing the computing resources necessary to map out the CMSSM.

\appendix
\section{CMSSM Map Making Kit}
\label{Appendix: Map Making Kit}
This appendix  provides the details for utilizing the data files which are contained within the tarball for this paper on the arXiv.  These files contain one CMSSM input point  per cell (which will yield a Higgs mass in the range $122 \GeV \lsim m_h \lsim 128 \GeV$ and relic density in the range $0.08 \lsim \Omega\,h^2 \lsim 0.14$) for all figures which require explicit parameters.  Our goal is for anyone with a working version of \texttt{SoftSUSY} and \texttt{DarkSUSY} to explore the CMSSM themselves in order to find their own benchmark points.

\tocless\subsection{\texttt{SoftSUSY} Input Card.}
This is the input card we use to compute the spectra using \texttt{SoftSUSY} v3.3.7:

\vspace{5pt}
\includegraphics[width=0.9\textwidth]{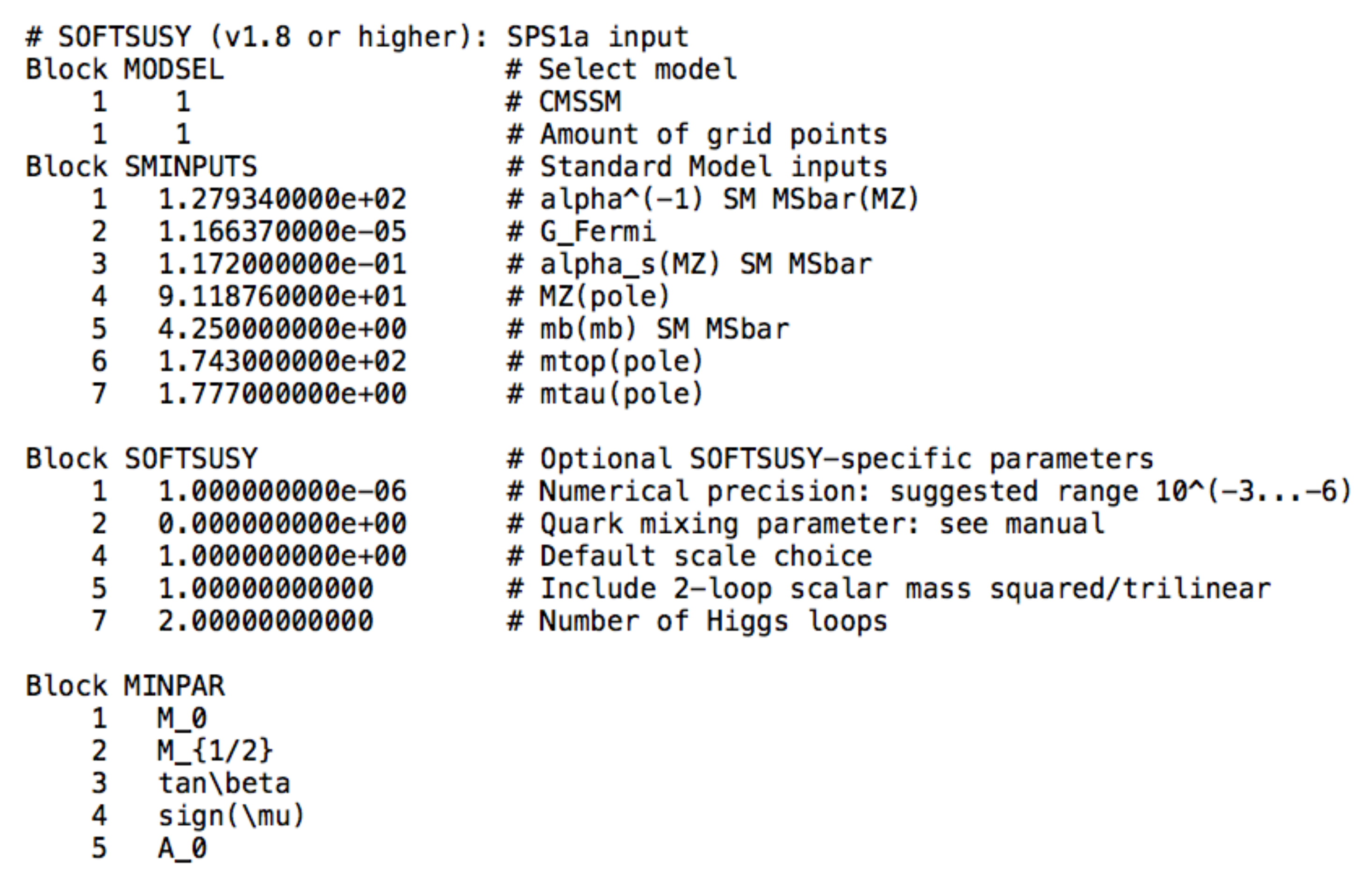}

The CMSSM inputs are specified by replacing the variables with the desired input values in the ``MINPAR" block.

\tocless\subsection{Format of Data Files}
Data files are available in the tarball on arXiv in connection to the preprint of this paper.  We have generated a data file corresponding to each region for all figures.  The files are named by the dark matter classification and the corresponding figure number.  The data is organized so that each row is one set of CMSSM inputs, separated by a comma and a space.  The order is
\be
M_0,\,M_\half,\,A_0,\,\text{sign}(\mu),\,\tan\beta \nonumber
\ee
The mass spectrum computed with \texttt{SoftSUSY} for all benchmarks, along with the cross sections computed with \texttt{Prospino}, and the decay tables computed with \texttt{SUSYHIT} for all LHC13 observable examples are also available on the arXiv.
\vspace{2pt}
\bibliography{CMSSM}% Produces the bibliography via BibTeX.

\providecommand{\href}[2]{#2}\begingroup\raggedright\begin{thebibliography}{100}

\bibitem{Aad:2012tfa}
{\bf ATLAS} Collaboration, G.~Aad {\em et al.}, ``{Observation of a new
  particle in the search for the Standard Model Higgs boson with the ATLAS
  detector at the LHC},''
  \href{http://dx.doi.org/10.1016/j.physletb.2012.08.020}{{\em Phys.Lett.} {\bf
  B716} (2012)  1--29},
\href{http://arxiv.org/abs/1207.7214}{{\tt arXiv:1207.7214 [hep-ex]}}.
%%CITATION = ARXIV:1207.7214;%%.

\bibitem{Chatrchyan:2012ufa}
{\bf CMS} Collaboration, S.~Chatrchyan {\em et al.}, ``{Observation of a new
  boson at a mass of 125 GeV with the CMS experiment at the LHC},''
  \href{http://dx.doi.org/10.1016/j.physletb.2012.08.021}{{\em Phys.Lett.} {\bf
  B716} (2012)  30--61},
\href{http://arxiv.org/abs/1207.7235}{{\tt arXiv:1207.7235 [hep-ex]}}.
%%CITATION = ARXIV:1207.7235;%%.

\bibitem{Ade:2013lta}
{\bf Planck} Collaboration, P.~Ade {\em et al.}, ``{Planck 2013 results. XVI.
  Cosmological parameters},''
\href{http://arxiv.org/abs/1303.5076}{{\tt arXiv:1303.5076 [astro-ph.CO]}}.
%%CITATION = ARXIV:1303.5076;%%.

\bibitem{Dimopoulos:1981zb}
S.~Dimopoulos and H.~Georgi, ``{Softly Broken Supersymmetry and SU(5)},''
\href{http://dx.doi.org/10.1016/0550-3213(81)90522-8}{{\em Nucl.Phys.} {\bf
  B193} (1981)  150}.
%%CITATION = NUPHA,B193,150;%%.

\bibitem{Dimopoulos:1981yj}
S.~Dimopoulos, S.~Raby, and F.~Wilczek, ``{Supersymmetry and the Scale of
  Unification},''
\href{http://dx.doi.org/10.1103/PhysRevD.24.1681}{{\em Phys.Rev.} {\bf D24}
  (1981)  1681--1683}.
%%CITATION = PHRVA,D24,1681;%%.

\bibitem{Drees:1992am}
M.~Drees and M.~M. Nojiri, ``{The Neutralino relic density in minimal $N=1$
  supergravity},'' \href{http://dx.doi.org/10.1103/PhysRevD.47.376}{{\em
  Phys.Rev.} {\bf D47} (1993)  376--408},
\href{http://arxiv.org/abs/hep-ph/9207234}{{\tt arXiv:hep-ph/9207234
  [hep-ph]}}.
%%CITATION = HEP-PH/9207234;%%.

\bibitem{Barger:1997kb}
V.~D. Barger and C.~Kao, ``{Relic density of neutralino dark matter in
  supergravity models},''
  \href{http://dx.doi.org/10.1103/PhysRevD.57.3131}{{\em Phys.Rev.} {\bf D57}
  (1998)  3131--3139},
\href{http://arxiv.org/abs/hep-ph/9704403}{{\tt arXiv:hep-ph/9704403
  [hep-ph]}}.
%%CITATION = HEP-PH/9704403;%%.

\bibitem{Okada:1990vk}
Y.~Okada, M.~Yamaguchi, and T.~Yanagida, ``{Upper bound of the lightest Higgs
  boson mass in the minimal supersymmetric standard model},''
\href{http://dx.doi.org/10.1143/PTP.85.1}{{\em Prog.Theor.Phys.} {\bf 85}
  (1991)  1--6}.
%%CITATION = PTPKA,85,1;%%.

\bibitem{Okada:1990gg}
Y.~Okada, M.~Yamaguchi, and T.~Yanagida, ``{Renormalization group analysis on
  the Higgs mass in the softly broken supersymmetric standard model},''
\href{http://dx.doi.org/10.1016/0370-2693(91)90642-4}{{\em Phys.Lett.} {\bf
  B262} (1991)  54--58}.
%%CITATION = PHLTA,B262,54;%%.

\bibitem{Haber:1993an}
H.~E. Haber and R.~Hempfling, ``{The Renormalization group improved Higgs
  sector of the minimal supersymmetric model},''
  \href{http://dx.doi.org/10.1103/PhysRevD.48.4280}{{\em Phys.Rev.} {\bf D48}
  (1993)  4280--4309},
\href{http://arxiv.org/abs/hep-ph/9307201}{{\tt arXiv:hep-ph/9307201
  [hep-ph]}}.
%%CITATION = HEP-PH/9307201;%%.

\bibitem{Haber:1996fp}
H.~E. Haber, R.~Hempfling, and A.~H. Hoang, ``{Approximating the radiatively
  corrected Higgs mass in the minimal supersymmetric model},''
  \href{http://dx.doi.org/10.1007/s002880050498}{{\em Z.Phys.} {\bf C75} (1997)
   539--554},
\href{http://arxiv.org/abs/hep-ph/9609331}{{\tt arXiv:hep-ph/9609331
  [hep-ph]}}.
%%CITATION = HEP-PH/9609331;%%.

\bibitem{Barate:2003sz}
{\bf LEP Working Group for Higgs boson searches, ALEPH , DELPHI, L3, OPAL}
  Collaboration, R.~Barate {\em et al.}, ``{Search for the standard model Higgs
  boson at LEP},'' \href{http://dx.doi.org/10.1016/S0370-2693(03)00614-2}{{\em
  Phys.Lett.} {\bf B565} (2003)  61--75},
\href{http://arxiv.org/abs/hep-ex/0306033}{{\tt arXiv:hep-ex/0306033
  [hep-ex]}}.
%%CITATION = HEP-EX/0306033;%%.

\bibitem{Hall:2011aa}
L.~J. Hall, D.~Pinner, and J.~T. Ruderman, ``{A Natural SUSY Higgs Near 126
  GeV},'' \href{http://dx.doi.org/10.1007/JHEP04(2012)131}{{\em JHEP} {\bf
  1204} (2012)  131},
\href{http://arxiv.org/abs/1112.2703}{{\tt arXiv:1112.2703 [hep-ph]}}.
%%CITATION = ARXIV:1112.2703;%%.

\bibitem{Carena:2011aa}
M.~Carena, S.~Gori, N.~R. Shah, and C.~E. Wagner, ``{A 125 GeV SM-like Higgs in
  the MSSM and the $\gamma \gamma$ rate},''
  \href{http://dx.doi.org/10.1007/JHEP03(2012)014}{{\em JHEP} {\bf 1203} (2012)
   014},
\href{http://arxiv.org/abs/1112.3336}{{\tt arXiv:1112.3336 [hep-ph]}}.
%%CITATION = ARXIV:1112.3336;%%.

\bibitem{Berger:2008cq}
C.~F. Berger, J.~S. Gainer, J.~L. Hewett, and T.~G. Rizzo, ``{Supersymmetry
  Without Prejudice},''
  \href{http://dx.doi.org/10.1088/1126-6708/2009/02/023}{{\em JHEP} {\bf 0902}
  (2009)  023},
\href{http://arxiv.org/abs/0812.0980}{{\tt arXiv:0812.0980 [hep-ph]}}.
%%CITATION = ARXIV:0812.0980;%%.

\bibitem{CahillRowley:2012rv}
M.~W. Cahill-Rowley, J.~L. Hewett, A.~Ismail, and T.~G. Rizzo, ``{The Higgs
  Sector and Fine-Tuning in the pMSSM},''
  \href{http://dx.doi.org/10.1103/PhysRevD.86.075015}{{\em Phys.Rev.} {\bf D86}
  (2012)  075015},
\href{http://arxiv.org/abs/1206.5800}{{\tt arXiv:1206.5800 [hep-ph]}}.
%%CITATION = ARXIV:1206.5800;%%.

\bibitem{Chamseddine:1982jx}
A.~H. Chamseddine, R.~L. Arnowitt, and P.~Nath, ``{Locally Supersymmetric Grand
  Unification},''
\href{http://dx.doi.org/10.1103/PhysRevLett.49.970}{{\em Phys.Rev.Lett.} {\bf
  49} (1982)  970}.
%%CITATION = PRLTA,49,970;%%.

\bibitem{Barbieri:1982eh}
R.~Barbieri, S.~Ferrara, and C.~A. Savoy, ``{Gauge Models with Spontaneously
  Broken Local Supersymmetry},''
\href{http://dx.doi.org/10.1016/0370-2693(82)90685-2}{{\em Phys.Lett.} {\bf
  B119} (1982)  343}.
%%CITATION = PHLTA,B119,343;%%.

\bibitem{Hall:1983iz}
L.~J. Hall, J.~D. Lykken, and S.~Weinberg, ``{Supergravity as the Messenger of
  Supersymmetry Breaking},''
\href{http://dx.doi.org/10.1103/PhysRevD.27.2359}{{\em Phys.Rev.} {\bf D27}
  (1983)  2359--2378}.
%%CITATION = PHRVA,D27,2359;%%.

\bibitem{Baer:2011ab}
H.~Baer, V.~Barger, and A.~Mustafayev, ``{Implications of a 125 GeV Higgs
  scalar for LHC SUSY and neutralino dark matter searches},''
  \href{http://dx.doi.org/10.1103/PhysRevD.85.075010}{{\em Phys.Rev.} {\bf D85}
  (2012)  075010},
\href{http://arxiv.org/abs/1112.3017}{{\tt arXiv:1112.3017 [hep-ph]}}.
%%CITATION = ARXIV:1112.3017;%%.

\bibitem{Baer:2012mv}
H.~Baer, V.~Barger, P.~Huang, D.~Mickelson, A.~Mustafayev, {\em et al.},
  ``{Post-LHC7 fine-tuning in the mSUGRA/CMSSM model with a 125 GeV Higgs
  boson},''
\href{http://arxiv.org/abs/1210.3019}{{\tt arXiv:1210.3019 [hep-ph]}}.
%%CITATION = ARXIV:1210.3019;%%.

\bibitem{Baer:2012uya}
H.~Baer, V.~Barger, and A.~Mustafayev, ``{Neutralino dark matter in
  mSUGRA/CMSSM with a 125 GeV light Higgs scalar},''
  \href{http://dx.doi.org/10.1007/JHEP05(2012)091}{{\em JHEP} {\bf 1205} (2012)
   091},
\href{http://arxiv.org/abs/1202.4038}{{\tt arXiv:1202.4038 [hep-ph]}}.
%%CITATION = ARXIV:1202.4038;%%.

\bibitem{Arbey:2011ab}
A.~Arbey, M.~Battaglia, A.~Djouadi, F.~Mahmoudi, and J.~Quevillon,
  ``{Implications of a 125 GeV Higgs for supersymmetric models},''
  \href{http://dx.doi.org/10.1016/j.physletb.2012.01.053}{{\em Phys.Lett.} {\bf
  B708} (2012)  162--169},
\href{http://arxiv.org/abs/1112.3028}{{\tt arXiv:1112.3028 [hep-ph]}}.
%%CITATION = ARXIV:1112.3028;%%.

\bibitem{Kadastik:2011aa}
M.~Kadastik, K.~Kannike, A.~Racioppi, and M.~Raidal, ``{Implications of the 125
  GeV Higgs boson for scalar dark matter and for the CMSSM phenomenology},''
  \href{http://dx.doi.org/10.1007/JHEP05(2012)061}{{\em JHEP} {\bf 1205} (2012)
   061},
\href{http://arxiv.org/abs/1112.3647}{{\tt arXiv:1112.3647 [hep-ph]}}.
%%CITATION = ARXIV:1112.3647;%%.

\bibitem{Buchmueller:2011ab}
O.~Buchmueller, R.~Cavanaugh, A.~De~Roeck, M.~Dolan, J.~Ellis, {\em et al.},
  ``{Higgs and Supersymmetry},''
  \href{http://dx.doi.org/10.1140/epjc/s10052-012-2020-3}{{\em Eur.Phys.J.}
  {\bf C72} (2012)  2020},
\href{http://arxiv.org/abs/1112.3564}{{\tt arXiv:1112.3564 [hep-ph]}}.
%%CITATION = ARXIV:1112.3564;%%.

\bibitem{Buchmueller:2012hv}
O.~Buchmueller, R.~Cavanaugh, M.~Citron, A.~De~Roeck, M.~Dolan, {\em et al.},
  ``{The CMSSM and NUHM1 in Light of 7 TeV LHC, $B_s$ to mu+mu- and XENON100
  Data},'' \href{http://dx.doi.org/10.1140/epjc/s10052-012-2243-3}{{\em
  Eur.Phys.J.} {\bf C72} (2012)  2243},
\href{http://arxiv.org/abs/1207.7315}{{\tt arXiv:1207.7315 [hep-ph]}}.
%%CITATION = ARXIV:1207.7315;%%.

\bibitem{Ellis:2012aa}
J.~Ellis and K.~A. Olive, ``{Revisiting the Higgs Mass and Dark Matter in the
  CMSSM},'' \href{http://dx.doi.org/10.1140/epjc/s10052-012-2005-2}{{\em
  Eur.Phys.J.} {\bf C72} (2012)  2005},
\href{http://arxiv.org/abs/1202.3262}{{\tt arXiv:1202.3262 [hep-ph]}}.
%%CITATION = ARXIV:1202.3262;%%.

\bibitem{Cao:2011sn}
J.~Cao, Z.~Heng, D.~Li, and J.~M. Yang, ``{Current experimental constraints on
  the lightest Higgs boson mass in the constrained MSSM},''
  \href{http://dx.doi.org/10.1016/j.physletb.2012.03.052}{{\em Phys.Lett.} {\bf
  B710} (2012)  665--670},
\href{http://arxiv.org/abs/1112.4391}{{\tt arXiv:1112.4391 [hep-ph]}}.
%%CITATION = ARXIV:1112.4391;%%.

\bibitem{Akula:2012kk}
S.~Akula, P.~Nath, and G.~Peim, ``{Implications of the Higgs Boson Discovery
  for mSUGRA},'' \href{http://dx.doi.org/10.1016/j.physletb.2012.09.007}{{\em
  Phys.Lett.} {\bf B717} (2012)  188--192},
\href{http://arxiv.org/abs/1207.1839}{{\tt arXiv:1207.1839 [hep-ph]}}.
%%CITATION = ARXIV:1207.1839;%%.

\bibitem{Fowlie:2012im}
A.~Fowlie, M.~Kazana, K.~Kowalska, S.~Munir, L.~Roszkowski, {\em et al.},
  ``{The CMSSM Favoring New Territories: The Impact of New LHC Limits and a 125
  GeV Higgs},'' \href{http://dx.doi.org/10.1103/PhysRevD.86.075010}{{\em
  Phys.Rev.} {\bf D86} (2012)  075010},
\href{http://arxiv.org/abs/1206.0264}{{\tt arXiv:1206.0264 [hep-ph]}}.
%%CITATION = ARXIV:1206.0264;%%.

\bibitem{Kowalska:2013hha}
K.~Kowalska, L.~Roszkowski, and E.~M. Sessolo, ``{Two ultimate tests of
  constrained supersymmetry},''
  \href{http://dx.doi.org/10.1007/JHEP06(2013)078}{{\em JHEP} {\bf 1306} (2013)
   078},
\href{http://arxiv.org/abs/1302.5956}{{\tt arXiv:1302.5956 [hep-ph]}}.
%%CITATION = ARXIV:1302.5956;%%.

\bibitem{Bertone:2011nj}
G.~Bertone, D.~G. Cerdeno, M.~Fornasa, R.~Ruiz~de Austri, C.~Strege, {\em et
  al.}, ``{Global fits of the cMSSM including the first LHC and XENON100
  data},'' \href{http://dx.doi.org/10.1088/1475-7516/2012/01/015}{{\em JCAP}
  {\bf 1201} (2012)  015},
\href{http://arxiv.org/abs/1107.1715}{{\tt arXiv:1107.1715 [hep-ph]}}.
%%CITATION = ARXIV:1107.1715;%%.

\bibitem{Strege:2011pk}
C.~Strege, G.~Bertone, D.~Cerdeno, M.~Fornasa, R.~Ruiz~de Austri, {\em et al.},
  ``{Updated global fits of the cMSSM including the latest LHC SUSY and Higgs
  searches and XENON100 data},''
  \href{http://dx.doi.org/10.1088/1475-7516/2012/03/030}{{\em JCAP} {\bf 1203}
  (2012)  030},
\href{http://arxiv.org/abs/1112.4192}{{\tt arXiv:1112.4192 [hep-ph]}}.
%%CITATION = ARXIV:1112.4192;%%.

\bibitem{Strege:2012bt}
C.~Strege, G.~Bertone, F.~Feroz, M.~Fornasa, R.~Ruiz~de Austri, {\em et al.},
  ``{Global Fits of the cMSSM and NUHM including the LHC Higgs discovery and
  new XENON100 constraints},''
  \href{http://dx.doi.org/10.1088/1475-7516/2013/04/013}{{\em JCAP} {\bf 1304}
  (2013)  013},
\href{http://arxiv.org/abs/1212.2636}{{\tt arXiv:1212.2636 [hep-ph]}}.
%%CITATION = ARXIV:1212.2636;%%.

\bibitem{Ghosh:2012dh}
D.~Ghosh, M.~Guchait, S.~Raychaudhuri, and D.~Sengupta, ``{How Constrained is
  the cMSSM?},'' \href{http://dx.doi.org/10.1103/PhysRevD.86.055007}{{\em
  Phys.Rev.} {\bf D86} (2012)  055007},
\href{http://arxiv.org/abs/1205.2283}{{\tt arXiv:1205.2283 [hep-ph]}}.
%%CITATION = ARXIV:1205.2283;%%.

\bibitem{Dighe:2013wfa}
A.~Dighe, D.~Ghosh, K.~M. Patel, and S.~Raychaudhuri, ``{Testing Times for
  Supersymmetry: Looking Under the Lamp Post},''
\href{http://arxiv.org/abs/1303.0721}{{\tt arXiv:1303.0721 [hep-ph]}}.
%%CITATION = ARXIV:1303.0721;%%.

\bibitem{Baltz:2004aw}
E.~A. Baltz and P.~Gondolo, ``{Markov chain Monte Carlo exploration of minimal
  supergravity with implications for dark matter},''
  \href{http://dx.doi.org/10.1088/1126-6708/2004/10/052}{{\em JHEP} {\bf 0410}
  (2004)  052},
\href{http://arxiv.org/abs/hep-ph/0407039}{{\tt arXiv:hep-ph/0407039
  [hep-ph]}}.
%%CITATION = HEP-PH/0407039;%%.

\bibitem{Kane:1993td}
G.~L. Kane, C.~F. Kolda, L.~Roszkowski, and J.~D. Wells, ``{Study of
  constrained minimal supersymmetry},''
  \href{http://dx.doi.org/10.1103/PhysRevD.49.6173}{{\em Phys.Rev.} {\bf D49}
  (1994)  6173--6210},
\href{http://arxiv.org/abs/hep-ph/9312272}{{\tt arXiv:hep-ph/9312272
  [hep-ph]}}.
%%CITATION = HEP-PH/9312272;%%.

\bibitem{Barbier:2004ez}
R.~Barbier, C.~Berat, M.~Besancon, M.~Chemtob, A.~Deandrea, {\em et al.},
  ``{R-parity violating supersymmetry},''
  \href{http://dx.doi.org/10.1016/j.physrep.2005.08.006}{{\em Phys.Rept.} {\bf
  420} (2005)  1--202},
\href{http://arxiv.org/abs/hep-ph/0406039}{{\tt arXiv:hep-ph/0406039
  [hep-ph]}}.
%%CITATION = HEP-PH/0406039;%%.

\bibitem{Moroi:1999zb}
T.~Moroi and L.~Randall, ``{Wino cold dark matter from anomaly mediated SUSY
  breaking},'' \href{http://dx.doi.org/10.1016/S0550-3213(99)00748-8}{{\em
  Nucl.Phys.} {\bf B570} (2000)  455--472},
\href{http://arxiv.org/abs/hep-ph/9906527}{{\tt arXiv:hep-ph/9906527
  [hep-ph]}}.
%%CITATION = HEP-PH/9906527;%%.

\bibitem{Gelmini:2006pw}
G.~B. Gelmini and P.~Gondolo, ``{Neutralino with the right cold dark matter
  abundance in (almost) any supersymmetric model},''
  \href{http://dx.doi.org/10.1103/PhysRevD.74.023510}{{\em Phys.Rev.} {\bf D74}
  (2006)  023510},
\href{http://arxiv.org/abs/hep-ph/0602230}{{\tt arXiv:hep-ph/0602230
  [hep-ph]}}.
%%CITATION = HEP-PH/0602230;%%.

\bibitem{Acharya:2009zt}
B.~S. Acharya, G.~Kane, S.~Watson, and P.~Kumar, ``{A Non-thermal WIMP
  Miracle},'' \href{http://dx.doi.org/10.1103/PhysRevD.80.083529}{{\em
  Phys.Rev.} {\bf D80} (2009)  083529},
\href{http://arxiv.org/abs/0908.2430}{{\tt arXiv:0908.2430 [astro-ph.CO]}}.
%%CITATION = ARXIV:0908.2430;%%.

\bibitem{ArkaniHamed:2006mb}
N.~Arkani-Hamed, A.~Delgado, and G.~Giudice, ``{The Well-tempered
  neutralino},'' \href{http://dx.doi.org/10.1016/j.nuclphysb.2006.02.010}{{\em
  Nucl.Phys.} {\bf B741} (2006)  108--130},
\href{http://arxiv.org/abs/hep-ph/0601041}{{\tt arXiv:hep-ph/0601041
  [hep-ph]}}.
%%CITATION = HEP-PH/0601041;%%.

\bibitem{Feng:1999mn}
J.~L. Feng, K.~T. Matchev, and T.~Moroi, ``{Multi - TeV scalars are natural in
  minimal supergravity},''
  \href{http://dx.doi.org/10.1103/PhysRevLett.84.2322}{{\em Phys.Rev.Lett.}
  {\bf 84} (2000)  2322--2325},
\href{http://arxiv.org/abs/hep-ph/9908309}{{\tt arXiv:hep-ph/9908309
  [hep-ph]}}.
%%CITATION = HEP-PH/9908309;%%.

\bibitem{Feng:2000gh}
J.~L. Feng, K.~T. Matchev, and F.~Wilczek, ``{Neutralino dark matter in focus
  point supersymmetry},''
  \href{http://dx.doi.org/10.1016/S0370-2693(00)00512-8}{{\em Phys.Lett.} {\bf
  B482} (2000)  388--399},
\href{http://arxiv.org/abs/hep-ph/0004043}{{\tt arXiv:hep-ph/0004043
  [hep-ph]}}.
%%CITATION = HEP-PH/0004043;%%.

\bibitem{Feng:2011aa}
J.~L. Feng, K.~T. Matchev, and D.~Sanford, ``{Focus Point Supersymmetry
  Redux},'' \href{http://dx.doi.org/10.1103/PhysRevD.85.075007}{{\em Phys.Rev.}
  {\bf D85} (2012)  075007},
\href{http://arxiv.org/abs/1112.3021}{{\tt arXiv:1112.3021 [hep-ph]}}.
%%CITATION = ARXIV:1112.3021;%%.

\bibitem{Feng:2012jfa}
J.~L. Feng and D.~Sanford, ``{A Natural 125 GeV Higgs Boson in the MSSM from
  Focus Point Supersymmetry with A-Terms},''
  \href{http://dx.doi.org/10.1103/PhysRevD.86.055015}{{\em Phys.Rev.} {\bf D86}
  (2012)  055015},
\href{http://arxiv.org/abs/1205.2372}{{\tt arXiv:1205.2372 [hep-ph]}}.
%%CITATION = ARXIV:1205.2372;%%.

\bibitem{Cohen:2010gj}
T.~Cohen, D.~J. Phalen, and A.~Pierce, ``{On the Correlation Between the
  Spin-Independent and Spin-Dependent Direct Detection of Dark Matter},''
  \href{http://dx.doi.org/10.1103/PhysRevD.81.116001}{{\em Phys.Rev.} {\bf D81}
  (2010)  116001},
\href{http://arxiv.org/abs/1001.3408}{{\tt arXiv:1001.3408 [hep-ph]}}.
%%CITATION = ARXIV:1001.3408;%%.

\bibitem{Cheung:2012qy}
C.~Cheung, L.~J. Hall, D.~Pinner, and J.~T. Ruderman, ``{Prospects and Blind
  Spots for Neutralino Dark Matter},''
\href{http://arxiv.org/abs/1211.4873}{{\tt arXiv:1211.4873 [hep-ph]}}.
%%CITATION = ARXIV:1211.4873;%%.

\bibitem{Draper:2013cka}
P.~Draper, J.~Feng, P.~Kant, S.~Profumo, and D.~Sanford, ``{Dark Matter
  Detection in Focus Point Supersymmetry},''
\href{http://arxiv.org/abs/1304.1159}{{\tt arXiv:1304.1159 [hep-ph]}}.
%%CITATION = ARXIV:1304.1159;%%.

\bibitem{Citron:2012fg}
M.~Citron, J.~Ellis, F.~Luo, J.~Marrouche, K.~Olive, {\em et al.}, ``{The End
  of the CMSSM Coannihilation Strip is Nigh},''
\href{http://arxiv.org/abs/1212.2886}{{\tt arXiv:1212.2886 [hep-ph]}}.
%%CITATION = ARXIV:1212.2886;%%.

\bibitem{Ellis:2001nx}
J.~R. Ellis, K.~A. Olive, and Y.~Santoso, ``{Calculations of neutralino stop
  coannihilation in the CMSSM},''
  \href{http://dx.doi.org/10.1016/S0927-6505(02)00151-2}{{\em Astropart.Phys.}
  {\bf 18} (2003)  395--432},
\href{http://arxiv.org/abs/hep-ph/0112113}{{\tt arXiv:hep-ph/0112113
  [hep-ph]}}.
%%CITATION = HEP-PH/0112113;%%.

\bibitem{Allanach:2005kz}
B.~Allanach and C.~Lester, ``{Multi-dimensional mSUGRA likelihood maps},''
  \href{http://dx.doi.org/10.1103/PhysRevD.73.015013}{{\em Phys.Rev.} {\bf D73}
  (2006)  015013},
\href{http://arxiv.org/abs/hep-ph/0507283}{{\tt arXiv:hep-ph/0507283
  [hep-ph]}}.
%%CITATION = HEP-PH/0507283;%%.

\bibitem{deAustri:2006pe}
R.~R. de~Austri, R.~Trotta, and L.~Roszkowski, ``{A Markov chain Monte Carlo
  analysis of the CMSSM},''
  \href{http://dx.doi.org/10.1088/1126-6708/2006/05/002}{{\em JHEP} {\bf 0605}
  (2006)  002},
\href{http://arxiv.org/abs/hep-ph/0602028}{{\tt arXiv:hep-ph/0602028
  [hep-ph]}}.
%%CITATION = HEP-PH/0602028;%%.

\bibitem{Akrami:2009hp}
Y.~Akrami, P.~Scott, J.~Edsjo, J.~Conrad, and L.~Bergstrom, ``{A Profile
  Likelihood Analysis of the Constrained MSSM with Genetic Algorithms},''
  \href{http://dx.doi.org/10.1007/JHEP04(2010)057}{{\em JHEP} {\bf 1004} (2010)
   057},
\href{http://arxiv.org/abs/0910.3950}{{\tt arXiv:0910.3950 [hep-ph]}}.
%%CITATION = ARXIV:0910.3950;%%.

\bibitem{Trotta:2008bp}
R.~Trotta, F.~Feroz, M.~P. Hobson, L.~Roszkowski, and R.~Ruiz~de Austri, ``{The
  Impact of priors and observables on parameter inferences in the Constrained
  MSSM},'' \href{http://dx.doi.org/10.1088/1126-6708/2008/12/024}{{\em JHEP}
  {\bf 0812} (2008)  024},
\href{http://arxiv.org/abs/0809.3792}{{\tt arXiv:0809.3792 [hep-ph]}}.
%%CITATION = ARXIV:0809.3792;%%.

\bibitem{Bridges:2010de}
M.~Bridges, K.~Cranmer, F.~Feroz, M.~Hobson, R.~R. de~Austri, {\em et al.},
  ``{A Coverage Study of the CMSSM Based on ATLAS Sensitivity Using Fast Neural
  Networks Techniques},'' \href{http://dx.doi.org/10.1007/JHEP03(2011)012}{{\em
  JHEP} {\bf 1103} (2011)  012},
\href{http://arxiv.org/abs/1011.4306}{{\tt arXiv:1011.4306 [hep-ph]}}.
%%CITATION = ARXIV:1011.4306;%%.

\bibitem{Feroz:2011bj}
F.~Feroz, K.~Cranmer, M.~Hobson, R.~Ruiz~de Austri, and R.~Trotta,
  ``{Challenges of Profile Likelihood Evaluation in Multi-Dimensional SUSY
  Scans},'' \href{http://dx.doi.org/10.1007/JHEP06(2011)042}{{\em JHEP} {\bf
  1106} (2011)  042},
\href{http://arxiv.org/abs/1101.3296}{{\tt arXiv:1101.3296 [hep-ph]}}.
%%CITATION = ARXIV:1101.3296;%%.

\bibitem{Buchmueller:2009fn}
O.~Buchmueller, R.~Cavanaugh, A.~De~Roeck, J.~Ellis, H.~Flacher, {\em et al.},
  ``{Likelihood Functions for Supersymmetric Observables in Frequentist
  Analyses of the CMSSM and NUHM1},''
  \href{http://dx.doi.org/10.1140/epjc/s10052-009-1159-z}{{\em Eur.Phys.J.}
  {\bf C64} (2009)  391--415},
\href{http://arxiv.org/abs/0907.5568}{{\tt arXiv:0907.5568 [hep-ph]}}.
%%CITATION = ARXIV:0907.5568;%%.

\bibitem{Allanach:2004rh}
B.~Allanach, A.~Djouadi, J.~Kneur, W.~Porod, and P.~Slavich, ``{Precise
  determination of the neutral Higgs boson masses in the MSSM},''
  \href{http://dx.doi.org/10.1088/1126-6708/2004/09/044}{{\em JHEP} {\bf 0409}
  (2004)  044},
\href{http://arxiv.org/abs/hep-ph/0406166}{{\tt arXiv:hep-ph/0406166
  [hep-ph]}}.
%%CITATION = HEP-PH/0406166;%%.

\bibitem{LEPSUSYWG/02-05.1}
{\bf LEP2 SUSY Working Group} Collaboration, S.~Ask {\em et al.}, ``Charginos,
  at large m0,'' Tech. Rep. LEPSUSYWG/02-05.1, CERN, 2004.

\bibitem{Allanach:2013cda}
B.~Allanach, D.~P. George, and B.~Gripaios, ``{The dark side of the $\mu$: on
  multiple solutions to renormalisation group equations, and why the CMSSM is
  not necessarily being ruled out},''
\href{http://arxiv.org/abs/1304.5462}{{\tt arXiv:1304.5462 [hep-ph]}}.
%%CITATION = ARXIV:1304.5462;%%.

\bibitem{Allanach:2001kg}
B.~Allanach, ``{SOFTSUSY: a program for calculating supersymmetric spectra},''
  \href{http://dx.doi.org/10.1016/S0010-4655(01)00460-X}{{\em
  Comput.Phys.Commun.} {\bf 143} (2002)  305--331},
\href{http://arxiv.org/abs/hep-ph/0104145}{{\tt arXiv:hep-ph/0104145
  [hep-ph]}}.
%%CITATION = HEP-PH/0104145;%%.

\bibitem{Pierce:1996zz}
D.~M. Pierce, J.~A. Bagger, K.~T. Matchev, and R.-j. Zhang, ``{Precision
  corrections in the minimal supersymmetric standard model},''
  \href{http://dx.doi.org/10.1016/S0550-3213(96)00683-9}{{\em Nucl.Phys.} {\bf
  B491} (1997)  3--67},
\href{http://arxiv.org/abs/hep-ph/9606211}{{\tt arXiv:hep-ph/9606211
  [hep-ph]}}.
%%CITATION = HEP-PH/9606211;%%.

\bibitem{Martin:2007pg}
S.~P. Martin, ``{Three-loop corrections to the lightest Higgs scalar boson mass
  in supersymmetry},'' \href{http://dx.doi.org/10.1103/PhysRevD.75.055005}{{\em
  Phys.Rev.} {\bf D75} (2007)  055005},
\href{http://arxiv.org/abs/hep-ph/0701051}{{\tt arXiv:hep-ph/0701051
  [hep-ph]}}.
%%CITATION = HEP-PH/0701051;%%.

\bibitem{Harlander:2008ju}
R.~Harlander, P.~Kant, L.~Mihaila, and M.~Steinhauser, ``{Higgs boson mass in
  supersymmetry to three loops},''
  \href{http://dx.doi.org/10.1103/PhysRevLett.101.039901,
  10.1103/PhysRevLett.100.191602}{{\em Phys.Rev.Lett.} {\bf 100} (2008)
  191602},
\href{http://arxiv.org/abs/0803.0672}{{\tt arXiv:0803.0672 [hep-ph]}}.
%%CITATION = ARXIV:0803.0672;%%.

\bibitem{Kant:2010tf}
P.~Kant, R.~Harlander, L.~Mihaila, and M.~Steinhauser, ``{Light MSSM Higgs
  boson mass to three-loop accuracy},''
  \href{http://dx.doi.org/10.1007/JHEP08(2010)104}{{\em JHEP} {\bf 1008} (2010)
   104},
\href{http://arxiv.org/abs/1005.5709}{{\tt arXiv:1005.5709 [hep-ph]}}.
%%CITATION = ARXIV:1005.5709;%%.

\bibitem{Gondolo:2004sc}
P.~Gondolo, J.~Edsjo, P.~Ullio, L.~Bergstrom, M.~Schelke, {\em et al.},
  ``{DarkSUSY: Computing supersymmetric dark matter properties numerically},''
  \href{http://dx.doi.org/10.1088/1475-7516/2004/07/008}{{\em JCAP} {\bf 0407}
  (2004)  008},
\href{http://arxiv.org/abs/astro-ph/0406204}{{\tt arXiv:astro-ph/0406204
  [astro-ph]}}.
%%CITATION = ASTRO-PH/0406204;%%.

\bibitem{Belanger:2010gh}
G.~Belanger, F.~Boudjema, P.~Brun, A.~Pukhov, S.~Rosier-Lees, {\em et al.},
  ``{Indirect search for dark matter with micrOMEGAs2.4},''
  \href{http://dx.doi.org/10.1016/j.cpc.2010.11.033}{{\em Comput.Phys.Commun.}
  {\bf 182} (2011)  842--856},
\href{http://arxiv.org/abs/1004.1092}{{\tt arXiv:1004.1092 [hep-ph]}}.
%%CITATION = ARXIV:1004.1092;%%.

\bibitem{Belanger:2004yn}
G.~Belanger, F.~Boudjema, A.~Pukhov, and A.~Semenov, ``{micrOMEGAs: Version
  1.3},'' \href{http://dx.doi.org/10.1016/j.cpc.2005.12.005}{{\em
  Comput.Phys.Commun.} {\bf 174} (2006)  577--604},
\href{http://arxiv.org/abs/hep-ph/0405253}{{\tt arXiv:hep-ph/0405253
  [hep-ph]}}.
%%CITATION = HEP-PH/0405253;%%.

\bibitem{Belanger:2001fz}
G.~Belanger, F.~Boudjema, A.~Pukhov, and A.~Semenov, ``{MicrOMEGAs: A Program
  for calculating the relic density in the MSSM},''
  \href{http://dx.doi.org/10.1016/S0010-4655(02)00596-9}{{\em
  Comput.Phys.Commun.} {\bf 149} (2002)  103--120},
\href{http://arxiv.org/abs/hep-ph/0112278}{{\tt arXiv:hep-ph/0112278
  [hep-ph]}}.
%%CITATION = HEP-PH/0112278;%%.

\bibitem{Djouadi:2006bz}
A.~Djouadi, M.~Muhlleitner, and M.~Spira, ``{Decays of supersymmetric
  particles: The Program SUSY-HIT (SUspect-SdecaY-Hdecay-InTerface)},'' {\em
  Acta Phys.Polon.} {\bf B38} (2007)  635--644,
\href{http://arxiv.org/abs/hep-ph/0609292}{{\tt arXiv:hep-ph/0609292
  [hep-ph]}}.
%%CITATION = HEP-PH/0609292;%%.

\bibitem{Martin:1997ns}
S.~P. Martin, ``{A Supersymmetry primer},''
\href{http://arxiv.org/abs/hep-ph/9709356}{{\tt arXiv:hep-ph/9709356
  [hep-ph]}}.
%%CITATION = HEP-PH/9709356;%%.

\bibitem{Casas:1997ze}
J.~A. Casas, ``{Charge and color breaking},''
\href{http://arxiv.org/abs/hep-ph/9707475}{{\tt arXiv:hep-ph/9707475
  [hep-ph]}}.
%%CITATION = HEP-PH/9707475;%%.

\bibitem{Abel:1998wr}
S.~Abel and T.~Falk, ``{Charge and color breaking in the constrained MSSM},''
  \href{http://dx.doi.org/10.1016/S0370-2693(98)01401-4}{{\em Phys.Lett.} {\bf
  B444} (1998)  427--434},
\href{http://arxiv.org/abs/hep-ph/9810297}{{\tt arXiv:hep-ph/9810297
  [hep-ph]}}.
%%CITATION = HEP-PH/9810297;%%.

\bibitem{Falk:1995cq}
T.~Falk, K.~A. Olive, L.~Roszkowski, and M.~Srednicki, ``{New constraints on
  superpartner masses},''
  \href{http://dx.doi.org/10.1016/0370-2693(95)01430-6}{{\em Phys.Lett.} {\bf
  B367} (1996)  183--187},
\href{http://arxiv.org/abs/hep-ph/9510308}{{\tt arXiv:hep-ph/9510308
  [hep-ph]}}.
%%CITATION = HEP-PH/9510308;%%.

\bibitem{Falk:1996zt}
T.~Falk, K.~A. Olive, L.~Roszkowski, A.~Singh, and M.~Srednicki, ``{Constraints
  from inflation and reheating on superpartner masses},''
  \href{http://dx.doi.org/10.1016/S0370-2693(97)00035-X}{{\em Phys.Lett.} {\bf
  B396} (1997)  50--57},
\href{http://arxiv.org/abs/hep-ph/9611325}{{\tt arXiv:hep-ph/9611325
  [hep-ph]}}.
%%CITATION = HEP-PH/9611325;%%.

\bibitem{Kusenko:1996jn}
A.~Kusenko, P.~Langacker, and G.~Segre, ``{Phase transitions and vacuum
  tunneling into charge and color breaking minima in the MSSM},''
  \href{http://dx.doi.org/10.1103/PhysRevD.54.5824}{{\em Phys.Rev.} {\bf D54}
  (1996)  5824--5834},
\href{http://arxiv.org/abs/hep-ph/9602414}{{\tt arXiv:hep-ph/9602414
  [hep-ph]}}.
%%CITATION = HEP-PH/9602414;%%.

\bibitem{Beenakker:1996ed}
W.~Beenakker, R.~Hopker, and M.~Spira, ``{PROSPINO: A Program for the
  production of supersymmetric particles in next-to-leading order QCD},''
\href{http://arxiv.org/abs/hep-ph/9611232}{{\tt arXiv:hep-ph/9611232
  [hep-ph]}}.
%%CITATION = HEP-PH/9611232;%%.

\bibitem{Beenakker:1996ch}
W.~Beenakker, R.~Hopker, M.~Spira, and P.~Zerwas, ``{Squark and gluino
  production at hadron colliders},''
  \href{http://dx.doi.org/10.1016/S0550-3213(97)00084-9}{{\em Nucl.Phys.} {\bf
  B492} (1997)  51--103},
\href{http://arxiv.org/abs/hep-ph/9610490}{{\tt arXiv:hep-ph/9610490
  [hep-ph]}}.
%%CITATION = HEP-PH/9610490;%%.

\bibitem{Beenakker:1997ut}
W.~Beenakker, M.~Kramer, T.~Plehn, M.~Spira, and P.~Zerwas, ``{Stop production
  at hadron colliders},''
  \href{http://dx.doi.org/10.1016/S0550-3213(98)00014-5}{{\em Nucl.Phys.} {\bf
  B515} (1998)  3--14},
\href{http://arxiv.org/abs/hep-ph/9710451}{{\tt arXiv:hep-ph/9710451
  [hep-ph]}}.
%%CITATION = HEP-PH/9710451;%%.

\bibitem{Beenakker:1999xh}
W.~Beenakker, M.~Klasen, M.~Kramer, T.~Plehn, M.~Spira, {\em et al.}, ``{The
  Production of charginos / neutralinos and sleptons at hadron colliders},''
  \href{http://dx.doi.org/10.1103/PhysRevLett.100.029901,
  10.1103/PhysRevLett.83.3780}{{\em Phys.Rev.Lett.} {\bf 83} (1999)
  3780--3783},
\href{http://arxiv.org/abs/hep-ph/9906298}{{\tt arXiv:hep-ph/9906298
  [hep-ph]}}.
%%CITATION = HEP-PH/9906298;%%.

\bibitem{Alwall:2008ve}
J.~Alwall, M.-P. Le, M.~Lisanti, and J.~G. Wacker, ``{Searching for Directly
  Decaying Gluinos at the Tevatron},''
  \href{http://dx.doi.org/10.1016/j.physletb.2008.06.065}{{\em Phys.Lett.} {\bf
  B666} (2008)  34--37},
\href{http://arxiv.org/abs/0803.0019}{{\tt arXiv:0803.0019 [hep-ph]}}.
%%CITATION = ARXIV:0803.0019;%%.

\bibitem{Alwall:2008va}
J.~Alwall, M.-P. Le, M.~Lisanti, and J.~G. Wacker, ``{Model-Independent Jets
  plus Missing Energy Searches},''
  \href{http://dx.doi.org/10.1103/PhysRevD.79.015005}{{\em Phys.Rev.} {\bf D79}
  (2009)  015005},
\href{http://arxiv.org/abs/0809.3264}{{\tt arXiv:0809.3264 [hep-ph]}}.
%%CITATION = ARXIV:0809.3264;%%.

\bibitem{Aprile:2002ef}
E.~Aprile, E.~Baltz, A.~Curioni, K.-L. Giboni, C.~Hailey, {\em et al.},
  ``{XENON: A 1 Tonne liquid xenon experiment for a sensitive dark matter
  search},''
\href{http://arxiv.org/abs/astro-ph/0207670}{{\tt arXiv:astro-ph/0207670
  [astro-ph]}}.
%%CITATION = ASTRO-PH/0207670;%%.

\bibitem{Li:2012zs}
H.~Gong, K.~Giboni, X.~Ji, A.~Tan, and L.~Zhao, ``{The Cryogenic System for the
  Panda-X Dark Matter Search Experiment},''
  \href{http://dx.doi.org/10.1088/1748-0221/8/01/P01002}{{\em JINST} {\bf 8}
  (2013)  P01002},
\href{http://arxiv.org/abs/1207.5100}{{\tt arXiv:1207.5100 [astro-ph.IM]}}.
%%CITATION = ARXIV:1207.5100;%%.

\bibitem{Akerib:2012ys}
{\bf LUX} Collaboration, D.~Akerib {\em et al.}, ``{The Large Underground Xenon
  (LUX) Experiment},'' \href{http://dx.doi.org/10.1016/j.nima.2012.11.135}{{\em
  Nucl.Instrum.Meth.} {\bf A704} (2013)  111--126},
\href{http://arxiv.org/abs/1211.3788}{{\tt arXiv:1211.3788 [physics.ins-det]}}.
%%CITATION = ARXIV:1211.3788;%%.

\bibitem{Brink:2005ej}
{\bf CDMS-II} Collaboration, P.~L. Brink {\em et al.}, ``{Beyond the CDMS-II
  dark matter search: SuperCDMS},'' {\em eConf} {\bf C041213} (2004)  2529,
\href{http://arxiv.org/abs/astro-ph/0503583}{{\tt arXiv:astro-ph/0503583
  [astro-ph]}}.
%%CITATION = ASTRO-PH/0503583;%%.

\bibitem{DMTools}
R.~Gaitskell and J.~Filippini, ``{DM Tools},''
{\em \url{http://dmtools.brown.edu/}}  .
%%CITATION = HEP-PH/9510308;%%.

\bibitem{Ellis:2008hf}
J.~R. Ellis, K.~A. Olive, and C.~Savage, ``{Hadronic Uncertainties in the
  Elastic Scattering of Supersymmetric Dark Matter},''
  \href{http://dx.doi.org/10.1103/PhysRevD.77.065026}{{\em Phys.Rev.} {\bf D77}
  (2008)  065026},
\href{http://arxiv.org/abs/0801.3656}{{\tt arXiv:0801.3656 [hep-ph]}}.
%%CITATION = ARXIV:0801.3656;%%.

\bibitem{Giedt:2009mr}
J.~Giedt, A.~W. Thomas, and R.~D. Young, ``{Dark matter, the CMSSM and lattice
  QCD},'' \href{http://dx.doi.org/10.1103/PhysRevLett.103.201802}{{\em
  Phys.Rev.Lett.} {\bf 103} (2009)  201802},
\href{http://arxiv.org/abs/0907.4177}{{\tt arXiv:0907.4177 [hep-ph]}}.
%%CITATION = ARXIV:0907.4177;%%.

\bibitem{Junnarkar:2013ac}
P.~Junnarkar and A.~Walker-Loud, ``{The Scalar Strange Content of the Nucleon
  from Lattice QCD},''
\href{http://arxiv.org/abs/1301.1114}{{\tt arXiv:1301.1114 [hep-lat]}}.
%%CITATION = ARXIV:1301.1114;%%.

\bibitem{Barbieri:1987fn}
R.~Barbieri and G.~Giudice, ``{Upper Bounds on Supersymmetric Particle
  Masses},''
\href{http://dx.doi.org/10.1016/0550-3213(88)90171-X}{{\em Nucl.Phys.} {\bf
  B306} (1988)  63}.
%%CITATION = NUPHA,B306,63;%%.

\bibitem{Arvanitaki:2012ps}
A.~Arvanitaki, N.~Craig, S.~Dimopoulos, and G.~Villadoro, ``{Mini-Split},''
  \href{http://dx.doi.org/10.1007/JHEP02(2013)126}{{\em JHEP} {\bf 1302} (2013)
   126},
\href{http://arxiv.org/abs/1210.0555}{{\tt arXiv:1210.0555 [hep-ph]}}.
%%CITATION = ARXIV:1210.0555;%%.

\bibitem{Aprile:2012nq}
{\bf XENON100} Collaboration, E.~Aprile {\em et al.}, ``{Dark Matter Results
  from 225 Live Days of XENON100 Data},''
  \href{http://dx.doi.org/10.1103/PhysRevLett.109.181301}{{\em Phys.Rev.Lett.}
  {\bf 109} (2012)  181301},
\href{http://arxiv.org/abs/1207.5988}{{\tt arXiv:1207.5988 [astro-ph.CO]}}.
%%CITATION = ARXIV:1207.5988;%%.

\bibitem{Aad:2012fqa}
{\bf ATLAS} Collaboration, G.~Aad {\em et al.}, ``{Search for squarks and
  gluinos with the ATLAS detector in final states with jets and missing
  transverse momentum using 4.7 fb$^{-1}$ of $\sqrt{s}=7$ TeV proton-proton
  collision data},'' \href{http://dx.doi.org/10.1103/PhysRevD.87.012008}{{\em
  Phys.Rev.} {\bf D87} (2013)  012008},
\href{http://arxiv.org/abs/1208.0949}{{\tt arXiv:1208.0949 [hep-ex]}}.
%%CITATION = ARXIV:1208.0949;%%.

\bibitem{Chatrchyan:2012qka}
{\bf CMS} Collaboration, S.~Chatrchyan {\em et al.}, ``{Search for physics
  beyond the standard model in events with a $Z$ boson, jets, and missing
  transverse energy in $pp$ collisions at $\sqrt{s}=7$ TeV},''
  \href{http://dx.doi.org/10.1016/j.physletb.2012.08.026}{{\em Phys.Lett.} {\bf
  B716} (2012)  260--284},
\href{http://arxiv.org/abs/1204.3774}{{\tt arXiv:1204.3774 [hep-ex]}}.
%%CITATION = ARXIV:1204.3774;%%.

\bibitem{CMS:2012yua}
{\bf CMS} Collaboration, S.~Chatrchyan {\em et al.}, ``{Search for
  supersymmetry with the razor variables at CMS},''
\href{http://arxiv.org/abs/CMS-PAS-SUS-12-005}{{\tt CMS-PAS-SUS-12-005}}.
%%CITATION = CMS-PAS-SUS-12-005 ETC.;%%.

\bibitem{Alves:2011sq}
D.~S. Alves, E.~Izaguirre, and J.~G. Wacker, ``{Where the Sidewalk Ends: Jets
  and Missing Energy Search Strategies for the 7 TeV LHC},''
  \href{http://dx.doi.org/10.1007/JHEP10(2011)012}{{\em JHEP} {\bf 1110} (2011)
   012},
\href{http://arxiv.org/abs/1102.5338}{{\tt arXiv:1102.5338 [hep-ph]}}.
%%CITATION = ARXIV:1102.5338;%%.

\bibitem{ATLAS:2013rla}
{\bf ATLAS} Collaboration, G.~Aad {\em et al.},
``{Search for direct production of charginos and neutralinos in events with
  three leptons and missing transverse momentum in 21$\,$fb$^{-1}$ of pp
  collisions at $\sqrt{s}=8\,$TeV with the ATLAS detector},''.
%%CITATION = ATLAS-CONF-2013-035 ETC.;%%.

\bibitem{Scott:2009jn}
P.~Scott, J.~Conrad, J.~Edsjo, L.~Bergstrom, C.~Farnier, {\em et al.},
  ``{Direct Constraints on Minimal Supersymmetry from Fermi-LAT Observations of
  the Dwarf Galaxy Segue 1},''
  \href{http://dx.doi.org/10.1088/1475-7516/2010/01/031}{{\em JCAP} {\bf 1001}
  (2010)  031},
\href{http://arxiv.org/abs/0909.3300}{{\tt arXiv:0909.3300 [astro-ph.CO]}}.
%%CITATION = ARXIV:0909.3300;%%.

\bibitem{Ripken:2010ja}
J.~Ripken, J.~Conrad, and P.~Scott, ``{Implications for constrained
  supersymmetry of combined H.E.S.S. observations of dwarf galaxies, the
  Galactic halo and the Galactic Centre},''
  \href{http://dx.doi.org/10.1088/1475-7516/2011/11/004}{{\em JCAP} {\bf 1111}
  (2011)  004},
\href{http://arxiv.org/abs/1012.3939}{{\tt arXiv:1012.3939 [astro-ph.HE]}}.
%%CITATION = ARXIV:1012.3939;%%.

\bibitem{Ackermann:2011wa}
{\bf Fermi-LAT} Collaboration, M.~Ackermann {\em et al.}, ``{Constraining Dark
  Matter Models from a Combined Analysis of Milky Way Satellites with the Fermi
  Large Area Telescope},''
  \href{http://dx.doi.org/10.1103/PhysRevLett.107.241302}{{\em Phys.Rev.Lett.}
  {\bf 107} (2011)  241302},
\href{http://arxiv.org/abs/1108.3546}{{\tt arXiv:1108.3546 [astro-ph.HE]}}.
%%CITATION = ARXIV:1108.3546;%%.

\bibitem{Hooper:2012sr}
D.~Hooper, C.~Kelso, and F.~S. Queiroz, ``{Stringent and Robust Constraints on
  the Dark Matter Annihilation Cross Section From the Region of the Galactic
  Center},''
\href{http://arxiv.org/abs/1209.3015}{{\tt arXiv:1209.3015 [astro-ph.HE]}}.
%%CITATION = ARXIV:1209.3015;%%.

\bibitem{Doro:2012xx}
{\bf CTA} Collaboration, M.~Doro {\em et al.}, ``{Dark Matter and Fundamental
  Physics with the Cherenkov Telescope Array},''
  \href{http://dx.doi.org/10.1016/j.astropartphys.2012.08.002}{{\em
  Astropart.Phys.} {\bf 43} (2013)  189--214},
\href{http://arxiv.org/abs/1208.5356}{{\tt arXiv:1208.5356 [astro-ph.IM]}}.
%%CITATION = ARXIV:1208.5356;%%.

\bibitem{Ripken:2012db}
J.~Ripken, A.~Cuoco, H.-S. Zechlin, J.~Conrad, and D.~Horns, ``{The sensitivity
  of Cherenkov telescopes to dark matter and astrophysically induced
  anisotropies in the diffuse gamma-ray background},''
\href{http://arxiv.org/abs/1211.6922}{{\tt arXiv:1211.6922 [astro-ph.HE]}}.
%%CITATION = ARXIV:1211.6922;%%.

\bibitem{Howe:2012xe}
K.~Howe and P.~Saraswat, ``{Excess Higgs Production in Neutralino Decays},''
  \href{http://dx.doi.org/10.1007/JHEP10(2012)065}{{\em JHEP} {\bf 1210} (2012)
   065},
\href{http://arxiv.org/abs/1208.1542}{{\tt arXiv:1208.1542 [hep-ph]}}.
%%CITATION = ARXIV:1208.1542;%%.

\bibitem{Arbey:2012fa}
A.~Arbey, M.~Battaglia, and F.~Mahmoudi, ``{Higgs Production in Neutralino
  Decays in the MSSM - The LHC and a Future e+e- Collider},''
\href{http://arxiv.org/abs/1212.6865}{{\tt arXiv:1212.6865 [hep-ph]}}.
%%CITATION = ARXIV:1212.6865;%%.

\bibitem{Baer:2012ts}
H.~Baer, V.~Barger, A.~Lessa, W.~Sreethawong, and X.~Tata, ``{Wh plus
  missing-$E_T$ signature from gaugino pair production at the LHC},''
  \href{http://dx.doi.org/10.1103/PhysRevD.85.055022}{{\em Phys.Rev.} {\bf D85}
  (2012)  055022},
\href{http://arxiv.org/abs/1201.2949}{{\tt arXiv:1201.2949 [hep-ph]}}.
%%CITATION = ARXIV:1201.2949;%%.

\bibitem{Ellis:1998kh}
J.~R. Ellis, T.~Falk, and K.~A. Olive, ``{Neutralino - Stau coannihilation and
  the cosmological upper limit on the mass of the lightest supersymmetric
  particle},'' \href{http://dx.doi.org/10.1016/S0370-2693(98)01392-6}{{\em
  Phys.Lett.} {\bf B444} (1998)  367--372},
\href{http://arxiv.org/abs/hep-ph/9810360}{{\tt arXiv:hep-ph/9810360
  [hep-ph]}}.
%%CITATION = HEP-PH/9810360;%%.

\bibitem{Arnowitt:2006jq}
R.~L. Arnowitt, B.~Dutta, T.~Kamon, N.~Kolev, and D.~A. Toback, ``{Detection of
  SUSY in the stau-neutralino coannihilation region at the LHC},''
  \href{http://dx.doi.org/10.1016/j.physletb.2006.05.090}{{\em Phys.Lett.} {\bf
  B639} (2006)  46--53},
\href{http://arxiv.org/abs/hep-ph/0603128}{{\tt arXiv:hep-ph/0603128
  [hep-ph]}}.
%%CITATION = HEP-PH/0603128;%%.

\bibitem{Aad:2012pra}
{\bf ATLAS} Collaboration, G.~Aad {\em et al.}, ``{Searches for heavy
  long-lived sleptons and R-Hadrons with the ATLAS detector in $pp$ collisions
  at $\sqrt{s}=7$ TeV},''
  \href{http://dx.doi.org/10.1016/j.physletb.2013.02.015}{{\em Phys.Lett.} {\bf
  B720} (2013)  277--308},
\href{http://arxiv.org/abs/1211.1597}{{\tt arXiv:1211.1597 [hep-ex]}}.
%%CITATION = ARXIV:1211.1597;%%.

\bibitem{Harz:2012fz}
J.~Harz, B.~Herrmann, M.~Klasen, K.~Kovarik, and Q.~L. Boulc'h,
  ``{Neutralino-stop co-annihilation into electroweak gauge and Higgs bosons at
  one loop},'' \href{http://dx.doi.org/10.1103/PhysRevD.87.054031}{{\em
  Phys.Rev.} {\bf D87} (2013)  054031},
\href{http://arxiv.org/abs/1212.5241}{{\tt arXiv:1212.5241 [hep-ph]}}.
%%CITATION = ARXIV:1212.5241;%%.

\bibitem{Drees:1992rr}
M.~Drees and M.~M. Nojiri, ``{New contributions to coherent neutralino -
  nucleus scattering},'' \href{http://dx.doi.org/10.1103/PhysRevD.47.4226}{{\em
  Phys.Rev.} {\bf D47} (1993)  4226--4232},
\href{http://arxiv.org/abs/hep-ph/9210272}{{\tt arXiv:hep-ph/9210272
  [hep-ph]}}.
%%CITATION = HEP-PH/9210272;%%.

\bibitem{Drees:1993bu}
M.~Drees and M.~Nojiri, ``{Neutralino - nucleon scattering revisited},''
  \href{http://dx.doi.org/10.1103/PhysRevD.48.3483}{{\em Phys.Rev.} {\bf D48}
  (1993)  3483--3501},
\href{http://arxiv.org/abs/hep-ph/9307208}{{\tt arXiv:hep-ph/9307208
  [hep-ph]}}.
%%CITATION = HEP-PH/9307208;%%.

\bibitem{Djouadi:2001kba}
A.~Djouadi, M.~Drees, P.~Fileviez~Perez, and M.~Muhlleitner, ``{Loop induced
  Higgs and Z boson couplings to neutralinos and implications for collider and
  dark matter searches},''
  \href{http://dx.doi.org/10.1103/PhysRevD.65.075016}{{\em Phys.Rev.} {\bf D65}
  (2002)  075016},
\href{http://arxiv.org/abs/hep-ph/0109283}{{\tt arXiv:hep-ph/0109283
  [hep-ph]}}.
%%CITATION = HEP-PH/0109283;%%.

\bibitem{D'Onofrio:2012ni}
M.~D'Onofrio, K.~Rummukainen, and A.~Tranberg, ``{The sphaleron rate at the
  electroweak crossover with 125 GeV Higgs mass},'' {\em PoS} {\bf LATTICE2012}
  (2012)  055,
\href{http://arxiv.org/abs/1212.3206}{{\tt arXiv:1212.3206 [hep-ph]}}.
%%CITATION = ARXIV:1212.3206;%%.

\bibitem{Cohen:2008nb}
T.~Cohen, D.~E. Morrissey, and A.~Pierce, ``{Changes in Dark Matter Properties
  After Freeze-Out},'' \href{http://dx.doi.org/10.1103/PhysRevD.78.111701}{{\em
  Phys.Rev.} {\bf D78} (2008)  111701},
\href{http://arxiv.org/abs/0808.3994}{{\tt arXiv:0808.3994 [hep-ph]}}.
%%CITATION = ARXIV:0808.3994;%%.

\bibitem{Kraml:2005kb}
S.~Kraml and A.~Raklev, ``{Same-sign top quarks as signature of light stops at
  the LHC},'' \href{http://dx.doi.org/10.1103/PhysRevD.73.075002}{{\em
  Phys.Rev.} {\bf D73} (2006)  075002},
\href{http://arxiv.org/abs/hep-ph/0512284}{{\tt arXiv:hep-ph/0512284
  [hep-ph]}}.
%%CITATION = HEP-PH/0512284;%%.

\end{thebibliography}\endgroup
\bibliographystyle{utphys}
\end{document}